\documentclass{aastex63}
\usepackage{subfigure}
\usepackage{url}
\usepackage{hyperref}
\usepackage{epsfig}
\usepackage{graphicx}
\usepackage{sidecap}
\usepackage{hanging}
\usepackage{color}
\usepackage{multirow}
\usepackage{enumerate}
\usepackage{natbib}
\usepackage{amssymb}
\usepackage{amsmath}
\usepackage[bottom]{footmisc}

\newcommand{\kepler}{{\it Kepler}}
\newcommand{\tess}{{\it TESS}}

\providecommand{\e}[1]{\ensuremath{\, \times \, 10^{#1}}}

\accepted{}
\shorttitle{\tess\ in the \kepler\ Field}
\shortauthors{}

\begin{document}

\title{\tess\ Observations of \kepler\ systems with Transit Timing Variations}
\correspondingauthor{Daniel Jontof-Hutter}
\email{djontofhutter@pacific.edu}

%%%%%%%%%%%%%%%%%%%%%%%%%%%%%%%%%%%%%%%%%%%%%%%%%%%%%%%%%%%%%%%%%%%%%%%%%%

\author[0000-0002-6227-7510]{Daniel Jontof-Hutter}
\affiliation{Department of Physics,
University of the Pacific,
Stockton, CA 95211, USA}

\author[0000-0002-4297-5506]{Paul A.\ Dalba} 
\altaffiliation{NSF Astronomy and Astrophysics Postdoctoral Fellow}
\affiliation{Department of Astronomy and Astrophysics, University of California, Santa Cruz, CA 95064, USA}
\affiliation{Department of Earth and Planetary Sciences, University of California, Riverside, CA 92521, USA}

\author[0000-0002-4881-3620]{John~H.~Livingston}
\affiliation{Department of Astronomy, University of Tokyo, 7-3-1 Hongo, Bunkyo-ku, Tokyo 113-0033, Japan}

%%%%%%%%%%%%%%%%%%%%%%%%%%%%%%%%%%%%%%%%%%%%%%%%%%%%%%%%%%%%%%%%%%%%%%%%%%

\begin{abstract}
We identify targets in the \kepler\ field that may be characterized by transit timing variations (TTVs) and are detectable by the Transiting Exoplanet Survey Satellite (\tess). Despite the reduced signal-to-noise ratio of \tess\ transits compared to \kepler, we recover 48 transits from 13 systems in Sectors 14, 15, 26, 40 and 41. We find strong evidence of a nontransiting perturber orbiting Kepler-396 (KOI-2672) and explore two possible cases of a third planet in that system that could explain the measured transit times. We update the ephemerides and mass constraints where possible at KOI-70 (Kepler-20), KOI-82 (Kepler-102), KOI-94 (Kepler-89), KOI-137 (Kepler-18), KOI-244 (Kepler-25), KOI-245 (Kepler-37), KOI-282 (Kepler-130), KOI-377 (Kepler-9), KOI-620 (Kepler-51), KOI-806 (Kepler-30), KOI-1353  (Kepler-289) and KOI-1783 (Kepler-1662).
\end{abstract}
%%%%%%%%%%%%%%%%%%%%%%%%%%%%%%%%%%%%%%%%%%%%%%%%%%%%%%%%%%%%%%%%%%%%%%%%%%
\section{Introduction}\label{sec:intro}
Transit timing has been of extraordinary value in characterizing the masses of exoplanets. Before the launch of the Transiting Exoplanet Survey Satellite \citep[(\tess;)][]{Ricker2015}, the majority of low-mass exoplanets with measured sizes and masses were among multitransiting systems characterized by observed transit timing variations (TTVs; \citealt{Hadden2014,Jontof-Hutter2016,Hadden2017,Jontof-Hutter2019,Jontof-Hutter2021}. These are due to coherent interactions between planets that cause measurable deviations from a constant orbital period \citep[e.g.,][]{Agol2005,Holman2005,Ford2011,Fabrycky2012,Mazeh2013,Steffen2013}. This windfall accrued from a combination of theoretical advances (\citealt{Nesvorny2008,Nesvorny2009,Lithwick2012,Agol2016}) and the abundance of compact multitransiting systems discovered by the \kepler\ mission (\citealt{Lissauer2011b,Ford2012,Fabrycky2014}). 

The signal-to-noise ratio (S/N) of TTVs increases with orbital period, which biases planetary mass detections to longer orbital periods if long-baseline light curves are taken \citep{Steffen2016}. Hence, the \kepler\ mission \citep{Borucki2010}, with its 4 yr near-continuous photometric observations, enabled the majority of low-mass exoplanet characterizations to date. Such characterizations require both interacting neighbors to be transiting to enable planetary masses and radii to be measured. Near first-order mean motion resonances, where orbital periods $P_{in}$  and $P_{out}$ are near the ratio $j:j-1$ for an integer $j$, the TTVs are dominated by a sinusoidal signal of periodicity given by
\begin{equation}
 P_{ttv} = \left| \frac{j-1}{P_{in}} - \frac{j}{P_{out}} \right|^{-1}
\label{eqn:Pttv}
\end{equation}
The majority of TTV mass characterizations are among compact multiplanet systems in the \kepler\ field, where TTVs are detected at the expected TTV periodicity (or super-period) given the orbital periods of the planets (\citealt{Lithwick2012,Hadden2014}). 

Both the TTV periodicity and the amplitude of the signal increase as the orbital period ratio approaches commensurability \citep{Lithwick2012}. This adds value to increasing the baseline of photometric observations. Some \kepler\ targets with TTVs are detectable from ground-based observatories \citep{vonEssen2018}. However, due to the long orbital periods and long transit durations of many of the most interesting systems, ground-based transit timing is inefficient. In some cases, only one of ingress or egress is recoverable in one night \citep[e.g.,][]{Dalba2016}. In addition, some of the candidates with long orbital periods have few transits during the season when the \kepler\ field is observable. 

Nevertheless, additional data can significantly improve the constraints of TTV models. Arguably, the most valuable follow-up data are where the periodicity of a large TTV signal is comparable to or exceeds the \kepler\ baseline (e.g. Kepler-29 and Kepler-177; \citealt{Vissapragada2020}). 

\tess\ solves some of the problems of ground-based follow-up with its 27-day sectors. So long as a target transits within the relevant time frame, its ingress and egress will likely be covered, enabling precise mid-transit timing measurements. To date, \tess\ has observed the \kepler\ field in part or in whole in Sectors 14, 15, 26, 40, and 41. \tess\ may provide useful transit photometry for a significant number of planets even though the transit S/N is much lower than in the \kepler\ data set. These may improve planetary mass measurements and refine ephemerides \citep{Battley2021}.

Our aim is to identify targets among \kepler's multiplanet systems where there are known TTVs or an expectation of TTVs such that \tess\ data may further characterize planetary masses and orbital parameters. In this section, we compile a list of TTV targets in the Kepler field where there is an expectation of detecting a transit in TESS data. In \S \ref{sec:obs}, we explain our light curve analysis and results, including our measured transit times. In \S \ref{sec:TTVmodels}, we present our analysis of measured transit times alongside \kepler\ data and characterize the planetary parameters. We discuss the results in \S \ref{sec:disc}. 

\subsection{System Selection}
We identified planets with known TTVs according to the transit timing catalog of \citet{Holczer2016}, and from this sample we selected for planets within multitransiting systems where the TTVs could be attributable to an interacting neighbor near a first-order or second-order mean motion resonance. For this, we calculated the period ratios of all adjacent planet pairs and the period ratios of nonadjacent planet pairs with one intermediate transiting planet from \textit{Kepler}'s multitransiting systems. We considered planet pairs to be near first-order resonance if the period ratio fell within a range bounded by the nearest third-order resonances to the first-order resonance. Outside of this domain, we considered period ratios up to 4:1 as nearer to second-order resonances. We excluded pairs with period ratios greater than 4:1, since their interactions are unlikely to cause any observed TTVs. These were at KOI-289, KOI-353, KOI-464, KOI-564, KOI-872, KOI-1781 and KOI-1792. These systems have detectable TTVs, but the known planets are likely noninteracting and hence TTV modeling will not characterize the masses of the transiting planets. These are distinct from systems where prior studies have considered the known planets to be sufficient for TTV modeling, but additional information from TTV or radial velocity (RV) implies the presence of nontransiting perturbers (see \S 3.3). 

We also excluded KOI-676 which has a pair near the 13:4 commensurability. The TTVs detected by \citet{Holczer2016} at the inner planet, KOI-676.01, are most likely caused by a nontransiting perturber.

In addition to the TTV systems identified by \citet{Holczer2016}, we added planets with an expected TTV periodicity $>$100 days that have no prior detection of TTVs. The TTV signals of planet pairs near first-order resonances are generally dominated by a single periodicity, given by Equation~\ref{eqn:Pttv}. Among pairs near second-order resonances, TTV signals have more components. For a period ratio near $j:j-2$, we calculated a periodicity that satisfies
\begin{equation}
\frac{1}{P_{ttv}} = \left|\frac{j-2}{P_{in}} - \frac{j}{P_{out}}\right|.  
\end{equation}

We determined which of these planet candidates were likely detectable in \tess\ light curves at the 3$\sigma$ level as follows. For a given planet transit depth $\delta_p$ and transit duration in hours, $T_{dur}$, we anticipated that
\begin{equation}
S/N_{TESS}  \approx \frac{\delta_p \sqrt{T_{dur}}}{\delta_{\rm{1 \: hr}}} 
\end{equation}
where $\delta_{\rm{1 \: hr}}$ is the \tess\ uncertainty for a 1 hr observation of the target estimated using the Web TESS Viewing Tool\footnote{\url{https://heasarc.gsfc.nasa.gov/cgi-bin/tess/webtess/wtv.py}}. We adopted the transit depth and duration measurements from \kepler\ DR25 for this calculation \citep{Thompson2018}. KOI-377.02 was excluded from DR25, and we adopted the values from \citet{Mullally2015}.

Finally, from this list we selected targets where transits were expected in the \tess\ sectors. Our targets are in Table~\ref{tbl-systemselection}. 

 \begin{table}[h!]
  \begin{center}
    \begin{tabular}{|c|c|c|c|c|c|c|}
      \hline
KOI  & P (days)  &   $\delta$ (ppm)    &  \tess\ $\delta_{\rm{1 \: hr}}$   &  T$_{dur}$ (hr)  &  Exp. S/N$_{TESS}$ & P$_{ttv}$ (days)  \\   
 \hline
94.01 & 22.343 &   5610 &    526.9 &    6.66  &   27.5  &   156 \\  
94.02 & 10.424 &   777 &    526.9 &    5.23  &   3.4  &   156 \\  
94.03 & 54.320 &   1980 &    526.9 &    8.59  &   11.0  &   126 \\  
137.01 & 7.642 &   2270 &    1270.2 &    3.41  &   3.3  &   268 \\  
137.02 & 14.859 &   3270 &    1270.2 &    3.53  &   4.8  &   268 \\  
152.01 & 52.091 &   2890 &    1618.7 &    8.64  &   5.2  &   721 \\  
244.01 & 12.720 &   1180 &    245.2 &    2.73  &   8.0  &   326 \\  
244.02 & 6.239 &   402 &    245.2 &    3.53  &   3.1  &   326 \\  
% We exclude KOI-277.01, Kep36c since expected SNR = 2.55
282.01 & 27.509 &   639 &    428.9 &    5.93  &   3.6  &   482 \\  %
%289.02 & 296.638 &   2760 &    639.5 &    16.43  &   17.5  &   0.0 \\  % Exclude P2/P1 ~10
%353.01 & 152.106 &   3690 &    1117.3 &    7.45  &   9.0  &   0.0 \\  % Exclude P2/p1 ~ 5
377.01 & 19.271 &  6661 &    1424.0 &     4.13  &  9.5 & $\approx$ 2000 \\  
377.02 & 38.908 &  6159 &    1424.0 &     4.52  &  9.2 & $\approx$ 2000 \\ 
%464.01 & 58.362 &   5370 &    2010.1 &    6.34  &   6.7  &   0.0 \\  % Exclude P2/P1 ~ 10
%564.02 & 127.905 &   4020 &    3146.1 &    14.47  &   4.9  &   0.0 \\ % Exclude P2/P1 ~ 6
620.01 & 45.155 &   6210 &    2691.8 &    5.78  &   5.5  &   1112 \\  
620.02 & 130.178 &   11600 &    2691.8 &    8.45  &   12.5  &   $\approx$ 2500 \\  
%676.01 & 7.973 &   3140 &    1447.1 &    2.98  &   3.7  &   31.9 \\  % exclude because Holczer Pttv = 738 days.
806.01 & 143.206 &   10600 &    4712.5 &    8.93  &   6.7  &   383 \\  
806.02 & 60.325 &   20300 &    4712.5 &    6.62  &   11.1  &   383 \\  
1353.01 & 125.865 &   12400 &    1786.0 &    9.02  &   20.9  &   874 \\  
1426.02 & 74.928 &   4120 &    1880.2 &    4.74  &   4.8  &   $\approx$ 70,000 \\  
1426.03 & 150.019 &   4320 &    1880.2 &    4.38  &   4.8  &   $\approx$ 70,000 \\  
1783.01 & 134.479 &   4030 &    1694.7 &    5.93  &   5.8  &   $\approx$ 2500 \\  
2672.01 & 88.512 &   2520 &    421.0 &    6.82  &   15.6  &   1506 \\  
2672.02 & 42.992 &   1120 &    421.0 &    4.73  &   5.8  &   1506 \\  
\hline % below are DR25, Pttv > 100, not in Holczer.
70.01 & 10.854 &   1030 &    590.8 &    3.78  &   3.4  &   171 \\  
82.01 & 16.146 &   941 &    318.1 &    3.72  &   5.7  &   123 \\  
245.01 & 39.792 &   610 &    135.9 &    4.47  &   9.5  &   302 \\  
351.01 & 331.597 &   8320 &    1470.1 &    14.40  &   21.5  &   $\approx$ 2200 \\  
351.02 & 210.601 &   4160 &    1470.1 &    11.99  &   9.8  &   $\approx$ 2200 \\
\hline
    \end{tabular}    
    \caption{Selected KOIs for Transit Timing with \tess. Note. The columns list the KOI, orbital period in days, nominal transit depth (from DR 25), estimated 1 hr photometric precision with \tess , nominal transit duration in hours, expected transit S/N and expected TTV period in days. The top section lists candidates with known TTVs from \citet{Holczer2016}. The bottom section lists additional candidates with an expected TTV period $>$100 days.}\label{tbl-systemselection}
  \end{center}
\end{table}

Most of these targets were anticipated by prior authors (e.g. \citealt{Christ2019,Goldberg2019}). Beyond targets identified by \citet{Christ2019}, our criteria for inclusion adds planets at KOI-245 (Kepler-37), KOI-282 (Kepler-130), KOI-351 (Kepler-90), KOI-1353 (Kepler-289) and KOI-1783 (Kepler-441). 

%%%%%%%%%%%%%%%%%%%%%%%%%%%%%%%%%%%%%%%%%%%%%%%%%%%%%%%%%%%%%%%%%%%%%%%%%%

\section{Observations}\label{sec:obs}

To date, \tess\ has observed the \kepler\ field during Sectors 14, 15, 26, 40, and 41. \kepler's TTV systems received different baselines of observation based on their position relative to \tess's pointing and the location of CCD gaps. For the systems that were observed for at least one sector, we accessed the target pixel file (TPF) data from the Mikulski Archive for Space Telescopes through the \texttt{lightkurve}\footnote{\url{https://docs.lightkurve.org/}} package \citep{Lightkurve2018}. TPF data are background-subtracted cutouts with the electron flux in the pixels containing and surrounding the target star \citep{Jenkins2016}. Most \kepler\ TTV systems were observed with 2-minute cadence under the Cycle 2 \tess\ Guest Investigator Program G022149 (PI: Jontof-Hutter). In several cases, a system did not receive 2-minute cadence owing to its proximity to the edge of the detector.  

\subsection{Trapezoidal Transit Modeling with Pixel-level Decorrelation}
The \tess\ mission has an emphasis on bright, nearby star systems \citep{Ricker2015}. Many of the \kepler\ TTV systems, on the other hand, are relatively faint and distant. Therefore, in many cases, the anticipated transit depths for our \kepler\ host stars are on the order of the uncertainty in their relative flux measurements from \tess. This presents a critical challenge to measuring the \tess\ transit times, and hence a careful treatment of the noise in the \tess\ photometry is warranted.

We employ pixel-level decorrelation \citep[PLD;][]{Deming2015} to simultaneously model the noise and the transit in the \tess\ photometry. PLD has been successfully used to model instrumental systematics in light curves stemming from motion of the stellar point-spread function (PSF) and intrapixel sensitivity variations \citep{Ingalls2012,Deming2015,Luger2016}. In PLD, the systematic intensity variations are modeled by treating the individual pixel light curves as basis vectors, which are obtained by applying a normalization to remove the noninstrumental signal and linearly combined, i.e.,
\begin{align*}
    M_\mathrm{PLD}^t(\boldsymbol{\alpha}) = \frac{\sum_{i=1}^n c_i P_i^t}{\sum_{i=1}^n P_i^t},
\end{align*}
where $M_\mathrm{PLD}$ is the PLD model (parameterized by $\boldsymbol{\alpha}$) at time $t$, the $P_i$ are the pixel light curves, and the $\boldsymbol{\alpha} = \{c_1, c_2, ..., c_n\}$ are the parameters of the PLD model; the number of basis vectors $n$ is determined by the number of pixels in the optimal \tess\ aperture.

PLD was initially designed to account for noise in light curves from the {\it Spitzer Space Telescope} \citep{Deming2015}. It is similarly applicable to \tess\ observations in which large pixels undersample the stellar PSF. The PLD coefficients are fit to the data simultaneously with the parameters of the trapezoidal transit model, thus allowing for error propagation. In the following explanation of our methods, we will show the data from the Kepler-396 system as a representative example of the full population of \kepler\ TTV systems. 

\begin{figure}
    \centering
    \includegraphics[scale=0.5]{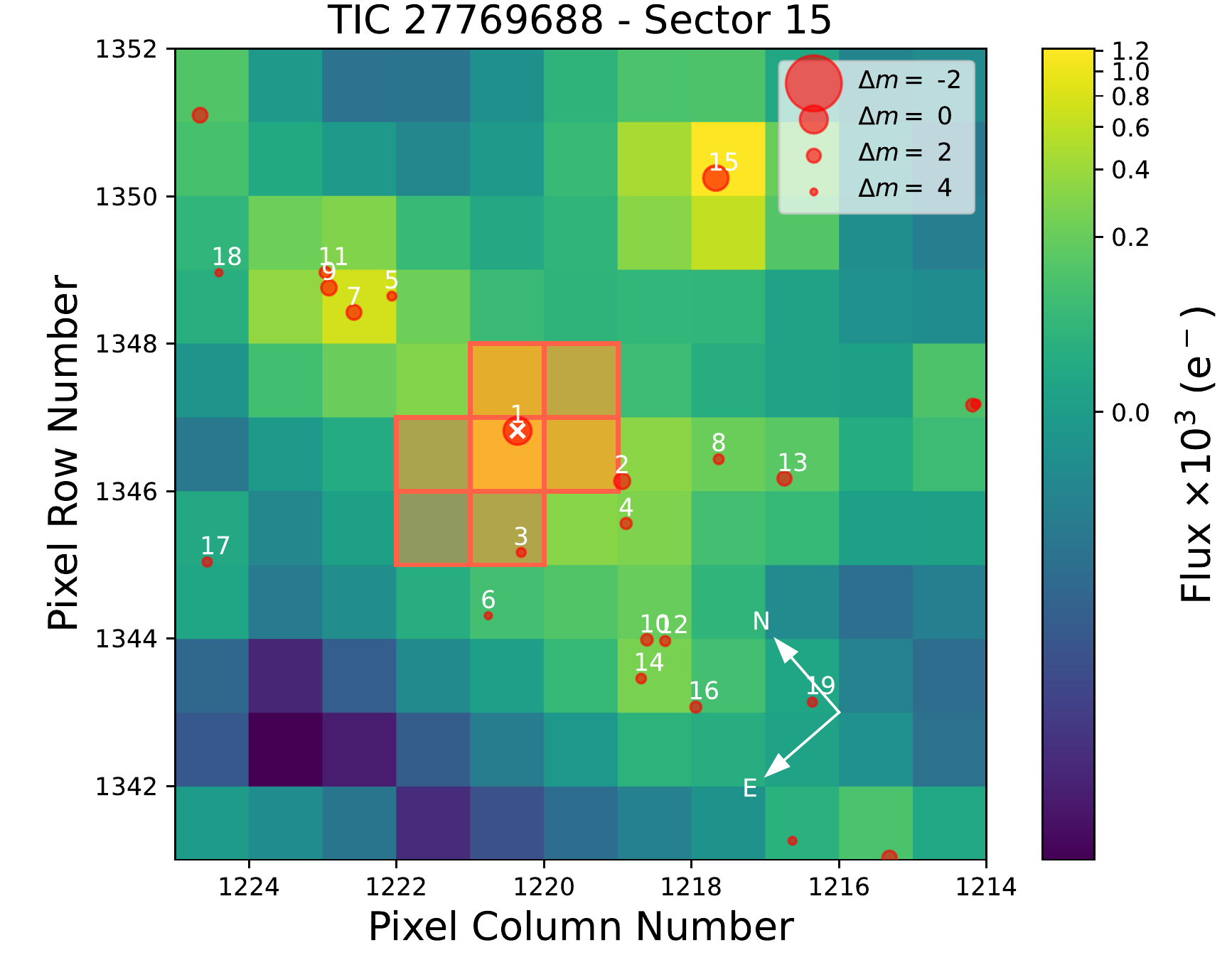}
    \caption{TPF for Kepler-396 from Sector~15 generated with \texttt{tpfplotter} \citep{Aller2020}. The numbered red circles are background stars identified by \textit{Gaia} with size corresponding to the magnitude difference compared to Kepler-396 (marked with a cross and numbered ``1''). The red squares identify the pixels in our custom aperture, which minimizes contamination from a nearby source. These pixels' light curves---normalized to their sum---formed the set of basis vectors for the PLD analysis of this system.}
    \label{fig:Kep396_tpf}
\end{figure}

For each target in each sector it was observed, we begin by identifying the pixels in the default photometric aperture as calculated by the Science Processing Operations Center (SPOC) data pipeline \citep{Jenkins2016}. The TPF for Kepler-396 from Sector~15 generated with the \texttt{tpfplotter}\footnote{\url{https://github.com/jlillo/tpfplotter}} tool \citep{Aller2020} is shown in Figure~\ref{fig:Kep396_tpf}. We inspected the aperture for dilution from nearby stars known to exist in the Gaia DR2 catalog \citep{Gaia2018}. In a few cases, we altered the default aperture to reduce contamination from background stars, especially if the transit depth we measured was much shallower than expected. 
 
After masking data during the expected transit time, we identified and removed outlier cadences from the TPF time series by generating a light curve using the default aperture photometry setting in \texttt{lightkurve}. For most systems, we trimmed the data to only include cadences within $\pm$0.5~days of the expected transit. This window was extended in a few cases (including Kepler-396~c) to allow for larger-than-expected TTVs. We then extracted the raw, background-subtracted light curves in each pixel using our photometric aperture. We normalized these light curves to their sum such that each cadence summed to unity. We also included an additional basis vector that was a linearly increasing function of time. The linear combination of the normalized basis vectors, weighted by coefficients $c_i$, where $i$ denotes the $i$th pixel, produced the noise model for each star's light curve. Since the aperture size varied by target and sector, the number of $c_i$ coefficients (and thus free parameters) was variable. 

The S/N in \tess\ data for any transit of a KOI is much lower than the S/N in \kepler. Since low S/N is especially prohibitive of detecting limb darkening, we chose to model the transit signal with a simple four-parameter trapezoid model: $T_C$ is the mid-transit time, $\delta$ is the transit depth, $T_{14}$ is the total transit time (first to fourth contact), and $T_{23}$ is the full transit time (second to third contact). The trapezoidal representation of the transit represented a sensible balance between simplicity and functionality and allowed us to place priors on the duration of each transit.

\subsection{Priors on Trapezoidal Transit Parameters}
We placed a uniform prior on the mid-transit time of width $\pm$0.25~days from the projected transit times of \citet{Jontof-Hutter2021} where available, or an extrapolation to a linear fit to the \kepler\ transit times. In some cases (such as Kepler-396c) the window for a detection in the \tess\ light curve was increased to $\pm$1~day. 

To detect the transits of known planets in \tess\ light curves, we imposed a strict prior on transit duration from \kepler\ data. Having a narrow prior on the duration of any transit meant that we did not generally improve on the duration measurements made from the \kepler\ data. However, given that our goal was to measure the mid-transit time and that the quality of the \kepler\ data for our faint targets always exceeded that of \tess, this was an acceptable choice.

Assuming low eccentricity, transit durations are given by 
\begin{equation}
T_{23} = \frac{P}{\pi}\sin^{-1}\left[\frac{R_{\star}}{a} \frac{\sqrt{(1-k)^2-b^2}}{\sin i}\right]
\label{eqn:T23} 
\end{equation}
and
\begin{equation}
T_{14} = \frac{P}{\pi}\sin^{-1}\left[\frac{R_{\star}}{a} \frac{\sqrt{(1+k)^2-b^2}}{\sin i}\right]
\label{eqn:T14} 
\end{equation}
where $P$ is the orbital period, $R_{\star}/a$ is the radius of the star scaled by the orbital semi-major axis, $k = R_{p}/R_{\star}$, $b$ is the impact parameter, and $i = \cos^{-1}(b R_{\star}/a)$ is the inclination \citep{Winn2010}. For our trapezoidal transit fits, our free parameters were $T_{14}$ and the ratio the transit durations $f_T\equiv T_{23}/T_{14}$, with priors based on published measurements of $R_{\star}/a$, $k$ and $b$ with their reported uncertainties (assumed to be Gaussian) from \textit{Kepler} DR25. 

For asymmetric errors, we adopted the larger of the two as the width of our priors. However, in many cases impact parameters are reported with asymmetric errors, with effectively an upper bound as the only constraint. For these, we assumed a uniform distribution between zero and the implied ``2$\sigma$" upper bound.

For example, DR25 reports the impact parameter of Kepler-25 c as $b = 0.787 ^{+0.051}_{-0.401}$, suggesting that while the bounds are uncertain, Kepler-25 c has a high impact parameter, moderately inconsistent with zero. A uniform distribution between zero and the upper limit would thus give too much weight to low values for the impact parameter. However, \kepler\ DR24 \citep{Coughlin2016} report $b = 0.1062 ^{+0.2045}_{-0.1062}$. We note that TTVs could cause erroneously high estimates of impact parameter since stacking the transits would increase the inferred ingress and egress time in the combined light curve. This is unlikely to be a factor for Kepler-25 c since the TTVs have an amplitude less than 5 minutes and are insignificant compared to the transit duration of $\approx 2.7$ hr. For this planet, we adopted a uniform distribution up to the upper bound given by DR25. 

Kepler-9 c (KOI-377.02) was not included in DR25 or DR24. We have adopted the relevant parameters from the Q1--Q16 \textit{Kepler} catalog of \citet{Mullally2015}.

In cases like KOI-94.03, where $R_{\star}/a$ is poorly constrained, we note that $(R_{\star}/a)^3 << 1$ and adopt a leading-order approximation for the inverse sine function, in which the ratio:
\begin{equation}
\frac{T_{23}}{T_{14}} \approx \sqrt{\frac{(1-k)^2-b^2}{(1+k)^2-b^2}}
\label{eqn:Tdur_approx}
\end{equation}
This eliminates the uncertainties from $R_{\star}/a$ and $\sin i$. 

For KOI-1426.03, the impact parameter reported in DR25 $b = 1.25 ^{+67.03}_{-0.17}$ is likely skewed by TTVs. This also affects the measured scaled planetary radius $k$. In estimating priors for transit duration, we assumed $k \approx \sqrt{\delta}$, and drew samples of $b \sim U[0,1]$.

For the PLD coefficients $c_i$, we allowed the sampler to explore all values between positive and negative infinity. Lastly, for transit depth $\delta$, we placed a uniform prior between 0 and 1. We generally did not attempt to correct for transit depth dilution except in a few cases in which we altered the photometric aperture. 

\subsection{Transit Recovery and Model Fitting}
We added the trapezoid model to the noise model to construct the final model light curve, which we fit to the data using the \texttt{emcee}\footnote{\url{https://github.com/dfm/emcee}} ensemble sampler \citep{ForemanMackey2013}. We initiated the fit with all PLD coefficients at zero and all transit parameters at the expected values based on \kepler\ DR25. We allowed 100 chains to explore the parameter space for 10,000 steps each for a total of $1\times10^6$ evaluations. We determined the burn-in by visual inspection, finding 2000 steps to usually be more than enough time for the chains to identify the region of highest likelihood. 

\begin{figure*}
    \centering
    \includegraphics[width=0.85\textwidth]{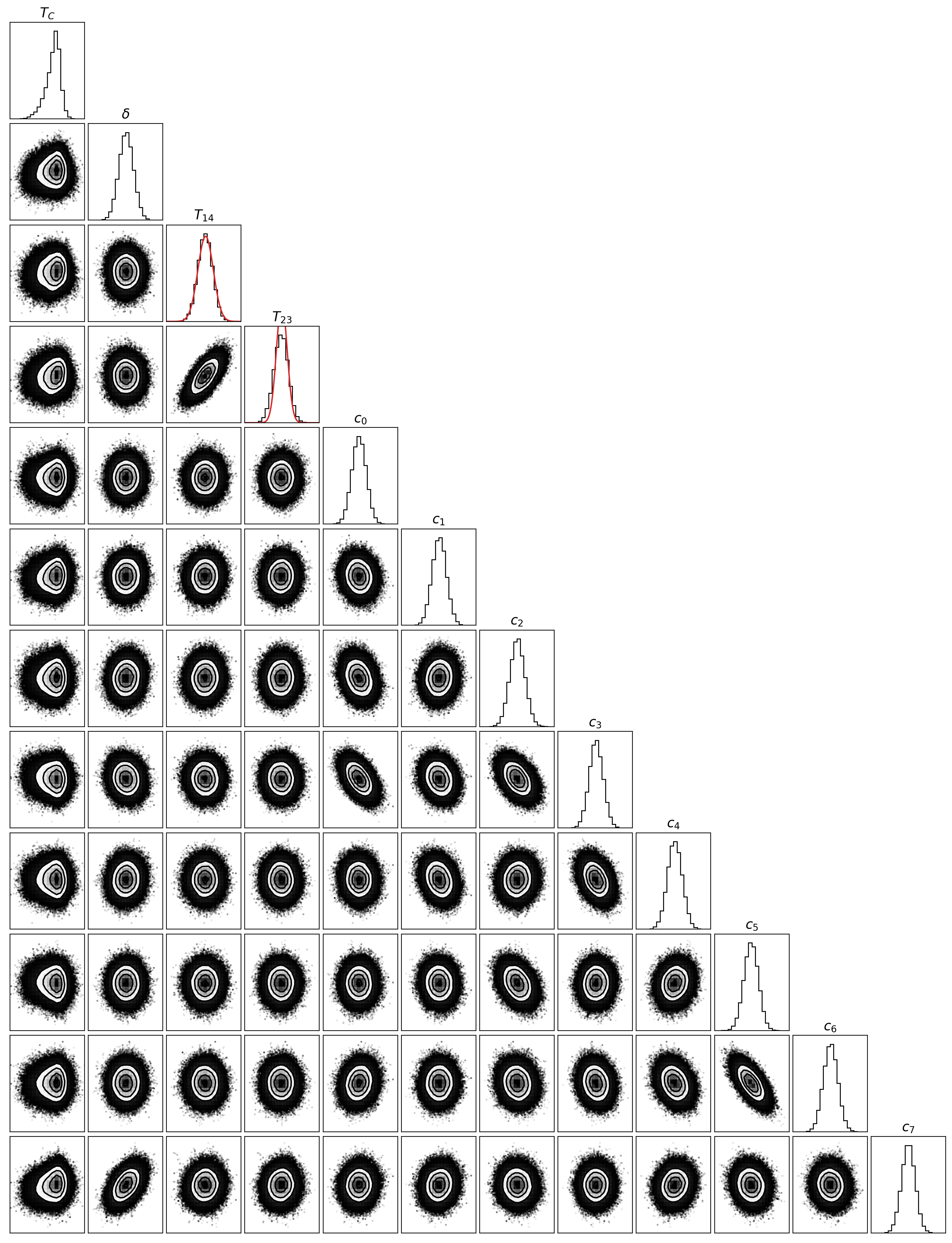}
    \caption{Posterior probability distributions for the PLD and trapezoidal transit fit to the Sector~15 Kepler-396~c transit. The first four columns are the trapezoid parameters, while the remaining columns are PLD basis vector coefficients. The red curves denote the normal priors on $T_{14}$ and $T_{23}$, the latter of which is transformed from the prior on $f_T$. These posteriors correspond to the fit shown in Figure~\ref{fig:Kep396c_models}.}
    \label{fig:Kep396c_post}
\end{figure*}

To illustrate our method for identifying and characterizing transit events in \tess\ data, we show our light-curve analysis for Kepler-396. In Figure~\ref{fig:Kep396c_post}, we show the full posterior distribution for all parameters modeled for Kepler-396~c. These posteriors are representative of other fits for which a clear transit detection was made. The posteriors for the PLD coefficients are normal distributions centered on zero. The transit duration parameters broadly follow their strict normal priors. The depth posterior is normal and well separated from zero. Lastly, the posterior for mid-transit time $T_{C}$ is unimodal and roughly normal, although it can be skewed as is the case for Kepler-396~c. 

In systems with weaker detections than shown for Kepler-396~c, the posterior on $T_C$ was usually broader, was less normal, and sometimes approached the edge of the uniform prior. In these cases, the depth posteriors were sometimes consistent with zero to $\sim3\sigma$. In systems with nondetections, the depth posteriors were typically within $2\sigma$ of zero and the $T_C$ posterior was either multimodal or flat but spanned the entire width of the prior. In weak detections and nondetections, the posteriors of the PLD coefficients were always normal (i.e., we could always fit a noise model even if a transit could not be detected). 

\begin{figure*}
  \centering
    \begin{tabular}{ccc}
      \includegraphics[width=5.6cm]{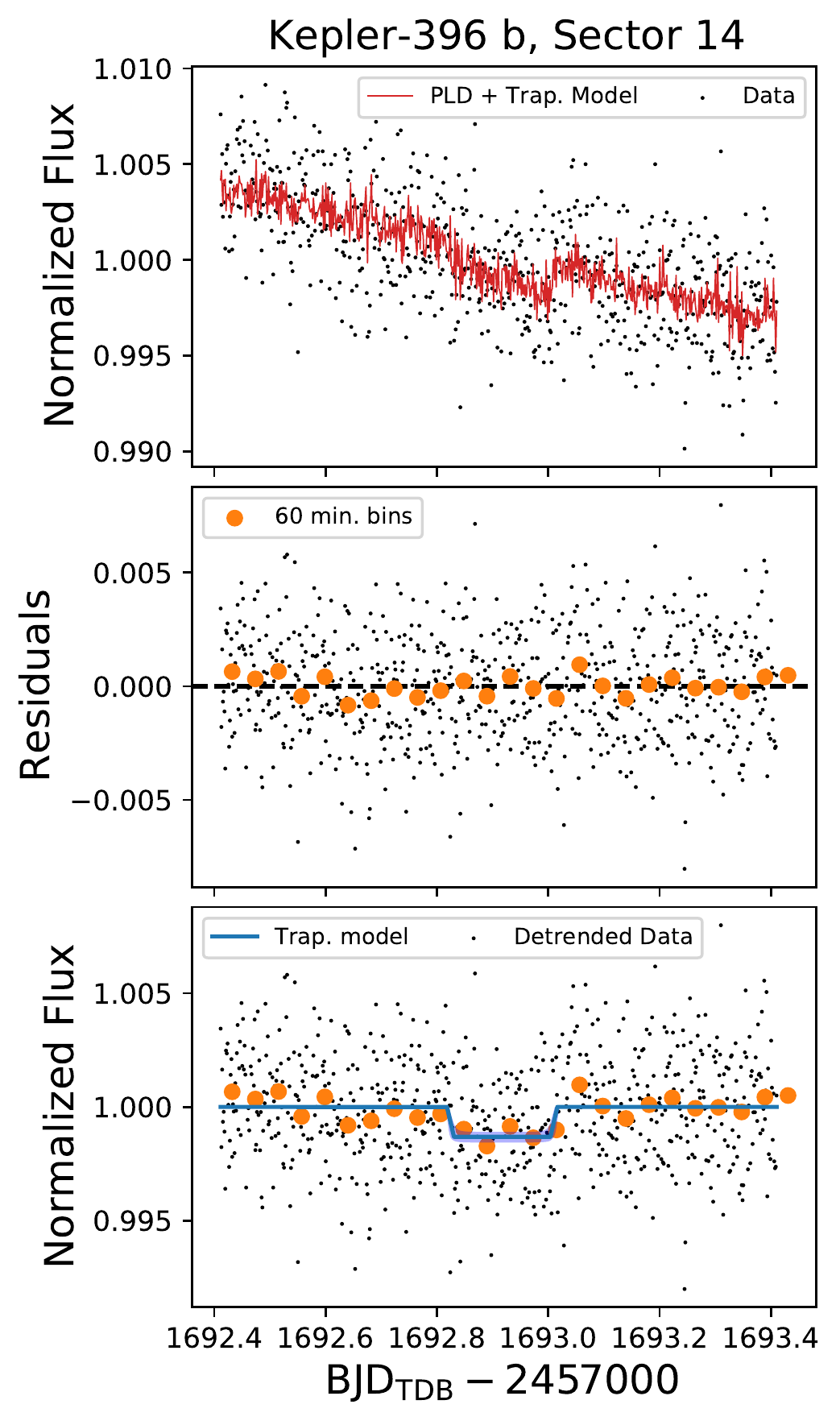} &
      \includegraphics[width=5.4cm]{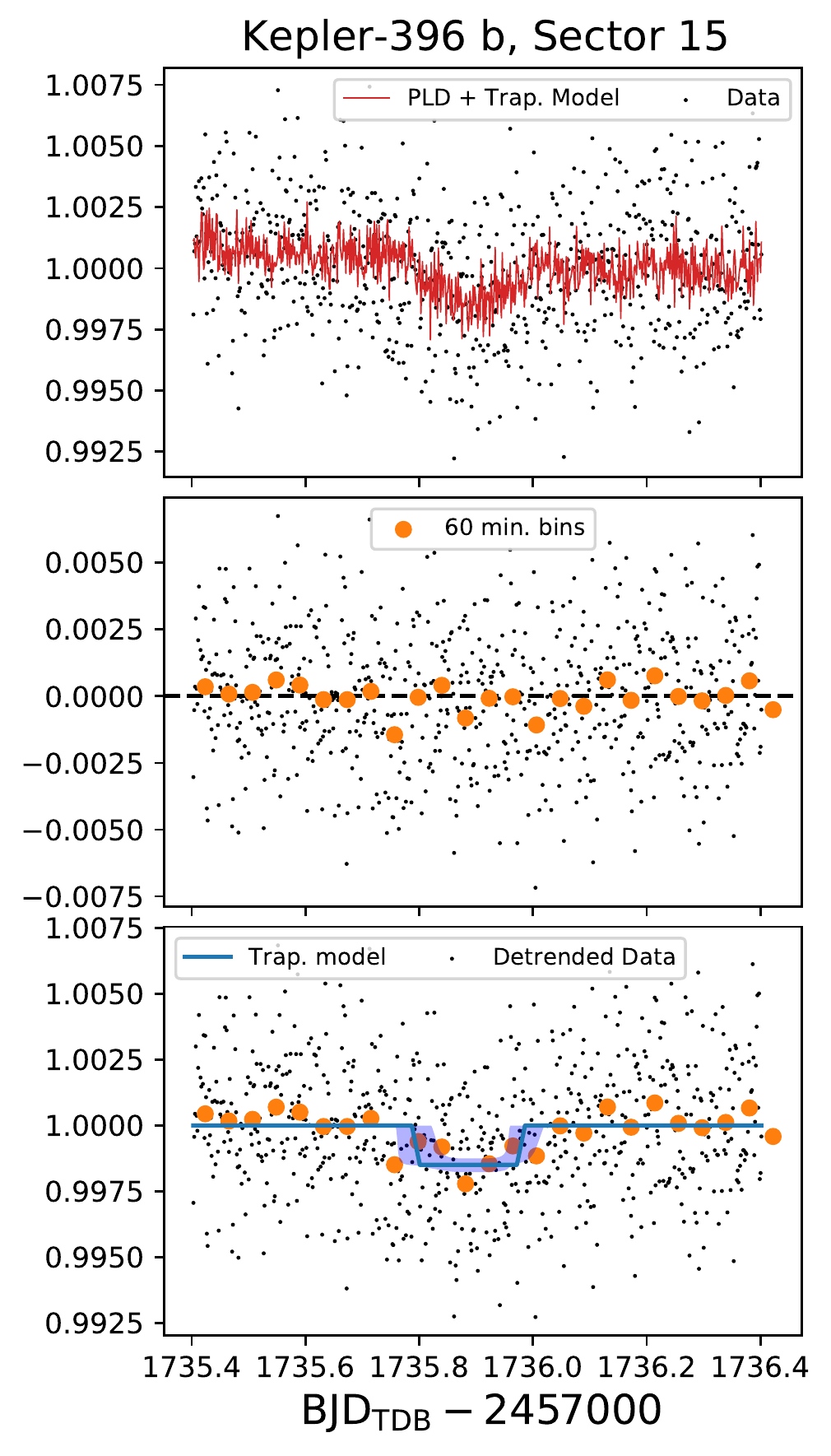} &
      \includegraphics[width=5.6cm]{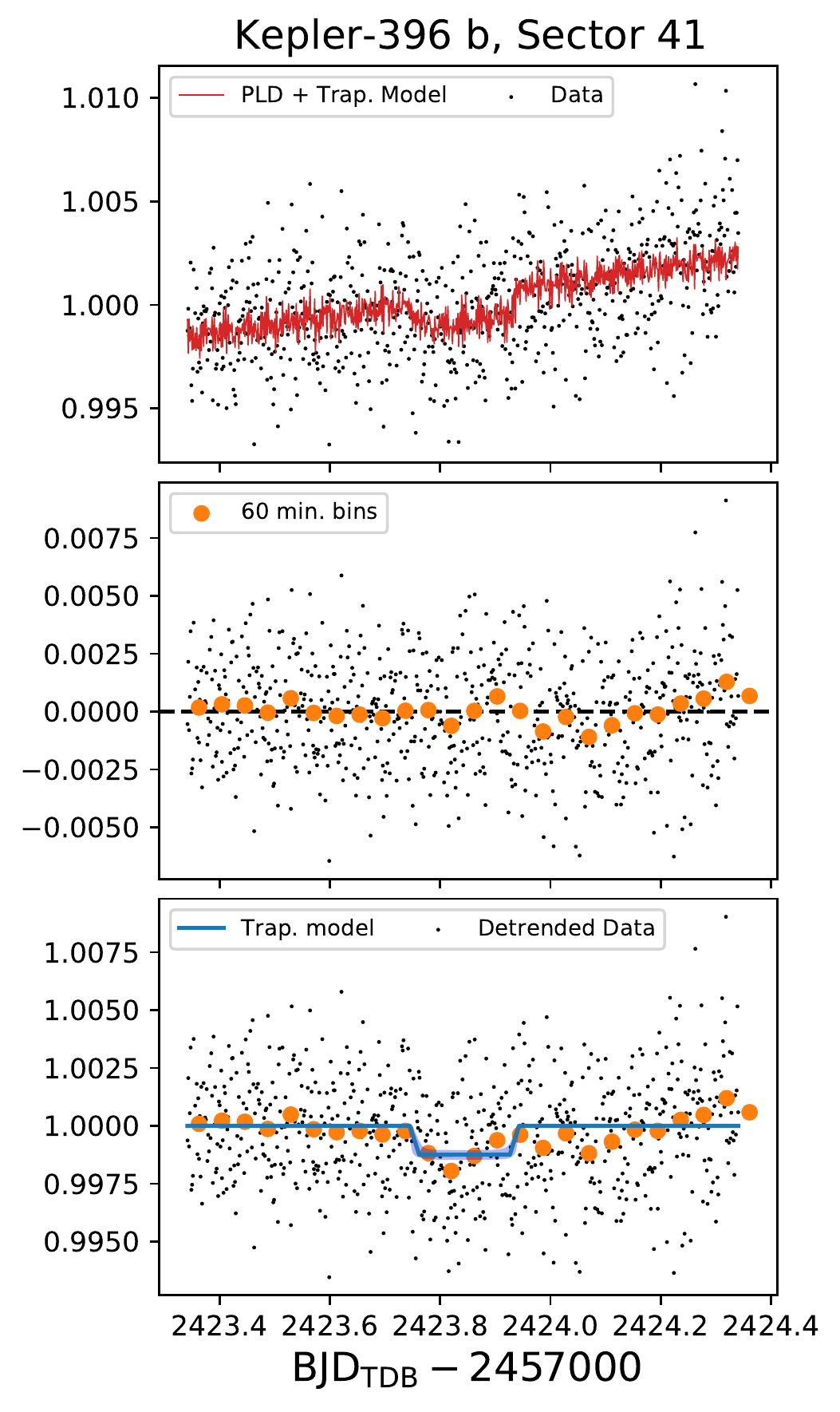} 
    \end{tabular}
  \caption{PLD and trapezoidal model fits to the three transits observed for Kepler-396~b. The top panel of each column shows normalized but not detrended flux from the default SPOC apertures. In each column, the data are centered on the expected transit timing based on TTV models of \kepler\ transit times for this system. The shaded region in the bottom panels shows the 68\% credible region for the trapezoidal fits, with the median in bold. The credible region for Kepler-396~b is less constrained in Sector 15 owing to correlated noise that is poorly modeled by PLD.}
  \label{fig:Kep396b_models}
\end{figure*}

\begin{figure*}
  \centering
    \begin{tabular}{cc}
      \includegraphics[width=5.6cm]{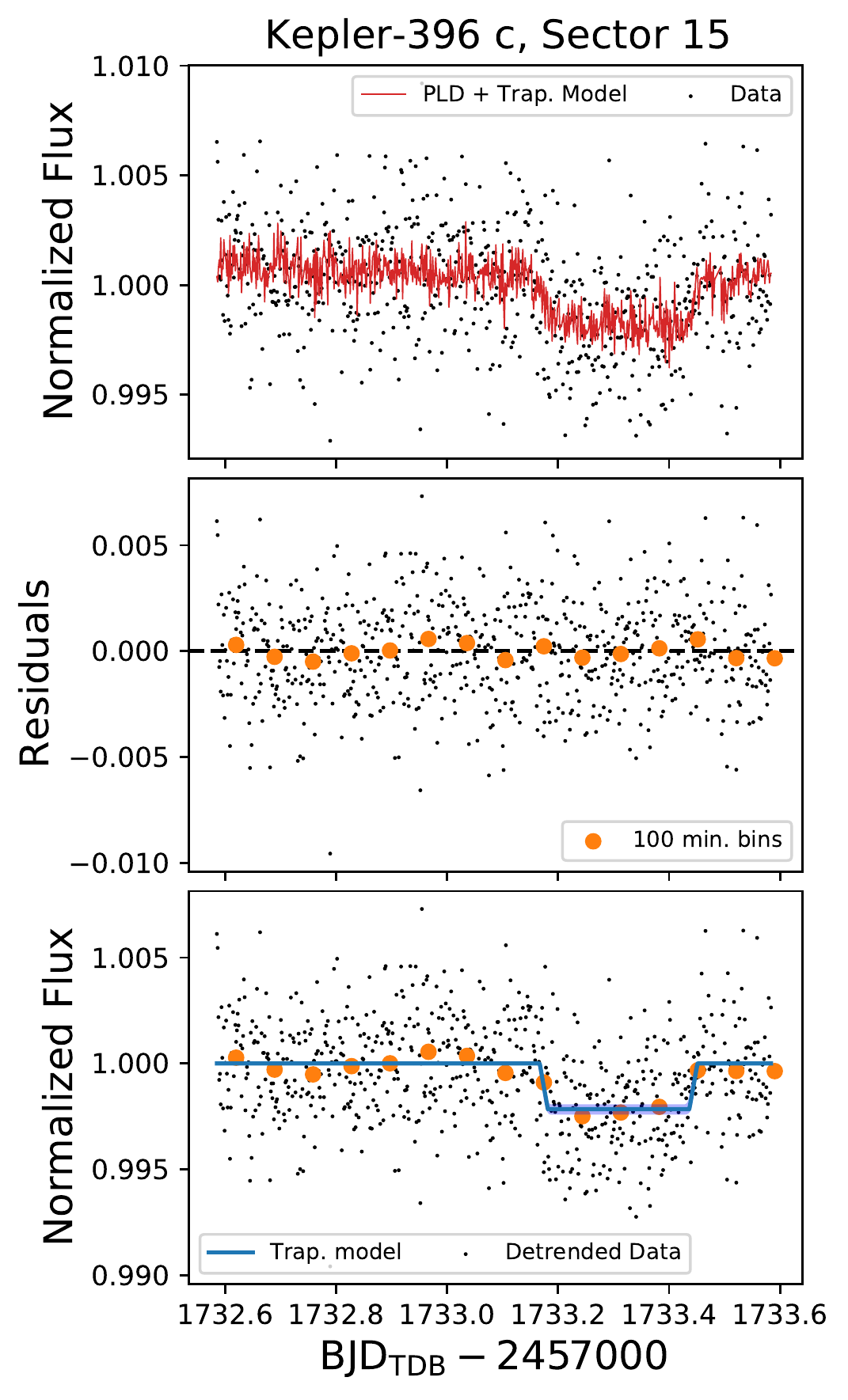} &
      \includegraphics[width=5.4cm]{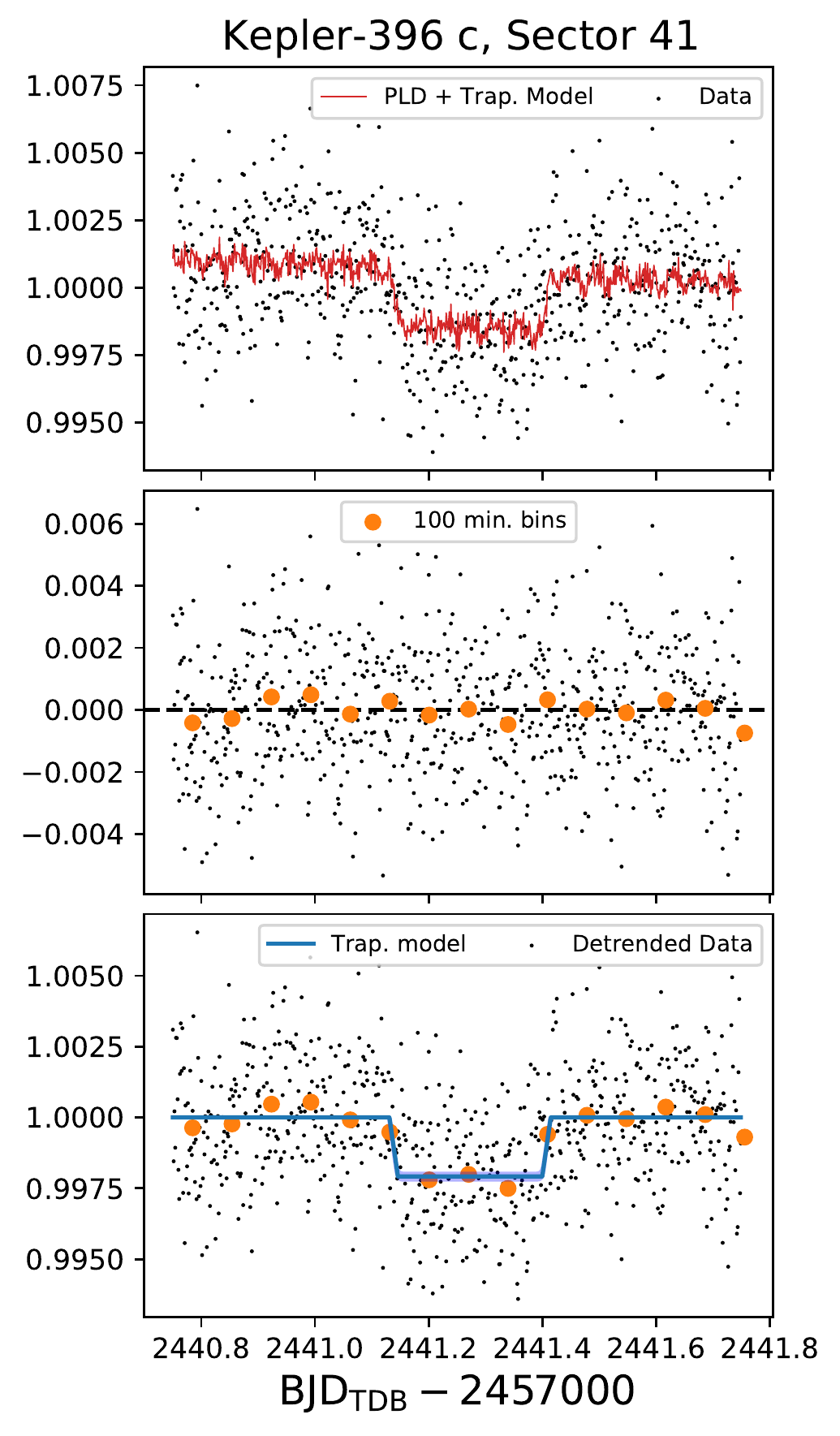}  
    \end{tabular}
  \caption{PLD and trapezoidal models fit to the two transits observed for Kepler-396~c. The style of this figure is identical to that of Figure~\ref{fig:Kep396b_models}. The Sector 15 transit occurs substantially later than expected.}
  \label{fig:Kep396c_models}
\end{figure*}

In Figures~\ref{fig:Kep396b_models} and \ref{fig:Kep396c_models}, we show the data and best-fit models for the three transits of Kepler-396~b in Sectors 14, 15, and 41 and the two transits of Kepler-396~c in Sectors 15 and 41. Each panel is centered on the expected mid-transit time. The Sector 15 transit of Kepler-396~c is clearly visible in the light curve and occurred substantially later than expected. The Sector 14 transit of Kepler-396~b is subtler and more representative of most other TTV systems that we modeled. Through the use of PLD, we are able to make an accurate measurement of this planet's transit even though the depth is on the order of the scatter in the data. The Sector 15 transit of Kepler-396~b, however, is a representative example of marginal detection. The timing of the transit is uncertain owing to correlated noise in the data that is not entirely captured by the PLD model. 

In Tables~\ref{tbl-transitsearch1}, ~\ref{tbl-transitsearch2} and~\ref{tbl-transitsearch3} we list all the transits that were sought from the \tess\ data. The majority of stars were observed with short (fast) cadence with exposures every 2 minutes. However, some stars were only observed with long (slow) cadence with exposures every 30 minutes. In the latter case, we accessed the full-frame image (FFI) data and manually chose photometric apertures to minimize contamination from background sources.

We required transit-like signatures to have a depth inconsistent with zero at the 2$\sigma$ level. In addition, despite our modeling of correlated noise, the risk remains of fitting low-S/N transits to noise features. We rejected the fitted central times of transit-like signatures that, by eye, appeared sensitive to residual correlated noise. The noise could affect the time of the transit model, in which the trapezoidal ingress or egress favors variations due to noise. In some cases we found low-S/N transits with mid-transit times that appear unlikely given the projected transit times following the \kepler\ data. In other cases, correlated noise caused multimodal transit timing posteriors.

The correlated noise can also affect the transit depth. We expect dilution to reduce the depth of transits in many \tess\ light curves compared to \kepler\ light curves of the same targets. Hence, we did not reject fits to \tess\ light curves with shallower transits. However, transits that were measured to be deeper than the \kepler\ model, while possible with different choices for apertures and background subtraction, are unlikely to be significant, and hence we rejected them as fitted red noise. KOI-137.02 (Kepler-18 d) is an exception to this. Our measured transit depth was 0.55$\% \pm 0.08\%$, significantly deeper than the \kepler\ DR25 measurement and discrepant at the 2.75$\sigma$ level. Kepler-18 is faint for a \tess\ target ($V=13.8$), and the flux in each pixel of the photometric aperture was on the order of the background signal. As a result, this system is especially sensitive to the background model, for which we used the SPOC pipeline default. The processed light curve for Sector 14 shows little evidence of time-correlated noise, and the transit is visible by eye at the expected time and with the expected duration. Since this light curve was particularly sensitive to background subtraction, we did not reject the strong detected signal due to its unexpected depth. Hence, we claim this transit as a detection.

In the case of KOI-137, the expected transits from planets c and d overlapped in Sector 15. We accounted for this by simultaneously fitting two trapezoid models with the duration priors of each planet. The result was a multimodal posterior to the mid-transit time for each planet, with transit depths that were larger than expected based on the \kepler\ data. Away from these transits, the light-curve residuals clearly exhibit time-correlated noise with features as large as or larger than the expected transits. For these reasons, we concluded that we were fitting noise features and claimed a nondetection for both transits.

 \begin{table}[h!]
  \begin{center}
    \begin{tabular}{|c|c|c|c|c|c|c|c|}
      \hline
Kep $\#$  & KOI  &   Sect.  & Pr. T$_{14}$ (hr) & Pr. T$_{23}$/T$_{34}$  &  Pred. Time  & Obs. Time &  status  \\ % col 4: 2.2f, col 5 1.3f
 \hline
20c & 70.01 &  14 & 3.78 $\pm$ 0.02 & 0.930 $\pm$ 0.004&  1683.709  $\pm$ 0.002  & & R1 \\
 & & 14 & & & 1705.417  $\pm$  0.002 &  & R2 \\
 & & 26 & & & 2400.079  $\pm$  0.002 & & R3 \\
 & & 26 & & & 2410.933  $\pm$ 0.002  &   2410.9509 $^{+0.026}_{-0.0087}$ &  A \\
 & & 26 & & & 2421.787  $\pm$ 0.002  &   2421.6798 $^{+0.096}_{-0.0096}$ & A \\
 & & 26 & & & 2432.641  $\pm$ 0.002 & & R2 \\
 & & 26 & & & 2443.495  $\pm$ 0.002 &  2443.515 $\pm$ 0.021 & A \\
 \hline
 102e & 82.01 &  14 & 3.72 $\pm$ 0.03 & 0.931  $\pm$0.006 & 1697.407  $\pm$      0.002 & 1697.416 $^{+0.021}_{-0.029}$  & A \\
 & & 26 & & & 2020.322 $\pm$ 0.003 & 2020.3142 $^{+0.0085}_{-0.018}$  &  A  \\
 & & 40 & & & 2391.673 $\pm$ 0.003 & 2391.678 $\pm$ 0.010  &  A  \\
 & & 40 & & & 2407.819 $\pm$ 0.003 & 2407.817 $\pm$ 0.025  &  A \\
 & & 41 & & & 2423.963 $\pm$ 0.003 & & R4 \\
 & & 41 & & & 2440.109 $\pm$ 0.003 & & R4 \\
 \hline
 89d & 94.01 &  14 & 6.664$\pm$0.004 & 0.851$\pm$0.001 & 1697.0182 $\pm$ 0.0005 & & R1 \\
 & & 15 & & & 1719.3614 $\pm$ 0.0005  & 1719.3595 $^{+0.0063}_{-0.0071}$ & A \\
% & & 40 & & & 2389.643708 &  \\
% & & 40 & & & 2411.987527 & \\
 & & 41 & & & 2434.3352 $\pm$ 0.0006 & 2434.3361 $^{+0.0041}_{-0.0037}$ & A \\
 \hline
 89c &  94.02 &14 & 5.23 $\pm$ 0.02 & 0.942 $\pm$ 0.001 & 1692.2613 $\pm$ 0.0021 & 1692.2954 $^{+0.0091}_{-0.0053}$ & A \\
 & & 14 & & & 1702.6851 $\pm$ 0.0021 & 1702.685 $^{+0.017}_{-0.041}$  & A \\
 & & 15 & & & 1713.1120 $\pm$ 0.0023  & 1713.098 $^{+0.035}_{-0.010}$ & A \\
 & & 15 & & & 1723.5370 $\pm$ 0.0025 & 1723.569 $^{+0.041}_{-0.023}$ & A \\
 & & 15 & & & 1733.9636 $\pm$ 0.0026 & & R3 \\
 & & 41 & & & 2432.3459 $\pm$ 0.0029 & & R3 \\
 & & 41 & & & 2442.7679 $\pm$ 0.0030 & & R3 \\
  \hline
 89e &  94.03 & 14 & 8.59 $\pm$ 0.02 & 0.907 $\pm$ 0.002 & 1688.0025 $\pm$ 0.0021  & 1687.950 $^{+0.021}_{-0.018}$ & A\\
 \hline
 18c & 137.01 & 14 & 3.41 $\pm$ 0.01 & 0.914 $\pm$ 0.003 & 1689.8538  $\pm$ 0.0007 & & R3  \\
 & & 14 & & & 1697.4947 $\pm$ 0.0007 & & R4 \\
 & & 14 & & & 1705.1357 $\pm$ 0.0007 & & R4 \\
 & & 15 & & & 1712.7768 $\pm$ 0.0007 & 1712.761 $^{+0.014}_{-0.013}$ & A \\
 & & 15 & & & 1720.4178 $\pm$ 0.0007 & & R4 \\
 & & 15 & & & 1728.0587 $\pm$ 0.0007 & & R3 \\
 & & 15 & & & 1735.6996 $\pm$ 0.0007 & 1735.681 $^{+0.042}_{-0.016}$ & A \\
\hline
    \end{tabular}    
    \caption{Transits Searched for and Recovered from \tess\ Sectors 14, 15, 26, 40, and 41: Part 1 of 3. Note. Priors in transit duration (Pr. T$_{14}$ and Pr. T$_{23}$/T$_{14}$; Columns (4) and (5)) correspond to the total duration from first to last contact, and the fraction of this duration in which the planet is entirely on the disk of the star respectively. The times listed in Column (6) are the predicted time following fits to \kepler\ data. Column (7) gives our measured transit time if it was detected. All times are in BJD$-$2,457,000. The last column lists whether we accepted the transit as a detection (A), or whether the transit was rejected for any of several reasons: a likely transit was totally or partially affected by a data gap (R1) the trapezoidal fit was a non-detection (R2), the posterior depth was consistent with zero (R3), or the fitted model appeared by inspection to be sensitive to correlated noise (R4). We note where short-cadence data were unavailable and we relied on FFIs (F).
    }\label{tbl-transitsearch1}
  \end{center}
\end{table}

 \begin{table}[h!]
  \begin{center}
    \begin{tabular}{|c|c|c|c|c|c|c|c|}
      \hline
Kep $\#$  & KOI  &   Sect.  & Prior T$_{14}$ (hr) & Prior T$_{23}$/T$_{34}$  &  Pred. Time  & Obs. Time &  status  \\ 
 \hline
  & & 40 & & & 2423.4433 $\pm$ 0.0009 & & R2 (F) \\
 & & 41 & & & 2431.0857 $\pm$ 0.0009 & & R2 (F) \\
 & & 41 & & & 2438.7275 $\pm$ 0.0009 & & R2 (F) \\
 \hline
  18d & 137.02 & 14 & 3.53 $\pm$ 0.01 & 0.898 $\pm$ 0.004 & 1690.7417 $\pm$ 0.0007 & 1690.7457 $^{+0.0044}_{-0.0037}$ & A \\
 & & 14 & & & 1705.6013 $\pm$ 0.0008 & 1705.6112 $^{+0.0060}_{-0.0044}$ & A \\ 
 & & 15 & & & 1720.4610 $\pm$ 0.0008  & & R4 \\
 & & 15 & & & 1735.3209 $\pm$ 0.0008 & & R4 \\
 & & 41 & & & 2433.6879 $\pm$ 0.0009 & & R1 \\
 \hline
79d &  152.01 & 14 & 8.64 $\pm$ 0.02 & 0.903 $\pm$ 0.002 & 1690.2027 $\pm$ 0.0041 & & R3 \\
  & & 41 & & & 2419.4699 $\pm$ 0.0044 & & R1 \\
  \hline
  25c &  244.01 & 14 & 2.73 $\pm$ 0.05 & 0.896 $\pm$ 0.049 & 1687.7166 $\pm$ 0.0006 & 1687.7126 $^{+0.0033}_{-0.0025}$ & A \\
 & & 14 & & & 1700.4368 $\pm$ 0.0006 & 1700.4451 $^{+0.0038}_{-0.0061}$  & A \\
 & & 40 & & & 2400.0563 $\pm$ 0.0007 & 2400.0483 $^{+0.0027}_{-0.0030}$ & A \\
 & & 40 & & & 2412.7763 $\pm$ 0.0007 & 2412.7788 $^{+0.0025}_{-0.0035}$ & A \\
 & & 40 & & & 2425.4964 $\pm$ 0.0007 & 2425.5071 $^{+0.0049}_{-0.0055}$ & A \\
 & & 41 & & & 2438.2165 $\pm$ 0.0007 & 2438.2144 $^{+0.0034}_{-0.0017}$ & A \\
 \hline
 25b & 244.02 & 14 & 3.53 $\pm$ 0.01 & 0.960 $\pm$ 0.004 & 1685.4406 $\pm$ 0.0016 & 1685.4358 $^{+0.017}_{-0.0055}$ & A\\
 & & 14 & & &  1691.6793 $\pm$ 0.0016 & & R2 \\
 & & 14 & & &  1697.9180 $\pm$ 0.0016 & 1697.9118 $^{+0.0061}_{-0.0038}$ & A\\
 & & 14 & & &  1704.1567 $\pm$ 0.0016 & 1704.1262 $^{+0.0079}_{-0.0092}$ & A\\
 & & 40 & & & 2390.3975 $\pm$ 0.0019 & & R2 \\
 & & 40 & & & 2396.6365 $\pm$ 0.0019 & 2396.546 $^{+0.026}_{-0.034}$ & A \\
 & & 40 & & & 2402.8754 $\pm$ 0.0019 & 2402.879 $^{+0.014}_{-0.036}$ & A \\
 & & 40 & & & 2409.1143 $\pm$ 0.0019 & & R2 \\
 & & 40 & & & 2415.3531 $\pm$ 0.0019 & & R2 \\
 & & 40 & & & 2421.5921 $\pm$ 0.0019 & & R2 \\
 & & 41 & & & 2427.8308 $\pm$ 0.0019 & & R2 \\
 & & 41 & & & 2434.0697 $\pm$ 0.0020  & & R2 \\
 & & 41 & & & 2440.3084 $\pm$ 0.0020 & & R2 \\
 & & 41 & & & 2446.5472 $\pm$ 0.0020 & & R1 \\
 \hline
  37d & 245.01 & 14 & 4.47 $\pm$ 0.02 & 0.947 $\pm$ 0.010 & 1708.9298 $\pm$ 0.0019 & 1708.9327 $^{+0.0037}_{-0.0031}$ &  A \\
 & & 26 & & & 2027.2674 $\pm$ 0.0021 & 2027.270 $^{+0.016}_{-0.020}$ &  A (F) \\
 & & 41 & & & 2425.1888 $\pm$ 0.0022 & 2425.1903 $^{+0.0039}_{-0.0033}$ & A \\
  \hline
  130c & 	282.01 &	14 &	5.93 $\pm$	0.04 & 	0.930 $\pm$	0.008 & 1701.2660 $\pm$ 0.0022 & & R2 (F) \\ %  & &   & & &   & \\
  & & 40  & & & 2416.4929 $\pm$ 0.0031 & 2416.4765 $^{+0.011}_{-0.0076 }$  & A\\
   & & 41  & & & 2444.0019 $\pm$ 0.0031 & & R3 \\
\hline
   90g &	351.02 &	40	& 11.99 $\pm$	0.03 &	0.886 $\pm$	0.005 &	2406.0349 $\pm$ 0.1228 & & R2 (F) \\
\hline
    \end{tabular}    
    \caption{Transits Searched for and Recovered from \tess\ Sectors 14, 15, 26, 40, and 41: Part 2 of 3. The format is the same as Table~\ref{tbl-transitsearch1}.
    }\label{tbl-transitsearch2}
  \end{center}
\end{table}

 \begin{table}[h!]
 \small
  \begin{center}
    \begin{tabular}{|c|c|c|c|c|c|c|c|}
      \hline
Kep $\#$   & KOI  &   Sect.  & Pr. T$_{14}$ (hr) & Pr. T$_{23}$/T$_{34}$  &  Pred. Time  & Obs. Time &  status  \\ % col 4: 2.2f, col 5 1.3f
 \hline
      9b &	377.01 &	14 &	4.13 $\pm$	0.01 &	0.858 $\pm$	0.004 &	1691.9272 $\pm$ 0.0102 &  &	R2 (F) \\
    & & 26  & & & 2018.8745 $\pm$ 0.0150  &  &	R2 (F) \\ 
     & & 40  & & & 2403.2777 $\pm$ 0.0126  & 2403.3141 $^{+0.0020}_{-0.0018}$ & A (F) \\
     & & 41  & & & 2422.5023 $\pm$ 0.0124 & 2422.5392 $^{+0.0057}_{-0.0029}$  & A (F) \\
     & & 41  & & & 2441.7270 $\pm$ 0.0122 & 2441.666 $^{+0.036}_{-0.14}$  & A (F) \\
        \hline
   9c &	377.02 &	14 &	4.53  $\pm$ 	0.04 &	0.743  $\pm$	0.018 &	1708.4681 $\pm$ 0.0235 & & R3 \\
     & &  26 & & & 2020.6351 $\pm$ 0.0335  & & R2 (F) \\
  & & 40  & & & 2411.3611 $\pm$ 0.0289 & 2411.4242 $^{+0.004}_{-0.049}$ & A (F)\\
   \hline
     51b &	620.01 &	14 &	5.78  $\pm$ 	0.02 &	0.858  $\pm$ 	0.007 & 1694.8509 $\pm$ 0.0023 &	1694.879 $^{+0.021}_{-0.011}$ &	A \\
  \hline
  51d &	620.02 &	14 &	8.45 $\pm$	0.02 &	0.815 $\pm$	0.005 & 1689.9844 $\pm$ 0.0192	& 1689.996 $^{+0.023}_{-0.013}$ &	A \\
  \hline
   30d &	806.01 &	41 &	8.93  $\pm$	0.10 &	0.811	  $\pm$ 0.018	& 2426.6872 $\pm$ 0.0045 & 2426.82$^{+0.10}_{-0.18}$ &	A (F) \\
    \hline
 30c &	806.02 &	14/15 &	6.62 $\pm$	0.03 &	0.762 $\pm$	0.009	& 1696.3701 $\pm$ 0.0011 & &	R1 \\
 &	 &	41 &	 &	 & 2420.2896 $\pm$ 0.0015 &	2420.295$^{+0.030}_{-0.021}$ &	A (F) \\
   \hline
 289c &	1353.01	& 15 &	9.02 $\pm$	0.02  &	0.815	 $\pm$ 0.001 & 1719.7421 $\pm$ 0.0020  &	1719.730$^{+0.012}_{-0.014}$ & A \\
   \hline
 297c &	1426.02 &	14 &	4.74  $\pm$	0.08 &	0.683	  $\pm$ 0.073  & 	1703.5308 $\pm$ 0.5527 & & R2 \\	
 \hline
  --- &	1426.03 &	40 &	4.38 $\pm$	0.05 &	0.739 $\pm$	0.236  &	2408.4621 $\pm$ 0.4476 & & R3 (F) \\
  \hline
   1662c &	1783.01	& 41 & 	5.93 $\pm$	0.06 &	0.139 $\pm$	0.083 & 2440.1137 $\pm$ 0.0460 &	2440.11$^{+0.11}_{-0.12}$ 	& A (F) \\
    \hline
 396c &	2672.01	 & 15 &	6.82 $\pm$	0.10 &	0.900 $\pm$	0.013 & 1733.0845 $\pm$ 0.0102 &	1733.3103$^{+0.0039}_{-0.0064}$	&	A \\
  & & 41  & & & 2441.1036 $\pm$ 0.0131 & 2441.2730$^{+0.0061}_{-0.0079}$  & A \\
  \hline
    396b & 	2672.02 & 	14 & 	4.73 $\pm$	0.11 & 	0.837 $\pm$	0.233 & 1692.9110 $\pm$ 0.0057 &	1692.9172$^{+0.0069}_{-0.0060}$  & A \\
   & & 15  & & & 1735.9019 $\pm$ 0.0058 & 1735.887$^{+0.047}_{-0.036}$  & A \\
   & &  41 & & & 2423.8334 $\pm$ 0.0070 &  2423.8437$^{+0.0077}_{-0.0070}$  & A \\
   \hline
    \end{tabular}    
    \caption{Transits searched for and recovered from \tess\ Sectors 14, 15, 26, 40 and 41: Part 3 of 3. The format is the same as Table~\ref{tbl-transitsearch1}.
    }\label{tbl-transitsearch3}
  \end{center}
\end{table}
 
We detected and measured 48 transit times from 90 expected transits overall. As can be seen in Tables~\ref{tbl-transitsearch1}, ~\ref{tbl-transitsearch2} and ~\ref{tbl-transitsearch3}, only KOI-94.03 (Kepler-89 e), KOI-244.01 (Kepler-25 c), KOI-245.01 (Kepler-37 d), KOI-620.01 (Kepler-51 b) KOI-620.02 (Kepler-51 d), KOI-2672.01 (Kepler-396 c), and KOI-2672.02 (Kepler-396 b) were detected at every expected opportunity. For KOI-152.01 (Kepler-79 d) and KOI-351.02 (Kepler-90 g), we did not detect any of the prospective transits. 

\citet{Goldberg2019} estimated which targets would benefit the most from transit times measured with \tess\ photometry. Of their 25 planets that could be the most improved through \tess\ observations, many were not on our list, since we did not anticipate a high enough S/N. Of those where we anticipated a detection, we accepted transit fits for Kepler-396 b, Kepler-396 c, Kepler-51 b, Kepler-51 d, Kepler-9 b, and Kepler-9 c. 

\section{Dynamical Fits and Planet Characterization} \label{sec:TTVmodels}
\subsection{Transit Timing Data Sets}
Given our measured transit times from \tess\ light curves and measured \kepler\ transit times from \citet{Rowe2015}, we fitted dynamical models to measured transit times. Two of our targets have additional transit times from other observing campaigns. \citet{Vissapragada2020} measured a transit time for KOI-1783.01 (Kepler-1662 b) using the Wide-field Infrared Camera on the Hale 200-inch telescope at Palomar Observatory, which we included in our dynamical fits for KOI-1783.

\citet{Libby-Roberts2020} observed a total of four transits of planets orbiting KOI-620 (Kepler-51) with the Hubble Space Telescope. We leave a detailed update of TTV constraints of Kepler-51 from post-Kepler observations to another study (J.H. Livingston et al. in preparation, including ground-based observations), and in this paper include \kepler\ and \tess\ data only. 

KOI-806.03 (Kepler-30 b) has a very large TTV amplitude of $\approx$ 1 day, with poorly fitted transit times in the catalog of \citep{Rowe2015} (also noted by \citealt{Panichi2018}). For this system, we used the transit times from \kepler\ photometry measured by \citet{Holczer2016}. 

\subsection{Transit Timing Model}
Our dynamical fits include five parameters per planet: dynamical mass, orbital period, first transit after epoch (BJD = 2,455,680), and eccentricity vector components e$\cos\omega$ and e$\sin\omega$. We neglect mutual inclinations since these are unlikely to be large enough to cause detectable TTVs (\citealt{Nesvorny2009,Fabrycky2014}), and we fix ascending nodes to zero and inclinations at 90$^{\circ}$. Our dynamical models use an eighth-order Prince-Dormand Runge-Kutta integrator \citep{Lissauer2011a} to simulate transit times. 

We sampled the posteriors of our free parameters with a differential evolution Markov Chain Monte Carlo (MCMC) algorithm. Our priors are uniform in orbital period $P$, and the time of the first transit after epoch $T_{0}$, uniform with positive-definite dynamical masses, and Gaussian in eccentricity vector components with a mean of zero and a standard deviation of 0.1. We initialized our walkers by randomly drawing from uniform distributions near a linear best-fit model to the transit times with circular orbits, with dynamical masses between 0 and 20 $M_{\oplus}/M_{\odot}$, $P$ and $T_{0}$ within 0.001 days of the linear model, and eccentricity vector components drawn uniformly from $[-0.01,0.01]$. We determined burn-in by visual inspection and calculated the effective sample size by measuring the autocorrelation length on our MCMC chains. We obtained a minimum of 1000 effective samples for all parameters.

Following dynamical fits, we characterized planetary masses using the stellar mass posteriors of \citet{Berger2020a} (listed in Table~\ref{tbl-stars}) and planetary bulk densities using the measured planetary radii of \citet{Berger2020b}. KOI-82.06 does not appear in \citet{Berger2020b}. For this planet, we have adopted the measured radius of DR25 \citep{Thompson2018} increased by 3.4\%, since the stellar radius measurement of \citet{Berger2020a} is that much larger than the published value in DR25. KOI-620.03 (Kepler-51 c) has a grazing transit with a poorly constrained radius in \citet{Berger2020b}. We adopt the measured radius for this planet from \citet{Libby-Roberts2020}. 

 \begin{table}[h!]
  \begin{center}
    \begin{tabular}{|c|c|c|}
      \hline
 KOI  & Kep \#  &     $M_{\star}$ ($M_{\odot}$) \\   
 \hline
 70 & 20 & 0.922 $^{+  0.043 }_{- 0.054  }$  \\
 82 & 102 & 0.766 $^{+ 	0.032 }_{-	0.038 }$  \\
 94 & 89  & 1.339$^{+	0.031}_{-	0.035}$   \\
 137 & 17 & 0.976$^{+	0.034}_{-	0.049}$  \\
 244 & 25 & 1.042$^{+	0.084 }_{-	0.073}$   \\
 245 & 37 & 0.77$^{+	0.047}_{-	0.036}$  \\
 282 & 130  &	1.026$^{+	0.069}_{-	0.07}$  \\
 377 & 9 & 1.024  $^{+0.033 }_{- 0.047}$  \\
 620 & 51 & 0.894$^{+	0.036}_{-	0.048}$   \\
 806 & 30  & 0.912$^{+	0.039}_{-	0.042}$   \\
 1353 & 289 & 1.051$^{+	0.031}_{-	0.042}$   \\
 1783 & 1662 & 1.133$^{+0.052}_{-0.066}$ \\
 2672 & 396 & 0.946$^{+	0.049}_{-	0.058}$  \\
\hline
    \end{tabular}    
    \caption{Adopted stellar masses for planet characterization \citep{Berger2020a}.}\label{tbl-stars}
  \end{center}
\end{table}

\subsection{Nontransiting Perturbers}
For two of our systems, KOI-2672 and KOI-94, no good fit to the transit times was found using just the transiting candidates. For these we added an extra planet in dynamical models to improve the fit to the transit times. While there are no unique solutions to the TTVs of these systems including nontransiting perturbers, we compared two suites of solutions for these systems: by including an additional planet beyond the transiting interacting planets, and including an additional planet in a large gap between the known planets. 

\subsubsection{KOI-2672, Kepler-396}
Kepler-396 has two transiting planets orbiting at 43.0 and 88.5 days, respectively. Their proximity to the 2:1 resonance causes an expected periodicity in the TTVs of $\sim$1520 days, close to the 4-yr baseline of \textit{Kepler} data. We found that a two-planet model is a poor fit to the \textit{Kepler} data (a reduced $\chi^2$ of 5.51), particularly for Kepler-396 c, which has several outlying residuals following dynamical fits. In this, our results differ from \citet{Battley2021}; in their figure 11, two fewer transit times for Kepler-396 c are included. The projected transit times for Kepler-396 b from \citet{Battley2021} are closely consistent with their two measured transit times from \tess\, and their fits including the two transit times from \tess\ for Kepler-396 b are almost indistinguishable from their \kepler-only fits. However, as shown in Figure~\ref{fig:Kep396c_models}, we clearly detect the transit of Kepler-396 c in Sector 15. The mid-transit time is about 5.4 hr after the expected time following two-planet fits to the \kepler\ data. Our measured transit depth (2070 ppm) is close to the \kepler\ DR25 value of 2519 ppm, with moderate dilution (see Figure~\ref{fig:Kep396_tpf}). Hence, we include this measured transit time from Sector 15 in our dynamical models. We recovered an additional transit of Kepler-396 c in Sector 41 and include both the measured transit times of Kepler-396 c in our dynamical models. 

We explored two plausible three-planet models, both of which significantly improve the model fits to the transit times. Since there are many possible solutions that would improve the goodness of fit with the addition of a planet, a thorough exploration of these possibilities is beyond the scope of this paper. However, we consider two suites of models to test the sensitivity of our posteriors to different possible solutions: one with an additional planet inserted with periods ranging from $\approx$117 to $\approx$152 days, near the 3:2 resonance exterior to Kepler-396 c and near 3:1 with Kepler-396 b (model A), and one with an added planet between the transiting planets, over a range of periods from $\approx$54 to $\approx$60 days, near the 3:2 resonance interior to Kepler-396 c and near the 4:3 resonance with Kepler-396 b (model B). The additional planet adds five parameters to the dynamical fits. These are the same as for the transiting planets and have the same priors, namely, uniform priors in orbital period, orbital phase and mass (with positive-definite masses), and Gaussian priors of width 0.1 in orbital eccentricity vector components. The goodness of fit for the best models is summarized in Table~\ref{tbl-koi2672summary}.

 \begin{table}[h!]
  \begin{center}
    \begin{tabular}{|c|c|c|c|c|c|}
      \hline
 model  & n  &     m    &     $\chi^2$  &  red.  $\chi^2$   &   $\Delta$BIC  \\   
 \hline
2pl & 52 & 10 & 549.4 & 13.08 & 0 \\ % bic = 588.912
3pl A  & 52 & 15 & 140.97  & 3.81  & -388.67 \\ % bic = 200.238
3pl B  & 52 & 15 & 193.85  & 5.24  & -335.79 \\ % bic = 253.118
\hline
    \end{tabular}    
    \caption{Summary of Models and Best-fit Solutions for Kepler-396. Note. Models listed in the first column include fits with two (2pl) or three planets (3pl), respectively. Model A (or B) refers to fits with a third planet initially inserted at an orbital period beyond that of Kepler-396 c (or between that of the transiting planets). The second and third columns list the number of data and free parameters, the fourth column lists the lowest $\chi^2$ found, the fifth column notes the reduced $\chi^2$, and the final column gives the change in BIC for each model compared to the two-planet model.}\label{tbl-koi2672summary}
  \end{center}
\end{table}

\begin{figure}[ht!]
\includegraphics [width = 2.5 in]{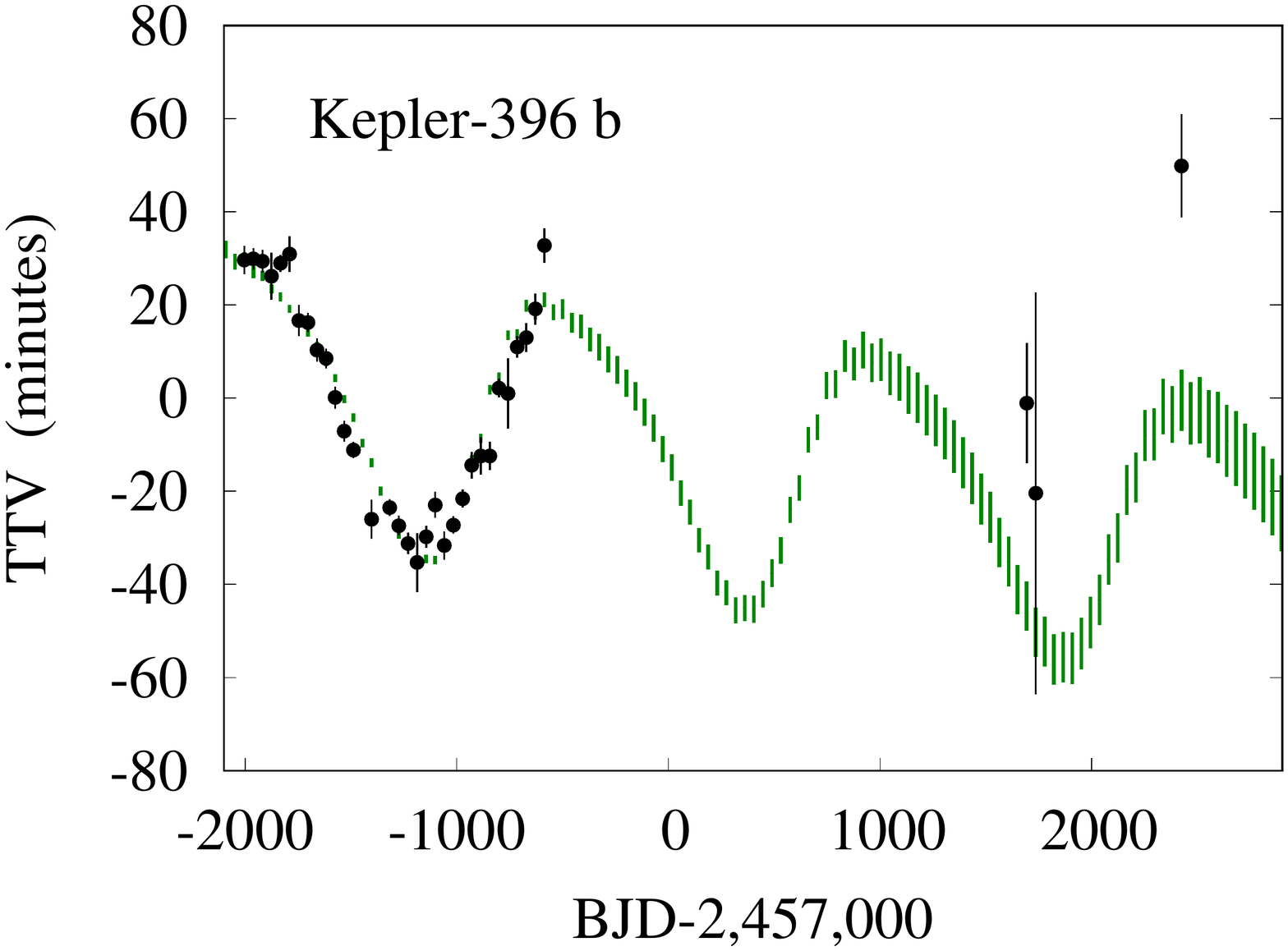}
\includegraphics [width = 2.5 in]{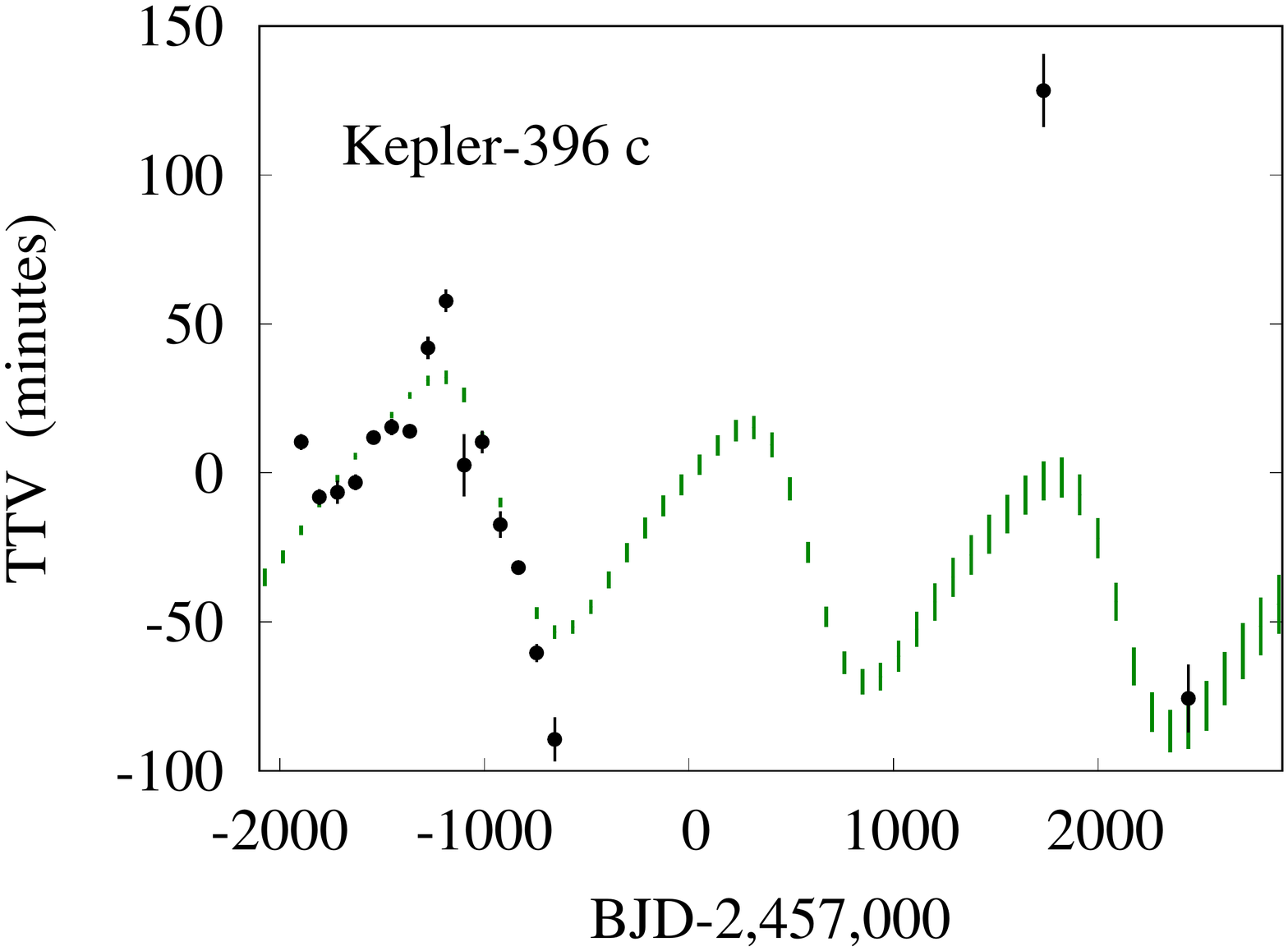}
\newline
\includegraphics [width = 2.5 in]{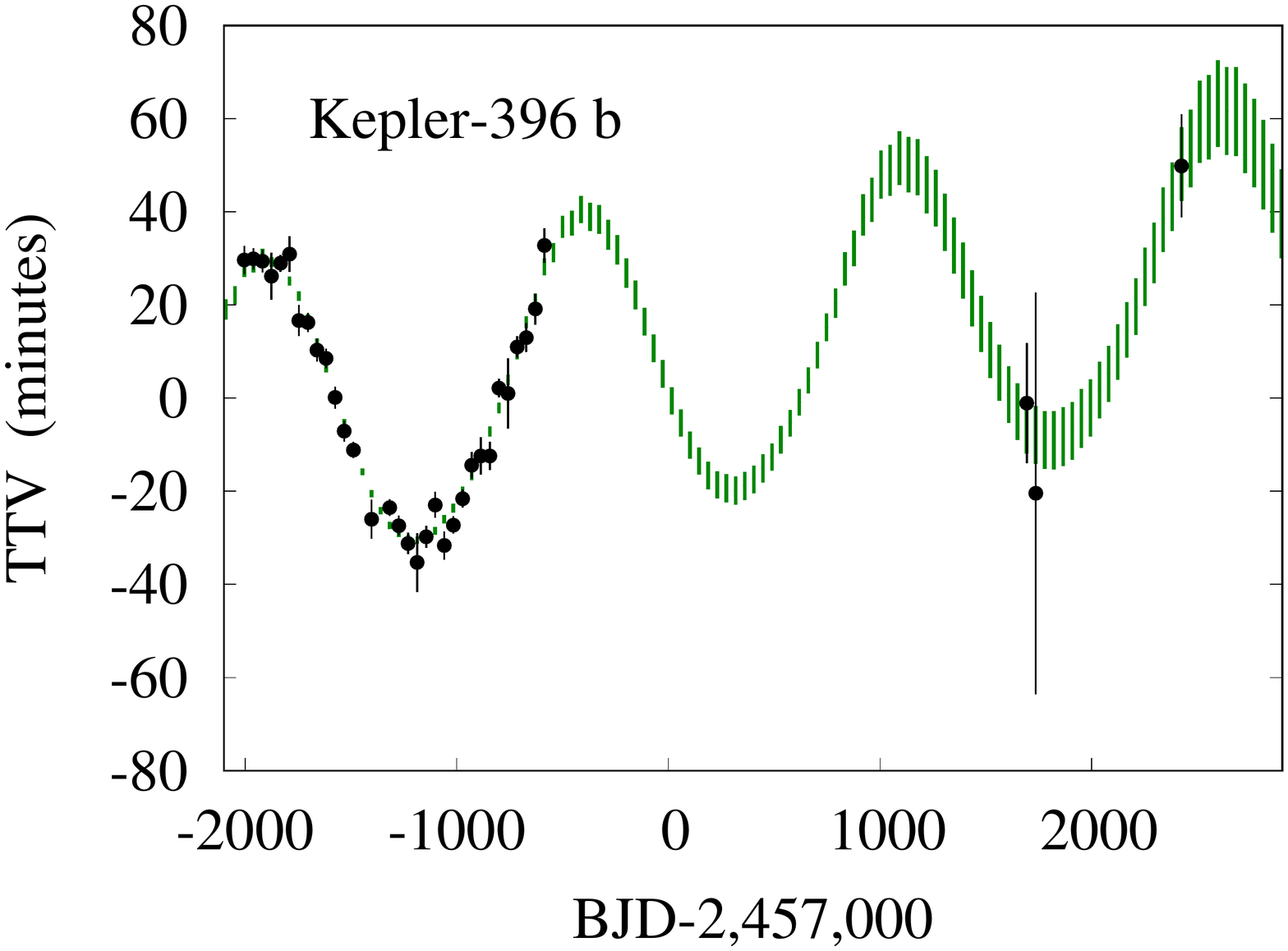}
\includegraphics [width = 2.5 in]{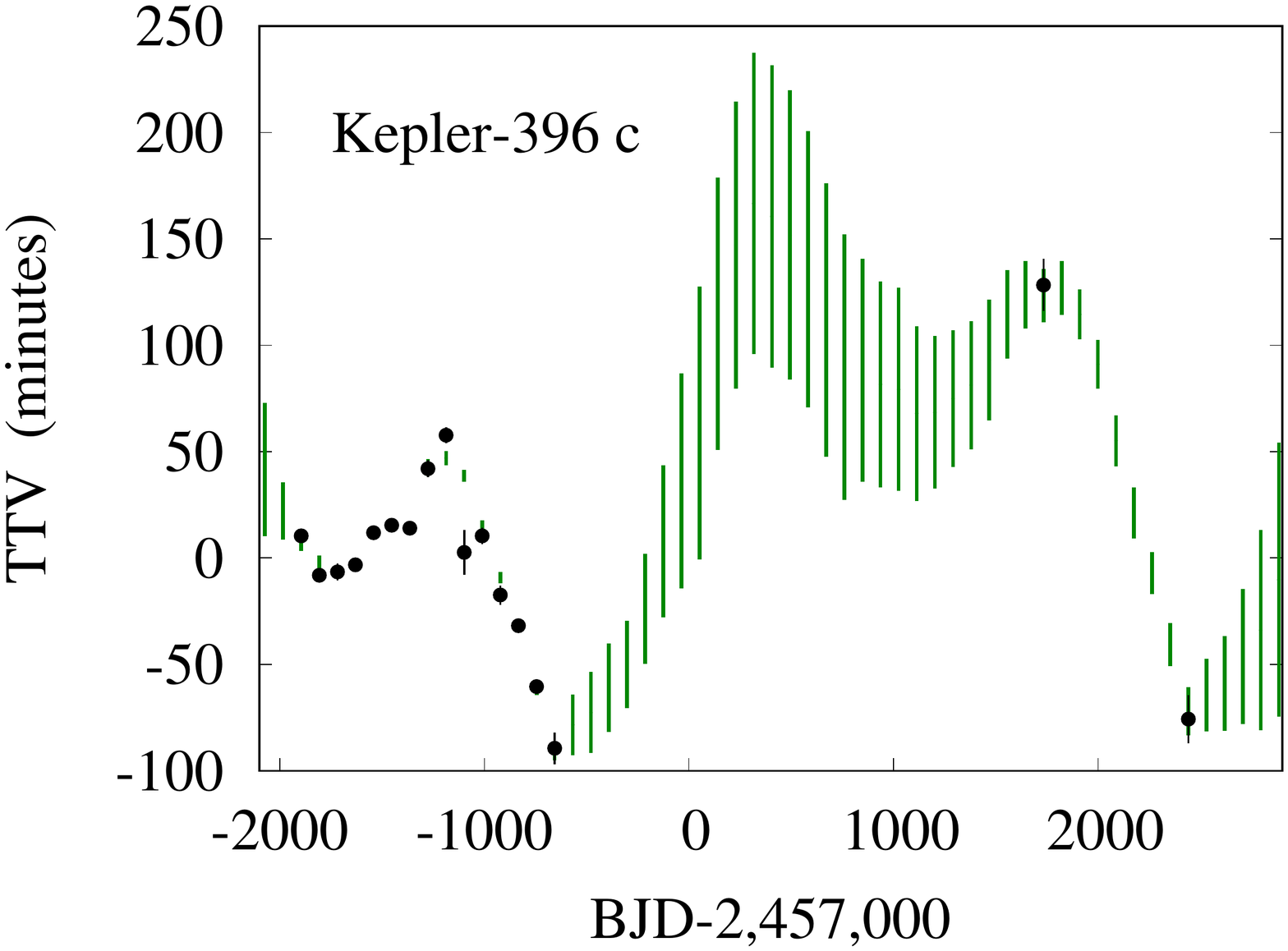}
\newline
\includegraphics [width = 2.5 in]{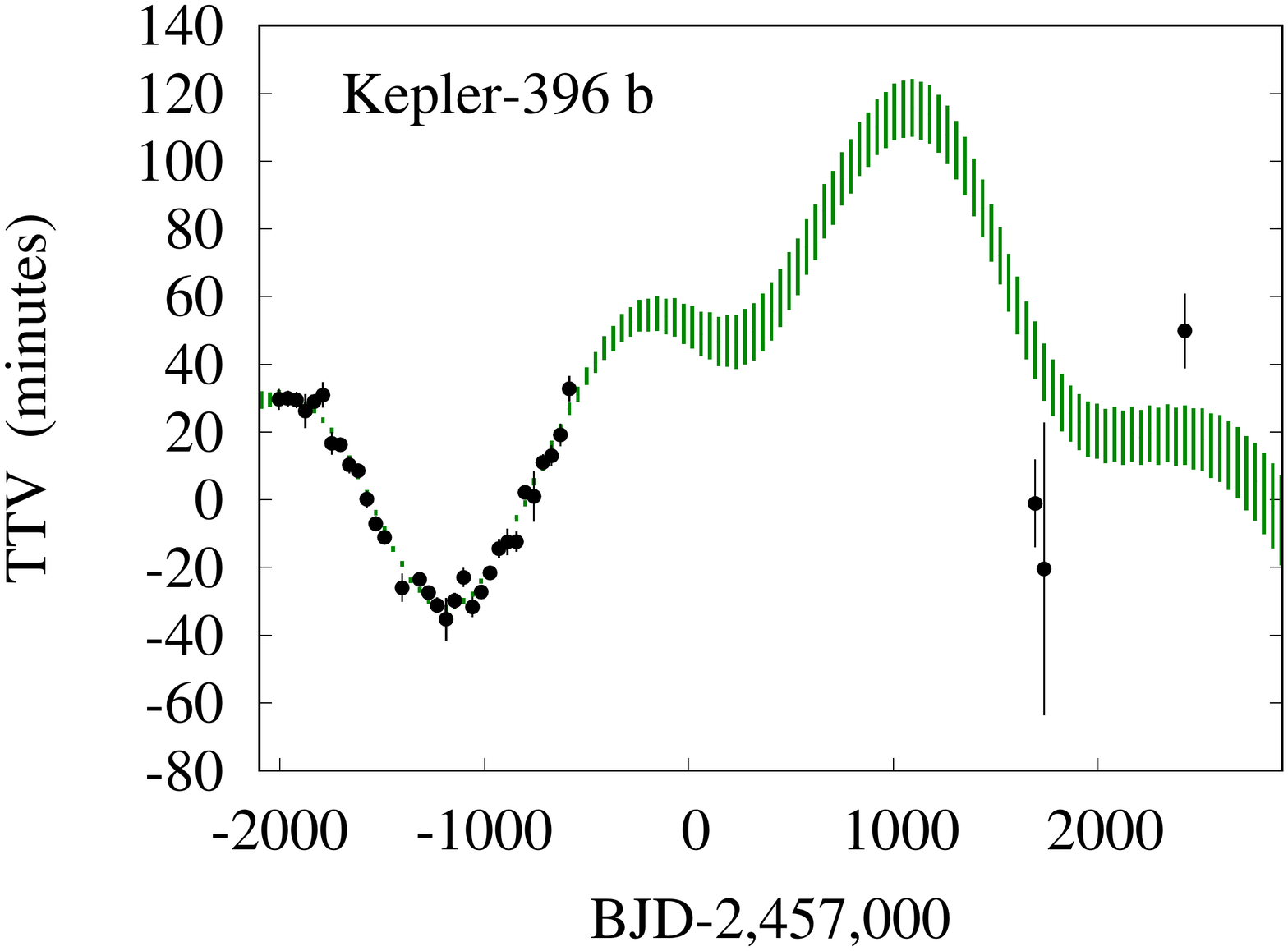}
\includegraphics [width = 2.5 in]{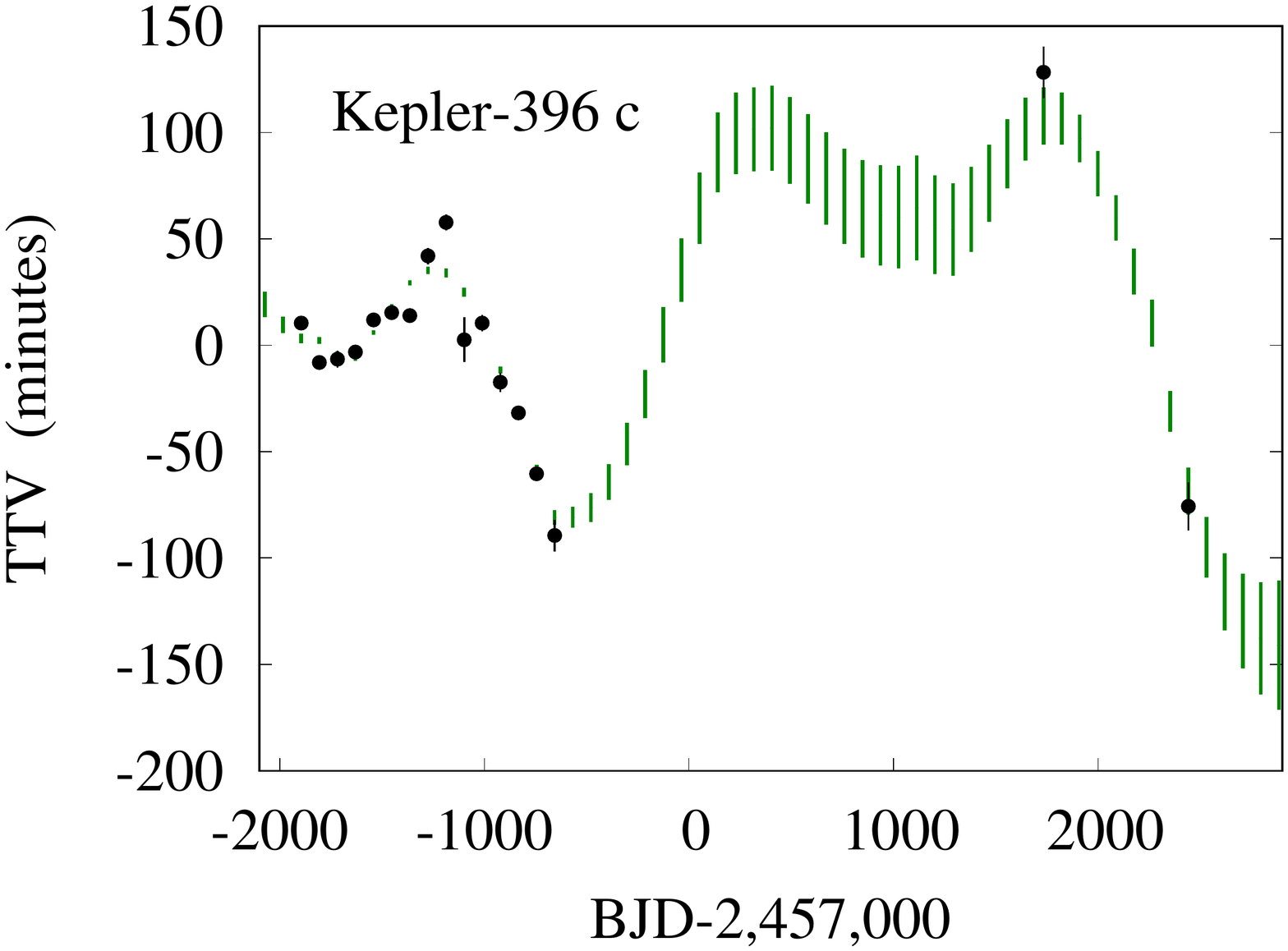}
\caption{TTV models ($O-C$, in minutes) for Kepler-396, with the combined \kepler /\tess\ transit times in black and 68\% confidence intervals with equal weight in the tails on simulated transit times in green. The top row corresponds to a two-planet model, the middle row includes a nontransiting planet orbiting beyond Kepler-396 c (model A), and the bottom row includes a nontransiting planet orbiting between Kepler-396 b and Kepler-396 c (model B).}
\label{fig:Kep396_ttv} 
\end{figure}

It is clear that a third planet is justified, with a substantial reduction in the Bayes information criterion (BIC) despite the penalty on additional parameters. Between the three-planet models, model A, with a planet inserted outside the orbit of Kepler-396 c, is a better fit to the data than model B. The TTVs of these models are plotted in Figure~\ref{fig:Kep396_ttv}. 

 \begin{table}[h!]
 \scriptsize
 %\tiny
  \begin{center}
    \begin{tabular}{|c|c|c|c|c|c|c|c|c|}
      \hline
KOI  & $P$ (days)  &      $T_0$ (days)    &      $e\cos\omega$   &  $e\sin\omega$     &     $ \frac{ M_p  }{M_{ \star} } \frac{M_{\odot} } {M_{\oplus} }  $ &  $ \frac{M_{p} }{M_{\oplus} }$ &  $ \frac{R_{p} }{R_{\oplus} }$ &  $\rho_{p}$ (g cm$^{-3}$) \\
 \hline
  2672.02 & 42.9921 $^{ + 0.0002  }_{ - 0.0002 }$  & -1316.6357 $^{ + 0.0007  }_{ - 0.0007 }$    & -0.007 $^{ + 0.002  }_{ - 0.003 }$    & -0.005 $^{ + 0.002  }_{ - 0.002 }$   & 10.1 $^{ + 1.1  }_{ - 1.1 }$  & 9.5 $^{ + 1.2  }_{ - 1.1 }$ &  3.23  $^{ + 0.50 }_{ - 0.19 }$   & 1.55 $^{ + 1.04  }_{ - 0.55 }$  \\ 
 2672.01   & 88.5168  $^{ + 0.0005 }_{ - 0.0005 }$  & -1276.2080  $^{ + 0.0011 }_{ - 0.0011 }$   & 0.059  $^{ + 0.006 }_{ - 0.006 }$   & 0.022  $^{ + 0.004 }_{ - 0.004 }$  &  12.6  $^{ + 1.7 }_{ - 1.6 }$  &  11.8  $^{ + 1.8 }_{ - 1.6 }$ &  4.58  $^{ + 0.13 }_{ - 0.12 }$  & 0.68  $^{ + 0.12 }_{ - 0.10 }$ \\
\hline
 2672.02   & 42.9926 $^{ + 0.0002  }_{ - 0.0002 }$  & -1316.6383 $^{ + 0.0007  }_{ - 0.0006 }$    & -0.095 $^{ + 0.025  }_{ - 0.032 }$    & -0.031 $^{ + 0.006  }_{ - 0.008 }$   & 5.92 $^{ + 2.93  }_{ - 1.57 }$  & 5.60 $^{ + 2.82  }_{ - 1.51 }$ &  3.23  $^{ + 0.50 }_{ - 0.19 }$   & 0.95 $^{ + 0.84  }_{ - 0.40 }$  \\ 
 2672.01   & 88.5231  $^{ + 0.0025 }_{ - 0.0015 }$  & -1276.1964  $^{ + 0.0026 }_{ - 0.0059 }$   & -0.035  $^{ + 0.016 }_{ - 0.021 }$   & -0.067  $^{ + 0.013 }_{ - 0.017 }$  &  4.91  $^{ + 1.62 }_{ - 1.18 }$  &  4.64  $^{ + 1.53 }_{ - 1.14 }$ &  4.58  $^{ + 0.13 }_{ - 0.12 }$  & 0.27  $^{ + 0.09 }_{ - 0.07 }$ \\ 
 2672x   & 140.9352  $^{ + 0.0781 }_{ - 0.0867 }$  & -1233.5344  $^{ + 8.5149 }_{ - 1.8773 }$  & -0.099  $^{ + 0.043 }_{ - 0.021 }$ &  -0.038  $^{ + 0.123 }_{ - 0.024 }$  & $<$ 6.18 & $<$ 5.99 &  ---  & ---  \\ 
 \hline
  2672.02  & 42.9919 $^{ + 0.0002  }_{ - 0.0002 }$  & -1316.6379 $^{ + 0.0006  }_{ - 0.0006 }$    & -0.062 $^{ + 0.009  }_{ - 0.010 }$    & 0.019 $^{ + 0.005  }_{ - 0.004 }$   & 6.93 $^{ + 0.93  }_{ - 0.88 }$  & 6.54 $^{ + 0.99  }_{ - 0.92 }$ &  3.23  $^{ + 0.50 }_{ - 0.19 }$   & 1.08 $^{ + 0.74  }_{ - 0.39 }$  \\ 
 2672x & 57.3666  $^{ + 0.0085 }_{ - 0.0082 }$  & -1288.0842  $^{ + 0.1530 }_{ - 0.1576 }$   & 0.019  $^{ + 0.007 }_{ - 0.007 }$   & -0.154  $^{ + 0.010 }_{ - 0.011 }$  &  0.20  $^{ + 0.03 }_{ - 0.03 }$  &  0.19  $^{ + 0.03 }_{ - 0.03 }$ &  ---  & ---\\ 
 2672.01   & 88.5173  $^{ + 0.0006 }_{ - 0.0006 }$ & -1276.2048  $^{ + 0.0011 }_{ - 0.0012 }$   & -0.021  $^{ + 0.016 }_{ - 0.016 }$  &  0.067  $^{ + 0.008 }_{ - 0.008 }$  & 5.35  $^{ + 0.58 }_{ - 0.57 }$  & 5.05  $^{ + 0.64 }_{ - 0.61 }$ &  4.58  $^{ + 0.13 }_{ - 0.12 }$   & 0.29  $^{ + 0.05 }_{ - 0.04 }$   \\ 
 \hline
    \end{tabular}    
    \caption{Posteriors summarized for KOI-2672 (Kepler-396) at the epoch BJD = 2,455,680, with Two Planets (Top Section), and fits to possible Three-planet Models in Lower Sections, with a Putative Nontransiting Planet named KOI-2672x in Lower Sections. Note. The columns denote the planet, orbital period, first transit after epoch given in BJD-2,457,000, e$\cos\omega$, e$\sin\omega$, dynamical mass scaled to the Earth-Sun system, mass and radius in Earth units, and bulk density in g cm$^{-3}$.}\label{tbl-koi2672}
  \end{center}
\end{table}

After burn-in, model fits with an additional planet beyond Kepler-396 c (model A) settled at an orbital period $\approx$140.9 days. In this case, the additional perturber interacts only with Kepler-396 c owing to the near 3:2 resonance with a TTV periodicity $\approx 770$ days, while its period ratio with Kepler-396 b is 3.3, far from any low-order resonance. By contrast, in model B, with the perturber between the known planets, TTVs with a periodicity $\approx 1027$ days are caused by the near 3:2 resonance with Kepler-396 c. In this case, the orbital periods of Kepler-396 b and the putative perturber at $\approx$57.4 days are consistent with a 4:3 mean motion resonance. 

Despite these different configurations, the masses of the transiting planets in the three-planet models agree closely (see Table~\ref{tbl-koi2672}). Between the two cases here, we note that model A is preferred over model B by a significant $\Delta$BIC. The projected transit times for the two models disagree, and hence more observations could further constrain this system. Predicted transit times for the two distinct models are listed in the appendix in Table~\ref{tbl-predictions}.

\subsubsection{KOI-94, Kepler-89}
Kepler-89 has four transiting planets. \citet{Weiss2013} found a strong signal in RV and measured the mass of Kepler-89 d at 106 $\pm$ 11 $M_{\oplus}$. In addition, they reported a weak detection of Kepler-89 b (10.5 $\pm$ 4.6 $M_{\oplus}$) and upper limits only on the other planets. 

\begin{figure}[ht!]
\includegraphics [width = 1.7 in]{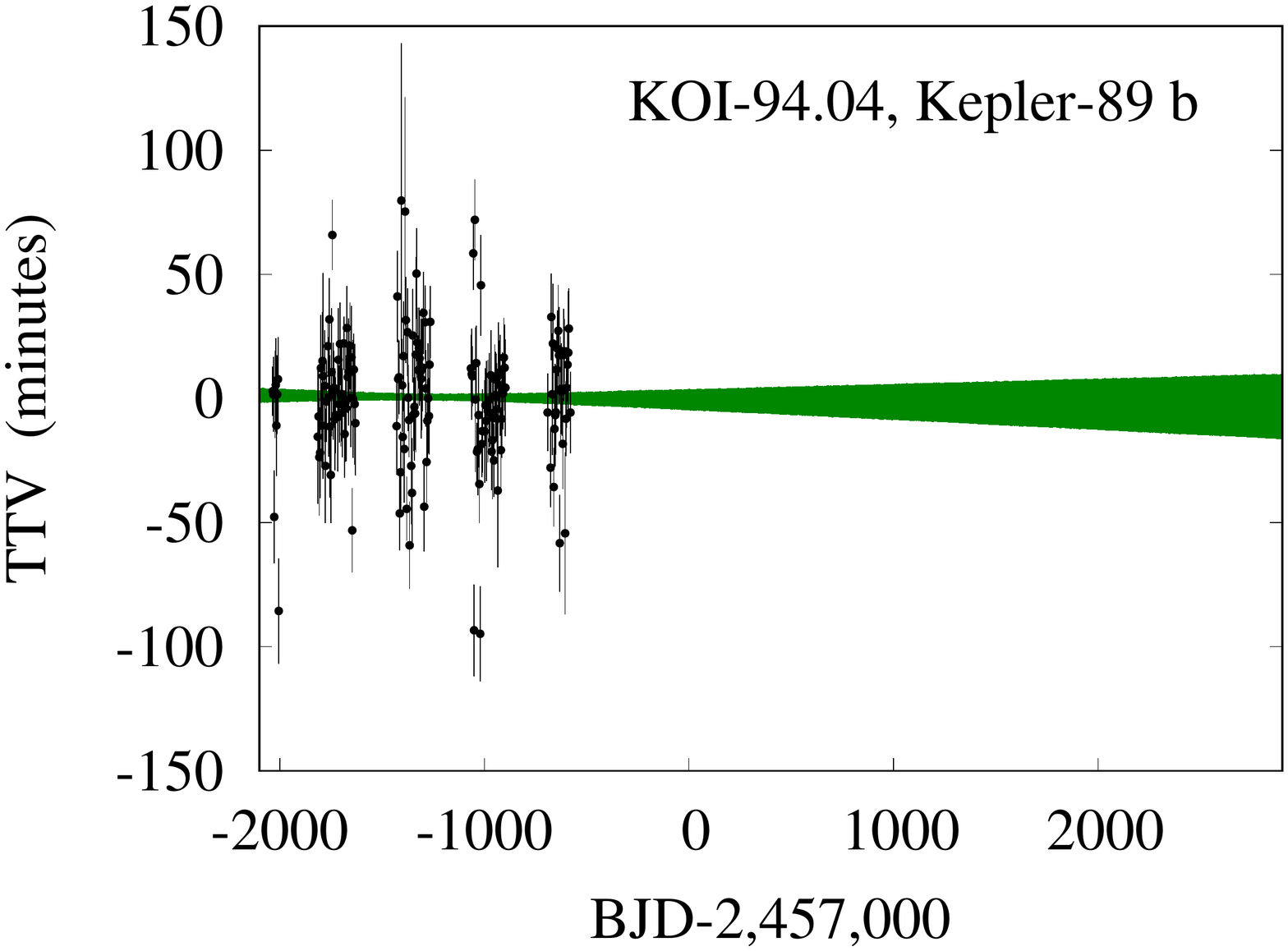}
\includegraphics [width = 1.7 in]{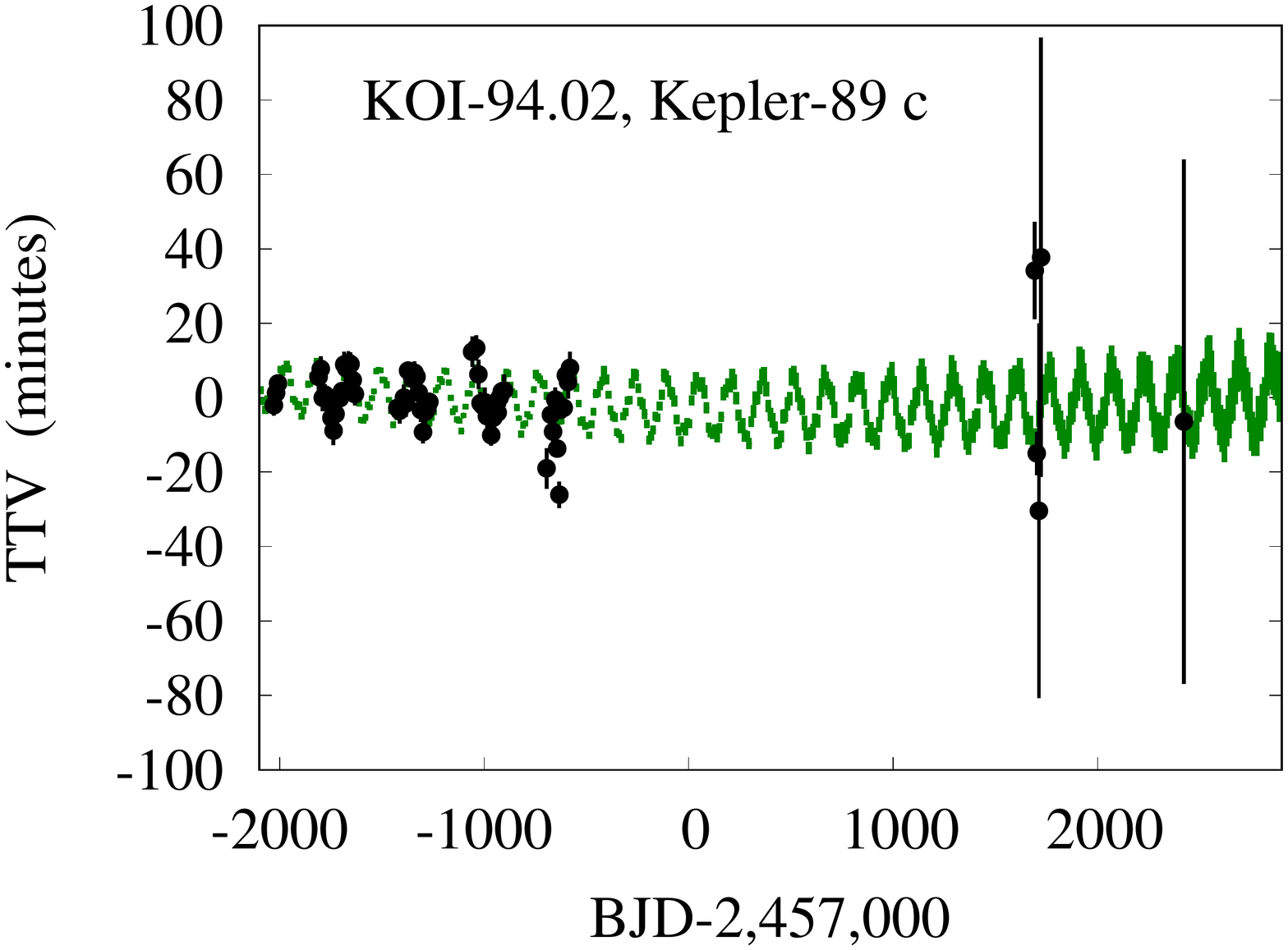}
\includegraphics [width = 1.7 in]{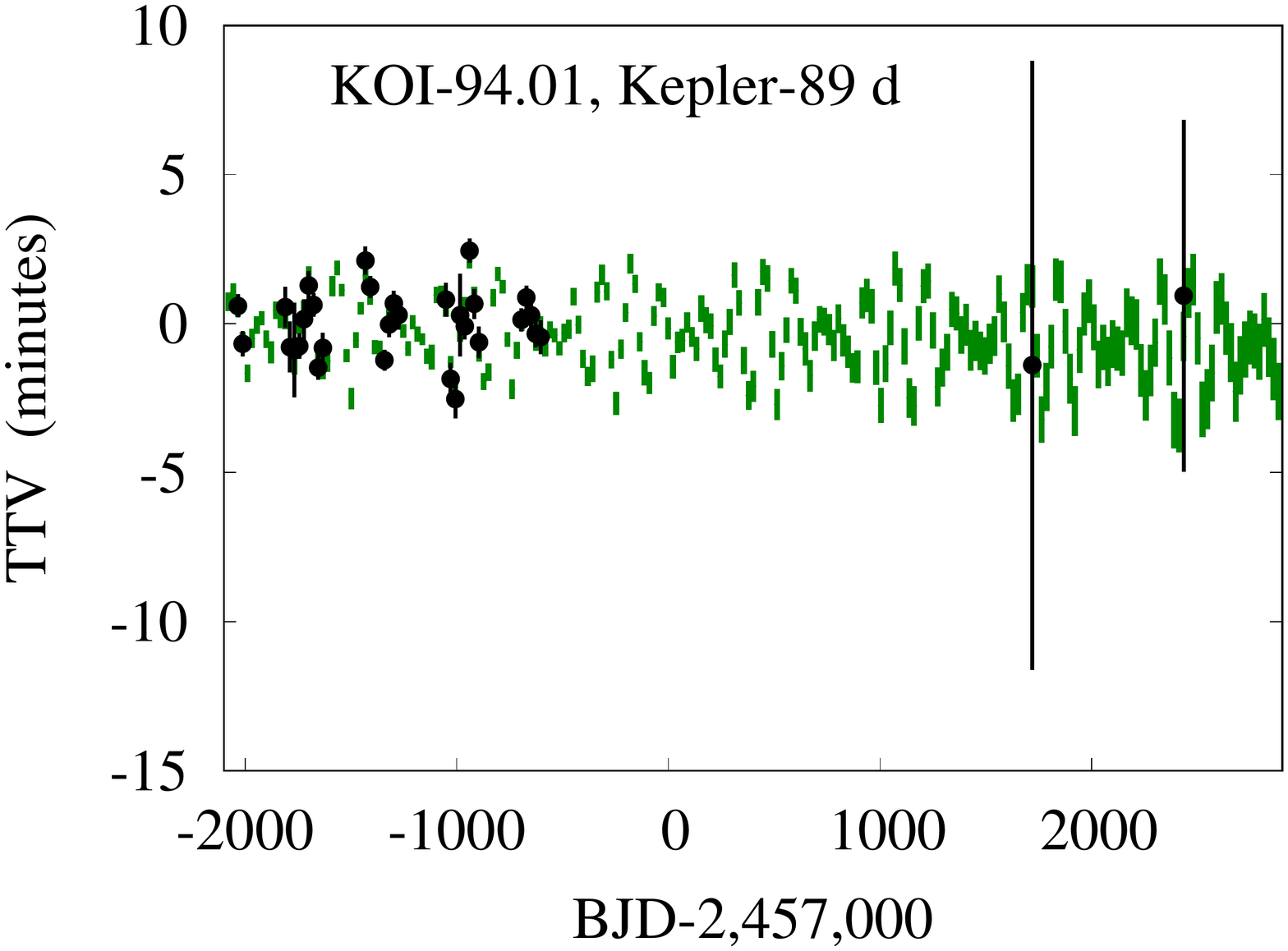}
\includegraphics [width = 1.7 in]{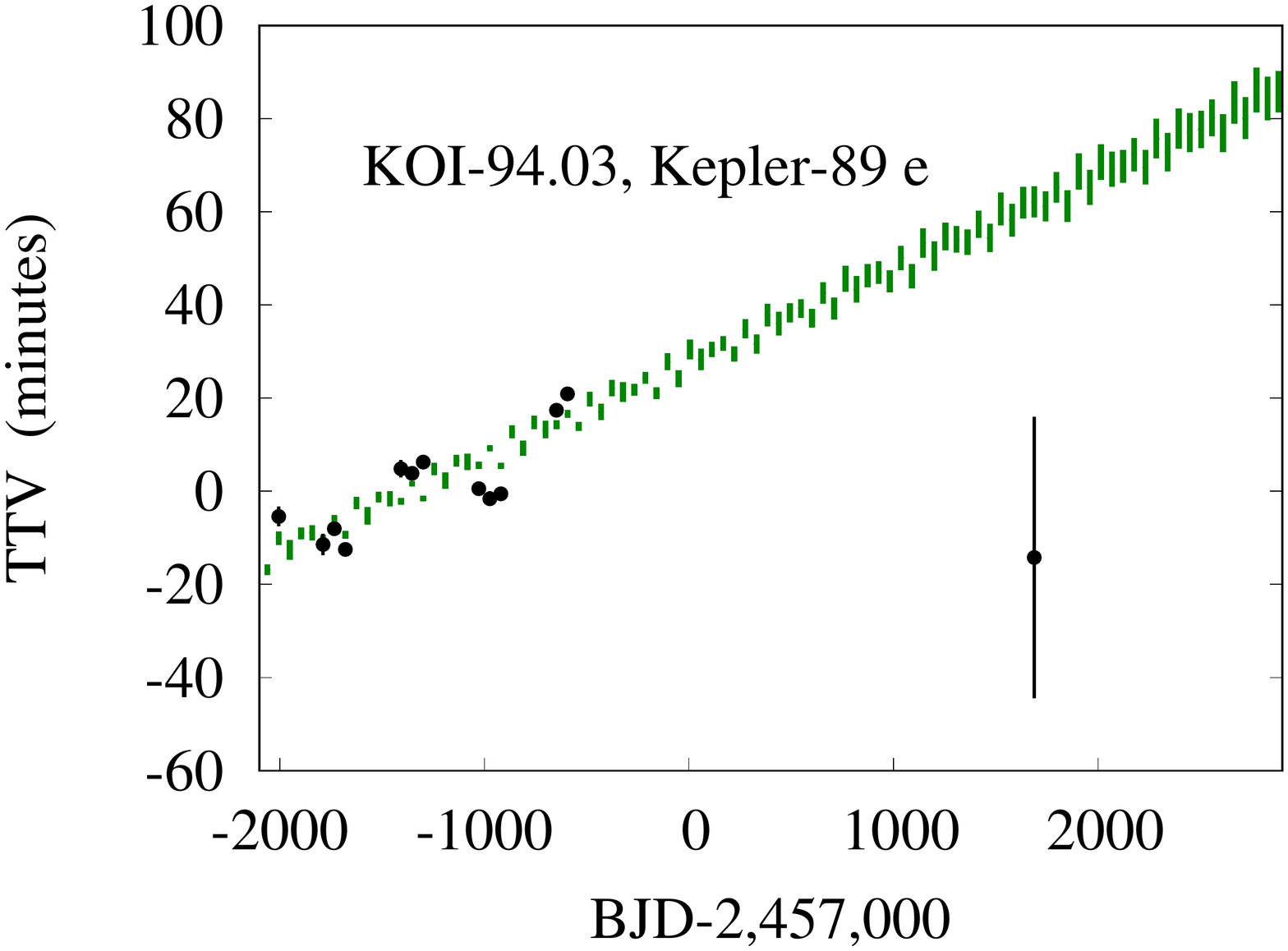}
\includegraphics [width = 1.7 in]{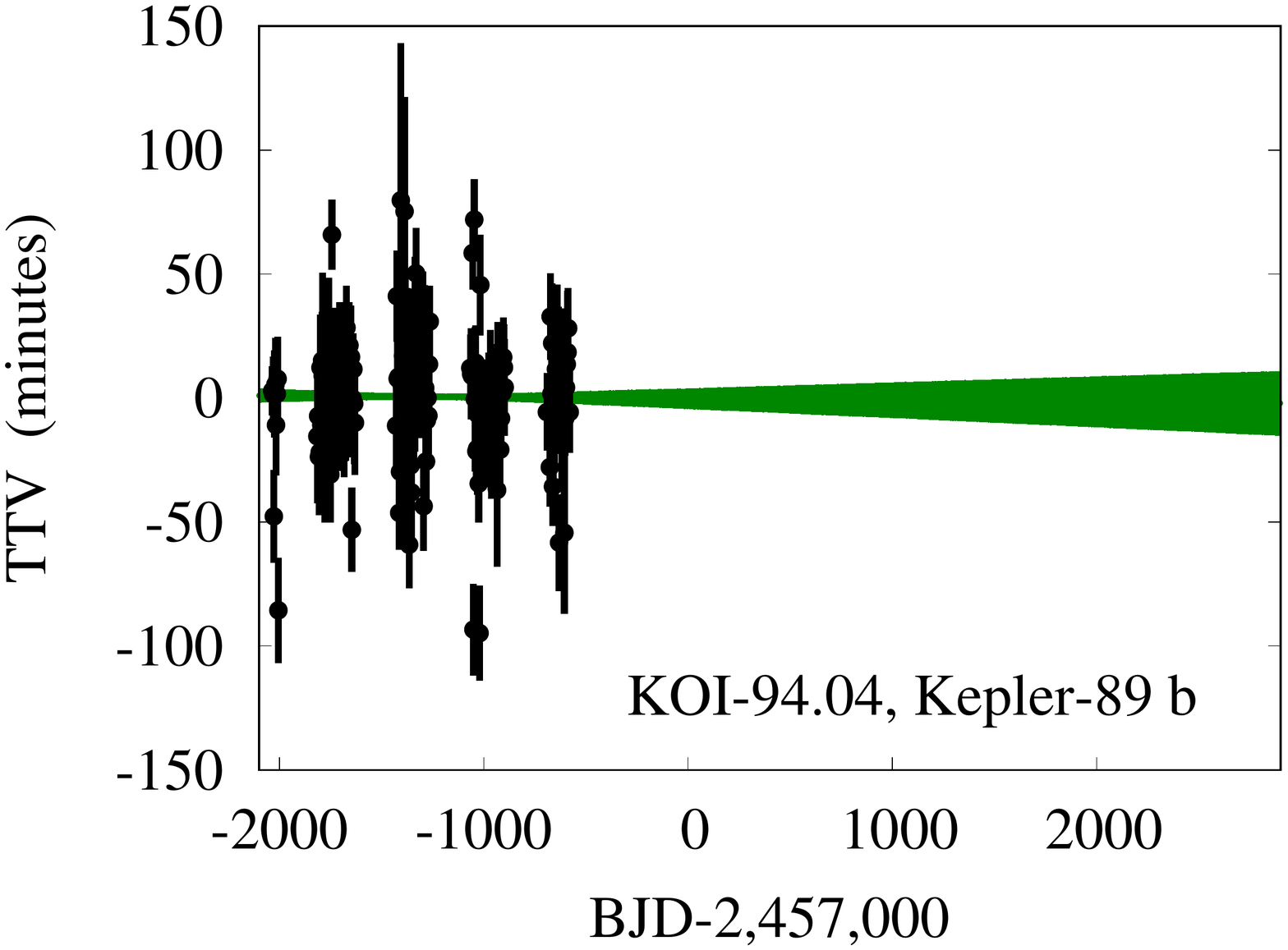}
\includegraphics [width = 1.7 in]{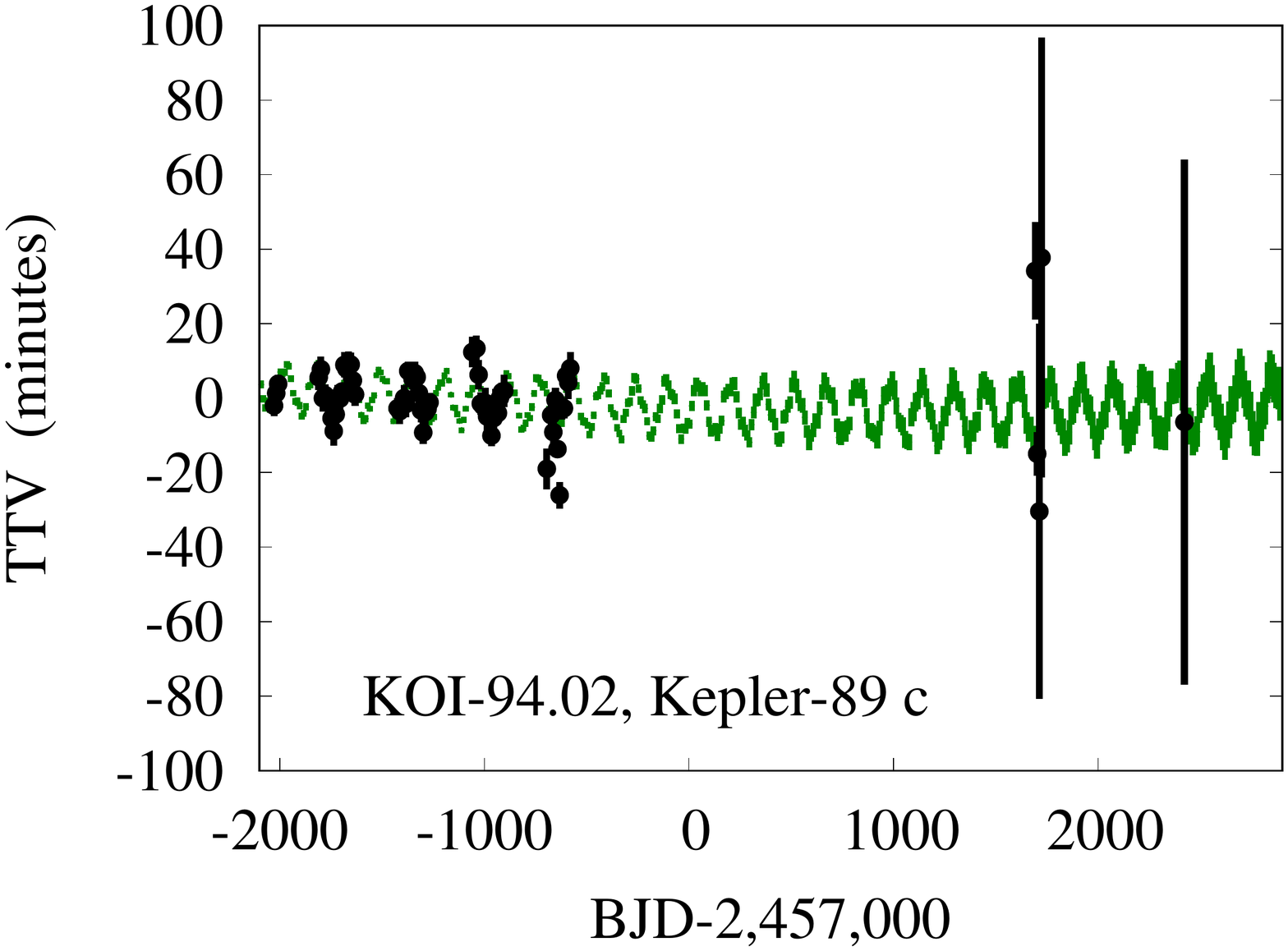}
\includegraphics [width = 1.7 in]{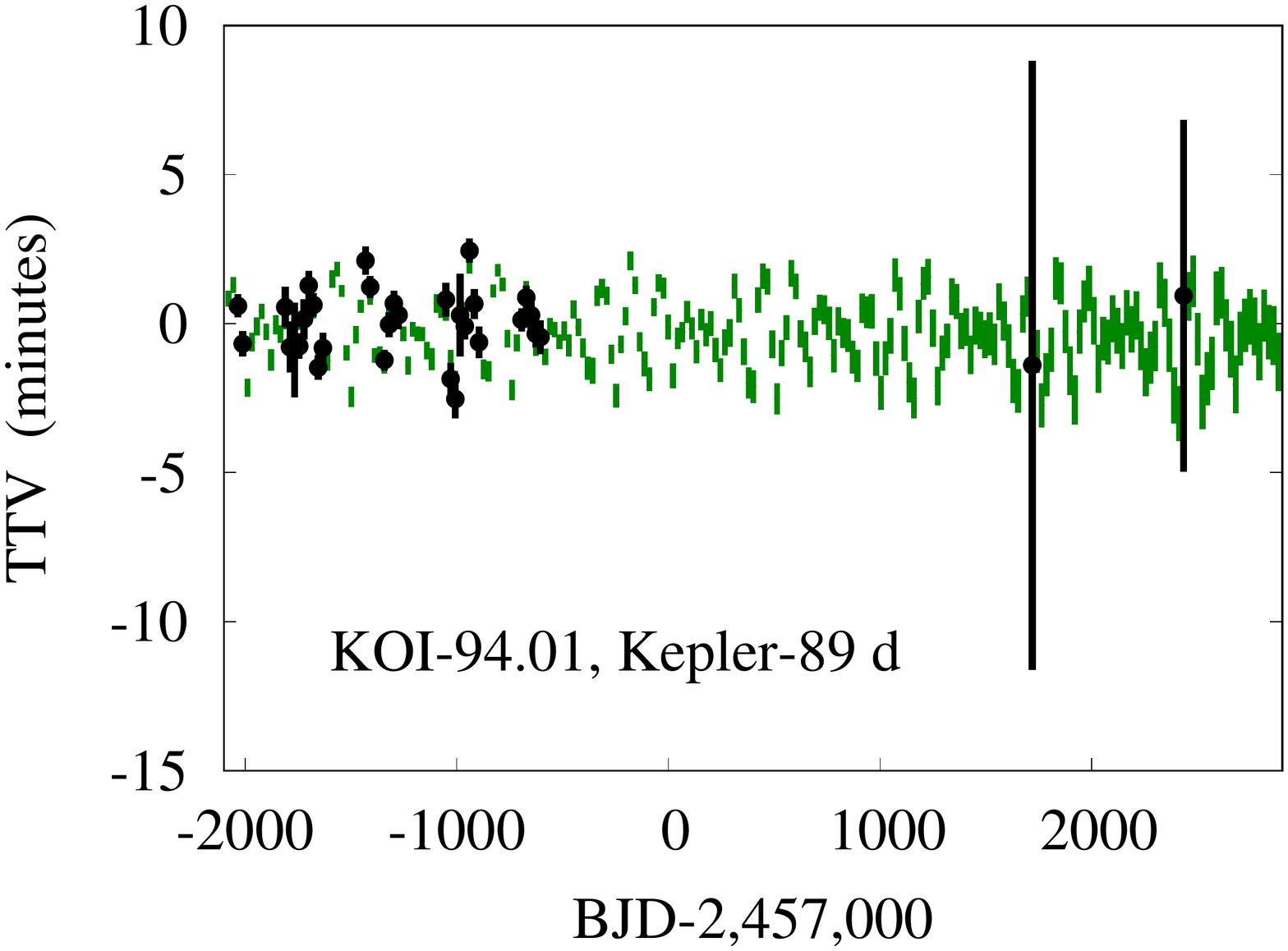}
\includegraphics [width = 1.7 in]{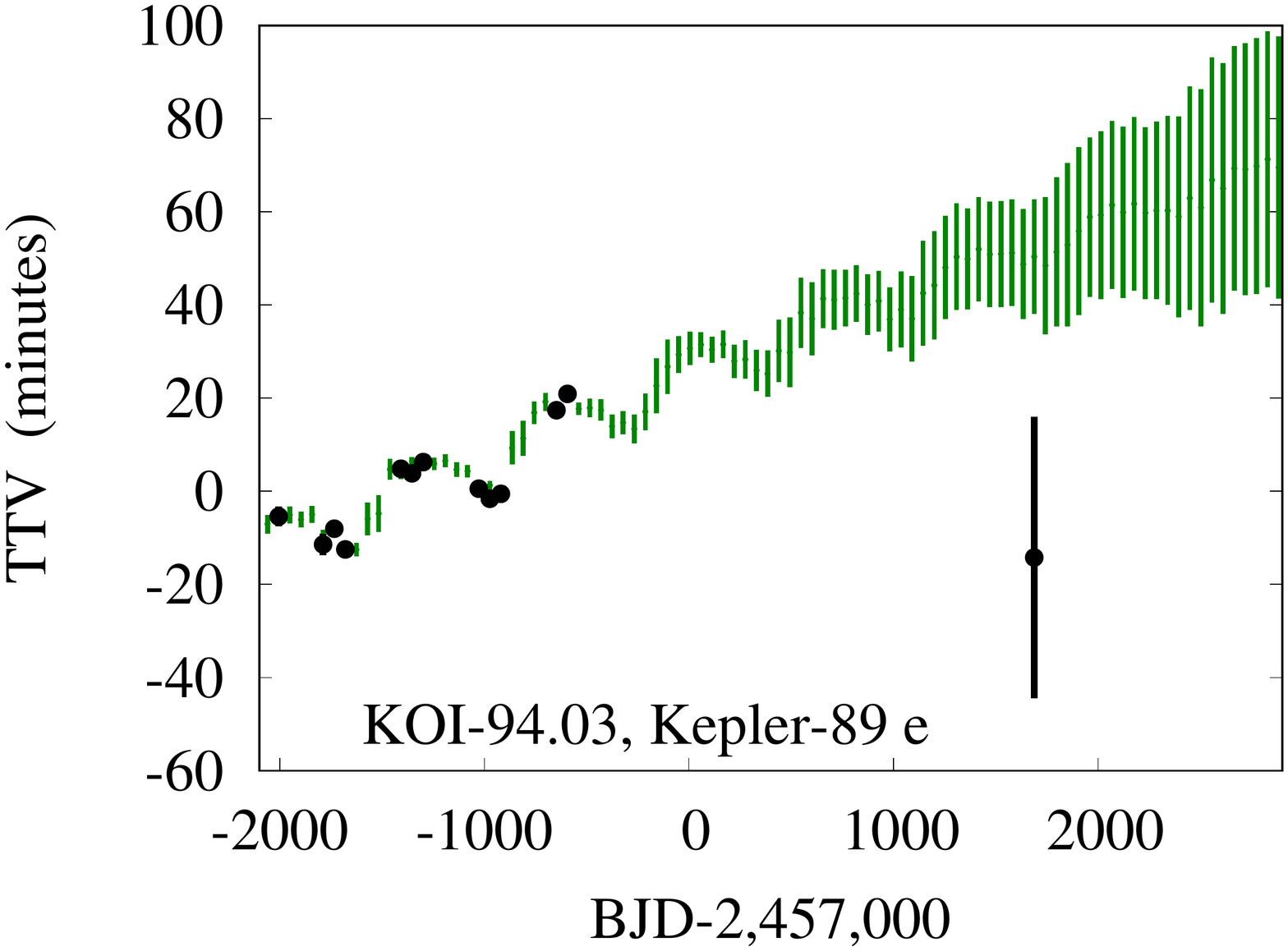}
\includegraphics [width = 1.7 in]{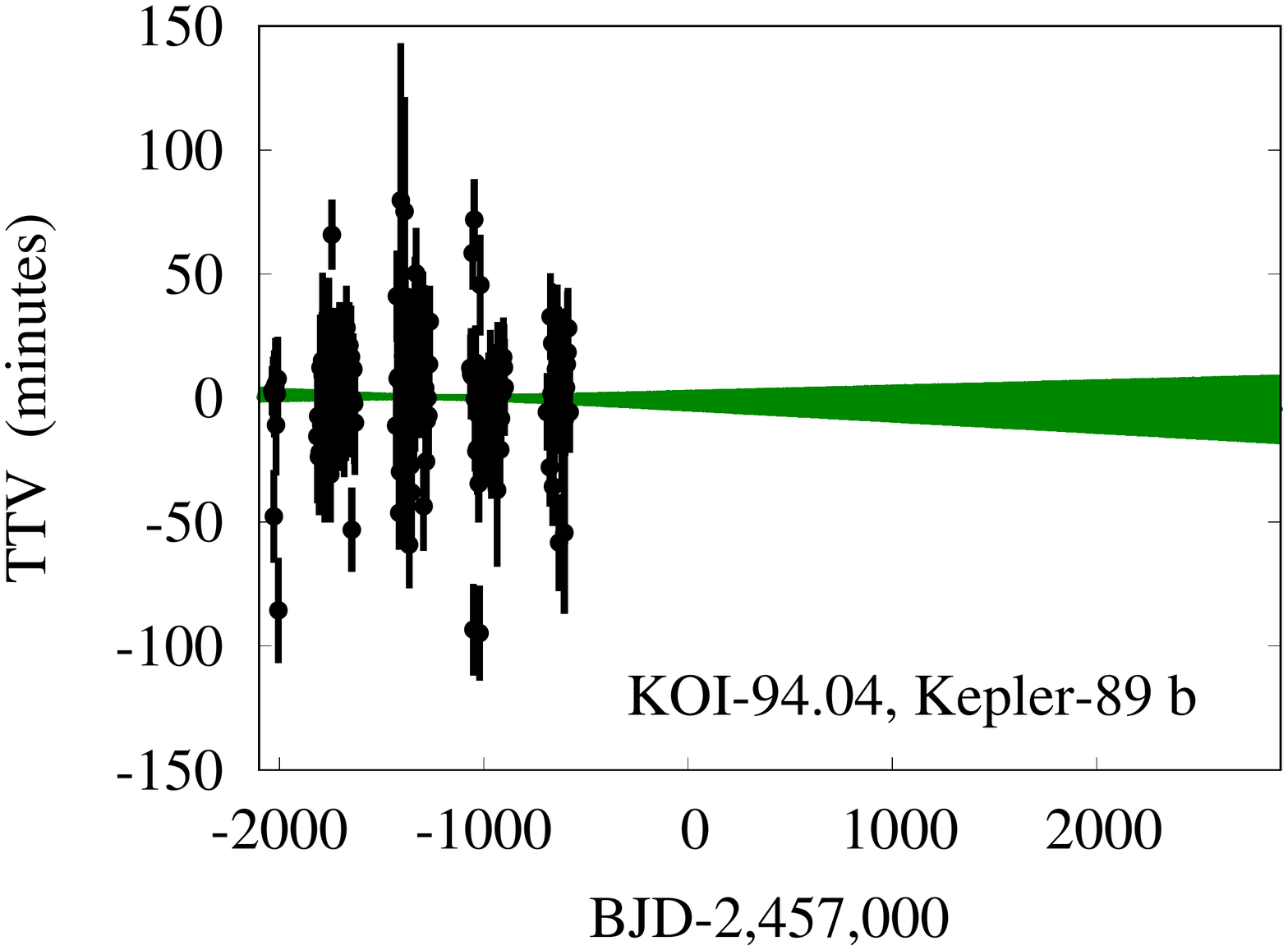}
\includegraphics [width = 1.7 in]{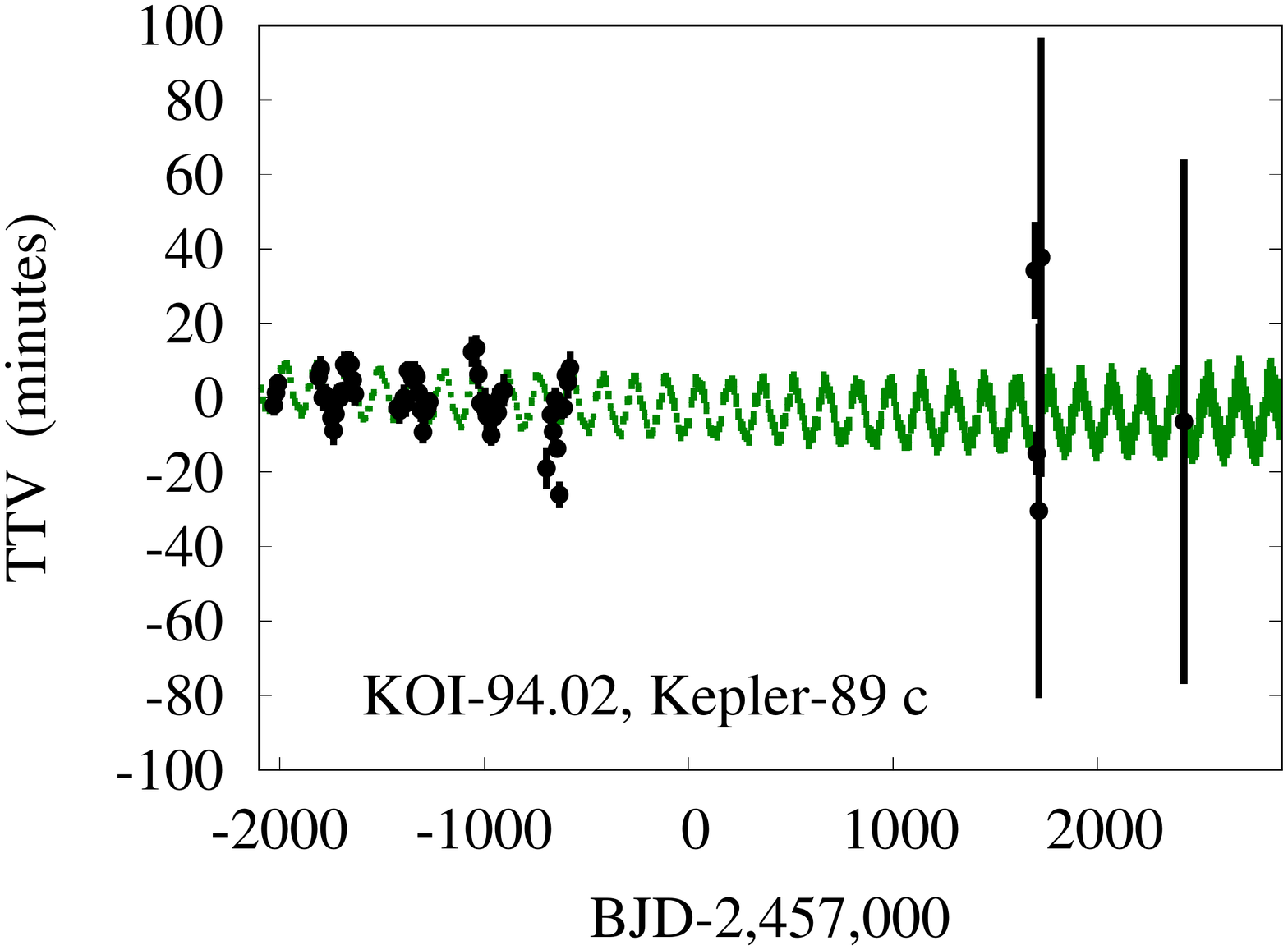}
\includegraphics [width = 1.7 in]{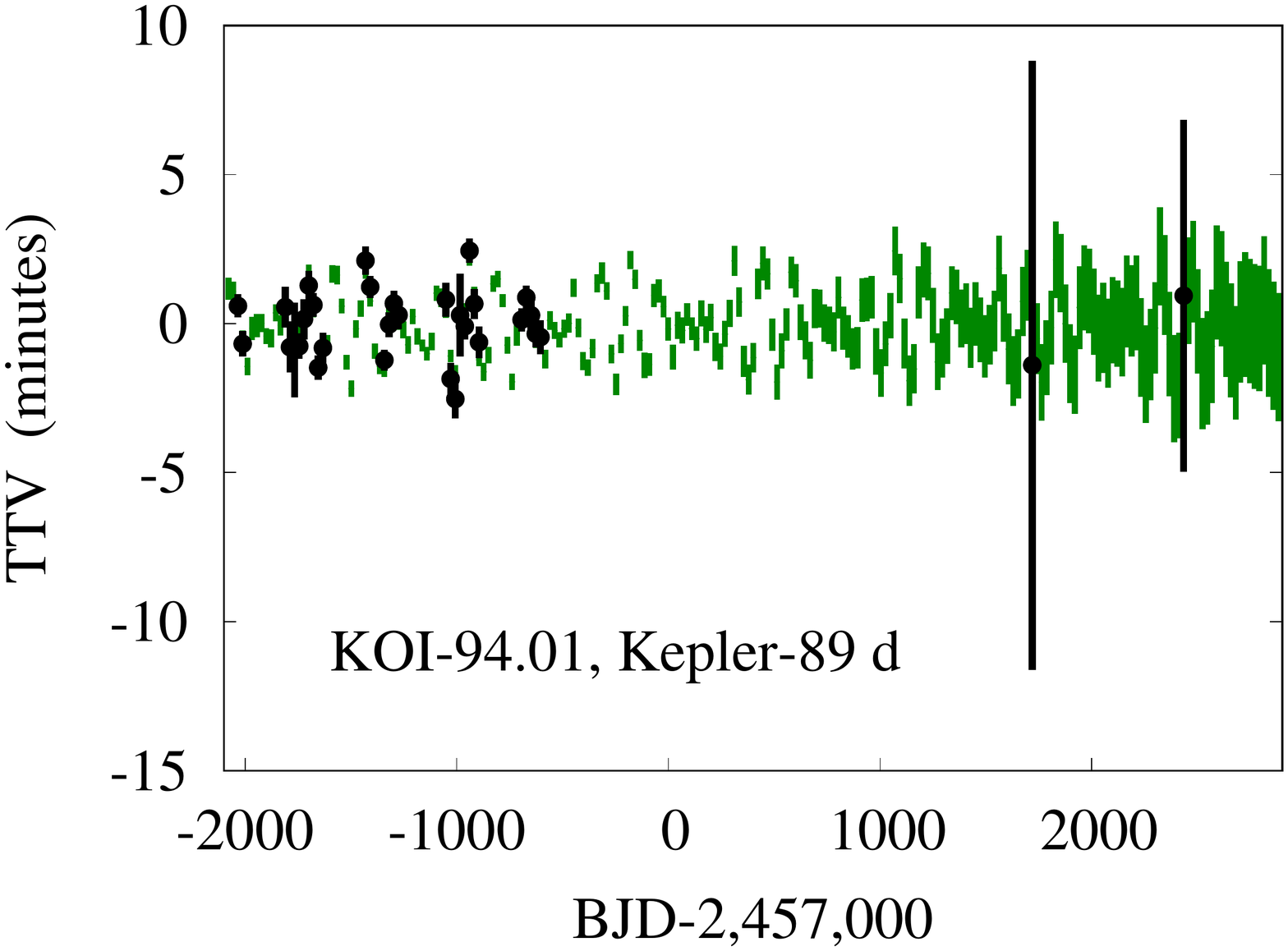}
\includegraphics [width = 1.7 in]{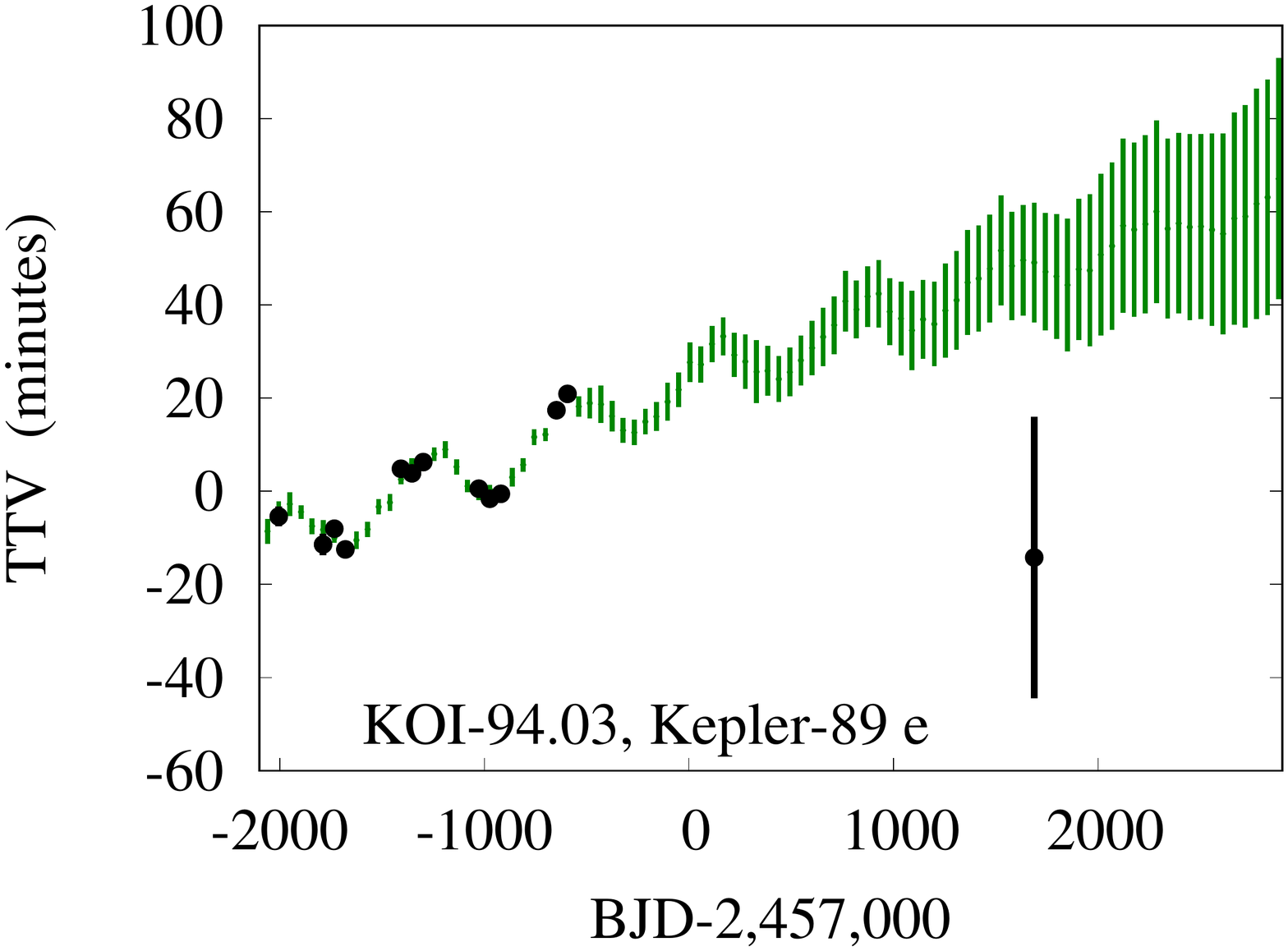}
\caption{
TTV models ($O-C$, in minutes) for KOI-94 (Kepler-89), with the combined \kepler /\tess\ transit times in black and 68\% confidence intervals with equal weight in the tails on simulated transit times in green. The top row corresponds to a four-planet model, the middle row includes a nontransiting planet orbiting beyond Kepler-89 e (model A), and the bottom row includes a nontransiting planet orbiting between Kepler-89 d and Kepler-89 e (model B).
}
\label{fig:Kep89_ttv} 
\end{figure}

\citet{Masuda2013} found a lower mass for Kepler-89 d from the TTVs ($52.1^{+6.9}_{-7.1}$ $M_{\oplus}$). The TTV signals seen in Kepler-89 c and Kepler-89 d (see Figure~\ref{fig:Kep89_ttv}) can be attributed to the $\sim$160-day periodicity from their proximity to the 2:1 resonance. Kepler-89 e has a strong TTV signal with a periodicity 0f $\sim$730 days. However, with large data gaps and missing transits in the transit timing catalog, the period, phase, and amplitude of this TTV signal are poorly constrained. In any case, the TTVs observed in Kepler-89 e are unlikely to be caused by any of the other transiting planets. It is far from any low-order resonance with other known planets; the period ratio between Kepler-89 e and its adjacent inner neighbor Kepler-89 d is 2.34. 

Since these earlier studies, the stellar mass of Kepler-89 has been revised slightly upward, from 1.28 $M_{\odot}$ to 1.34 $M_{\odot}$. This revision would cause the measured mass of Kepler-89 d from RV to be 111 $M_{\oplus}$.

With a four-planet model for the TTV at KOI-94 (Kepler-89), we find a mass for KOI-94.01 (Kepler-89 d) higher than those of \citet{Masuda2013} and \citet{Battley2021}. However, as shown in Table~\ref{tbl-koi94summary}, the best-fit TTV model with the four transiting planets does not provide a good fit overall, with a reduced $\chi^2 = 2.48$. 

We performed dynamical fits to the measured transit times with an additional planet included in the models. There are many possible solutions for an additional planet to explain the $\sim$730-day periodicity in the TTVs of Kepler-89 e, and it is beyond the purposes of this study to thoroughly explore these possibilities. Here we briefly explore two cases. In both cases, we retain the relative compactness of the Kepler-89 system, with the TTVs caused by near mean motion resonances. While there are systems of massive nontransiting planets causing TTVs near their orbital periods and confirmed with RV (e.g. Kepler-419 \citealt{Almenara2018}, Kepler-448 and Kepler-693, \citealt{Masuda2017}), in each of these cases the nontransiting perturber is significantly more massive than Jupiter. In the case of Kepler-89, \citet{Weiss2013} estimate an upper limit on planets at a distance of $\sim 1.7 $ AU (corresponding to a period $\sim$ 700 days) to be $\sim$ 0.1 $M_{\rm{Jup}}$. 

In model A, we inserted an additional planet outside the 54.32-day orbital period of Kepler-89 e, with our walkers initialized over a range of periods ranging from $\approx$93 to $\approx$124 days. For the nontransiting planet, we initialized $T_{0}$ randomly between our chosen epoch and one orbital period later. In this model, the fifth planet induces TTVs in Kepler-89 e but is too distant from the inner three planets to cause variations in their transit times. Hence, while this model is expected \textit{a priori} to improve the fit to the transit times of Kepler-89 e, it is unlikely to resolve the discrepancy between the RV and four-planet TTV model in measuring the mass of Kepler-89 d. 

In model B, we inserted a fifth planet interior to the orbit of Kepler-89 e with walkers initialized over a wide range of periods from $\approx$31 to $\approx$41 days. This is near the 2:3 commensurability with Kepler-89 e and near the 5:3 resonance with Kepler-89 d. Hence, in model B, the putative planet may interact with both Kepler-89 d and Kepler-89 e.

We summarize the best fits found for these four- and five-planet models in Table~\ref{tbl-koi94summary}. 
 \begin{table}[h!]
  \begin{center}
    \begin{tabular}{|c|c|c|c|c|c|}
      \hline
 model  & n  &     m    &     $\chi^2$  &  red.  $\chi^2$   &   $\Delta$BIC  \\   
 \hline
4pl    & 277 & 20  & 636.5 & 2.48 & 0 \\ % bic = 748.98
5pl A  & 277 & 25  & 474.7 & 1.88 & -133.7 \\  % bic = 615.30
5pl B  & 277 & 25  & 478.4 & 1.90 & -130.0 \\  % bic = 619.00
\hline 
    \end{tabular}    
    \caption{Summary of Models and Best-fit Solutions for Kepler-89. Note. Models listed in the first column include fits to \kepler\ and \tess\ transit times with four (4pl) or five planets (5pl), respectively. Model A (or B) refers to fits with a fifth planet initially inserted outside the orbit of Kepler-89 e (or between Kepler-89 d and e). The second and third columns list the number of data and free parameters, the fourth column lists the lowest $\chi^2$ found, the fifth column notes the reduced $\chi^2$, and the final column gives the change in BIC compared to the four-planet model.}\label{tbl-koi94summary}
  \end{center}
\end{table}

We found that an additional planet significantly improves the TTV model and that the additional five parameters are justified, but that there is no significant difference in the likelihood of the two different five-planet models. Our posteriors are summarized in Table~\ref{tbl-koi94}.

The TTVs plotted in Fig~\ref{fig:Kep89_ttv} show that an additional planet in the system improves the TTV fit for the outermost transiting planet, Kepler-89 e. Nevertheless, several outlying transit times remain, particularly for the second half of the Kepler data for Kepler-89 c.

 \begin{table}[h!]
 \tiny
  \begin{center}
    \begin{tabular}{|c|c|c|c|c|c|c|c|c|}
      \hline
KOI  & $P$ (days)  &      $T_0$ (days)    &      $e\cos\omega$   &  $e\sin\omega$     &     $ \frac{ M_p  }{M_{ \star} } \frac{M_{\odot} } {M_{\oplus} }  $ &  $ \frac{M_{p} }{M_{\oplus} }$ &  $ \frac{R_{p} }{R_{\oplus} }$ &  $\rho_{p}$ (g cm$^{-3}$) \\
\hline
 94.04 & 3.74316 $^{ + 0.00004  }_{ - 0.00004 }$  & -1316.6922 $^{ + 0.0009  }_{ - 0.0009 }$    & -0.063 $^{ + 0.038  }_{ - 0.044 }$    & 0.137 $^{ + 0.047  }_{ - 0.048 }$   & 43 $^{ + 19  }_{ - 16}$  & 58 $^{ + 25  }_{ - 21 }$ &  1.50  $^{ + 0.07 }_{ - 0.04 }$   & 95 $^{ + 46  }_{ - 36 }$  \\ 
 94.02 & 10.4284  $^{ + 0.0013 }_{ - 0.0012 }$  & -1309.7511  $^{ + 0.0011 }_{ - 0.0010 }$   & 0.033  $^{ + 0.009 }_{ - 0.009 }$   & 0.003  $^{ + 0.011 }_{ - 0.010 }$  &  7.0  $^{ + 2.0 }_{ - 1.6 }$  &  94  $^{ + 2.7 }_{ - 2.2 }$ &  3.71  $^{ + 0.08 }_{ - 0.07 }$  & 1.01  $^{ + 0.30 }_{ - 0.24 }$ \\ 
 94.01 & 22.3424  $^{ + 0.0004 }_{ - 0.0004 }$ & -1319.2829  $^{ + 0.0001 }_{ - 0.0001 }$   & 0.023  $^{ + 0.022 }_{ - 0.018 }$  &  0.026  $^{ + 0.020 }_{ - 0.017 }$  & 52.2  $^{ + 9.7 }_{ - 8.7 }$  & 70  $^{ + 13 }_{ - 12 }$ &  9.98  $^{ + 0.19 }_{ - 0.20 }$   & 0.39  $^{ + 0.08 }_{ - 0.07 }$   \\ 
 94.03 & 54.3261  $^{ + 0.0015 }_{ - 0.0015 }$ & -1299.5999  $^{ + 0.0012 }_{ - 0.0010 }$   & 0.024  $^{ + 0.040 }_{ - 0.039 }$  &  0.008  $^{ + 0.033 }_{ - 0.022 }$  & 13.9  $^{ + 2.5 }_{ - 2.2 }$  & 19  $^{ + 3 }_{ - 3 }$  &  6.07  $^{ + 0.12 }_{ - 0.18 }$  & 0.46  $^{ + 0.09 }_{ - 0.08 }$   \\ 
\hline
 94.04 & 3.74318 $^{ + 0.00002  }_{ - 0.00002 }$  & -1316.6922 $^{ + 0.0009  }_{ - 0.0009 }$    & -0.084 $^{ + 0.040  }_{ - 0.045 }$    & 0.115 $^{ + 0.037  }_{ - 0.036 }$   & 37 $^{ + 15  }_{ - 12 }$  & 50 $^{ + 19  }_{ - 16 }$ &  1.50  $^{ + 0.07 }_{ - 0.04 }$   & 81 $^{ + 36  }_{ - 28 }$  \\
 94.02 & 10.4254  $^{ + 0.0009 }_{ - 0.0010 }$  & -1309.7536  $^{ + 0.0008 }_{ - 0.0007 }$   & 0.046  $^{ + 0.014 }_{ - 0.010 }$   & -0.001  $^{ + 0.011 }_{ - 0.009 }$  &  4.2  $^{ + 1.3 }_{ - 1.2 }$  &  5.6  $^{ + 1.8 }_{ - 1.6 }$ &  3.71  $^{ + 0.08 }_{ - 0.07 }$  & 0.60  $^{ + 0.20 }_{ - 0.18 }$ \\ 
 94.01 & 22.3427  $^{ + 0.0003 }_{ - 0.0003 }$ & -1319.2828  $^{ + 0.0001 }_{ - 0.0001 }$   & 0.025  $^{ + 0.012 }_{ - 0.010 }$  &  -0.007  $^{ + 0.014 }_{ - 0.012 }$  & 37  $^{ + 8 }_{ - 8 }$  & 49  $^{ + 11 }_{ - 11 }$ &  9.98  $^{ + 0.19 }_{ - 0.20 }$   & 0.27  $^{ + 0.06 }_{ - 0.06 }$   \\ 
 94.03 & 54.3240  $^{ + 0.0013 }_{ - 0.0013 }$ & -1299.5951  $^{ + 0.0007 }_{ - 0.0008 }$   & 0.025  $^{ + 0.017 }_{ - 0.018 }$  &  -0.005  $^{ + 0.020 }_{ - 0.019 }$  & 19  $^{ + 4 }_{ - 3 }$  & 25.2  $^{ + 5.1 }_{ - 4.2 }$  &  6.07  $^{ + 0.12 }_{ - 0.18 }$  & 0.62  $^{ + 0.13 }_{ - 0.11 }$   \\ 
 94x & 118.0  $^{ + 0.4 }_{ - 0.6 }$ & -1312.3  $^{ + 8.5 }_{ - 5.9 }$   & 0.040  $^{ + 0.057 }_{ - 0.087 }$  &  -0.082  $^{ + 0.043 }_{ - 0.039 }$  & 7.0  $^{ + 2.6 }_{ - 1.8 }$  & 9.4  $^{ + 3.5 }_{ - 2.4 }$  &  ---  & ---   \\ 
\hline
 94.04   & 3.74316  $^{ + 0.00004 }_{ - 0.00005 }$ & -1316.6922 $^{ + 0.0009  }_{ - 0.0009 }$   & -0.054 $^{ + 0.067  }_{ - 0.068 }$  & 0.112 $^{ + 0.054  }_{ - 0.063 }$  & $<$ 65 & $<$ 87 &  1.50  $^{ + 0.07 }_{ - 0.04 }$  & $<$ 153  \\ 
 94.02 & 10.4265  $^{ + 0.0014 }_{ - 0.0014 }$  & -1309.7529  $^{ + 0.0013 }_{ - 0.0012 }$   & 0.025  $^{ + 0.011 }_{ - 0.007 }$   & 0.006  $^{ + 0.012 }_{ - 0.008 }$  &  5.2  $^{ + 1.9 }_{ - 1.7 }$  &  6.9  $^{ + 2.5 }_{ - 2.3 }$ &  3.71  $^{ + 0.08 }_{ - 0.07 }$  & 0.74  $^{ + 0.27 }_{ - 0.25 }$ \\ 
 94.01 & 22.3423  $^{ + 0.0004 }_{ - 0.0003 }$ & -1319.2827  $^{ + 0.0002 }_{ - 0.0002 }$   & -0.011  $^{ + 0.013 }_{ - 0.012 }$  &  0.009  $^{ + 0.016 }_{ - 0.013 }$  & 43  $^{ + 10 }_{ - 10 }$  & 57  $^{ + 13 }_{ - 14 }$ &  9.98  $^{ + 0.19 }_{ - 0.20 }$   & 0.32  $^{ + 0.08 }_{ - 0.08 }$   \\ 
 94x & 39.99  $^{ + 0.03 }_{ - 0.03 }$ & -1307.5  $^{ + 0.9 }_{ - 0.6 }$   & 0.020  $^{ + 0.034 }_{ - 0.030 }$  &  -0.023  $^{ + 0.030 }_{ - 0.027 }$  & 0.53  $^{ + 0.23 }_{ - 0.19 }$  & 0.71  $^{ + 0.31 }_{ - 0.25 }$  &  --- & ---  \\ 
 94.03 & 54.3268  $^{ + 0.0018 }_{ - 0.0018 }$ & -1299.5960 $^{ + 0.0008 }_{ - 0.0008 }$   & -0.025  $^{ + 0.024 }_{ - 0.022 }$  &  0.000  $^{ + 0.022 }_{ - 0.021 }$  & 15.5  $^{ + 3.0 }_{ - 2.7 }$  & 20.7  $^{ + 4.0 }_{ - 3.6 }$  &  6.07  $^{ + 0.12 }_{ - 0.18 }$  & 0.51  $^{ + 0.11 }_{ - 0.09 }$   \\
\hline
    \end{tabular}    
    \caption{Posteriors Summarized for Kepler-89 at the Epoch BJD = 2,455,680,  with Four Planets (Top Section), and Fits to Possible Five-planet Models in Lower Sections, with a Putative Nontransiting Planet Named KOI-94x. Note. The columns denote the planet, orbital period, first transit after epoch given in BJD-2,457,000, e$\cos\omega$, e$\sin\omega$, dynamical mass scaled to the Earth-Sun system, mass and radius in Earth units, and bulk density in g cm$^{-3}$.}\label{tbl-koi94}
  \end{center}
\end{table}

We find similar mass constraints on KOI-94.01 (Kepler-89 d) in both five-planet models, although the results are even more discrepant with the RV data than the four-planet model is. The mass of KOI-94.02 (Kepler-89 c) is lower in the five-planet models than the four-planet model, and the two solutions are consistent with each other. 

Most of the improvement in the fit from an additional planet appears in the TTVs of Kepler-89 e, which has a $\sim$730-day periodicity that cannot be explained by near-resonant interactions with the other transiting planets. In both of our five-planet models, the additional planet induces TTVs at this periodicity and improves the fit. The measured transit times from \tess\ for Kepler-89 c and Kepler-89 d are consistent with prior models, although their uncertainties are too large to have an important effect on the model fits. However, none of our models closely fit the measured transit time from \tess\ for Kepler-89 e. 

In summary, the \tess\ photometry does not appear to resolve the discrepancy between the TTV masses and the RV data, and while our tests of two different five-planet models improved the fit to the \kepler\ data, neither provided a good fit to the measured transit time of Kepler-89 e from \tess. 

\subsection{Other systems}
\begin{figure}[ht!]
\includegraphics [width = 2.3 in]{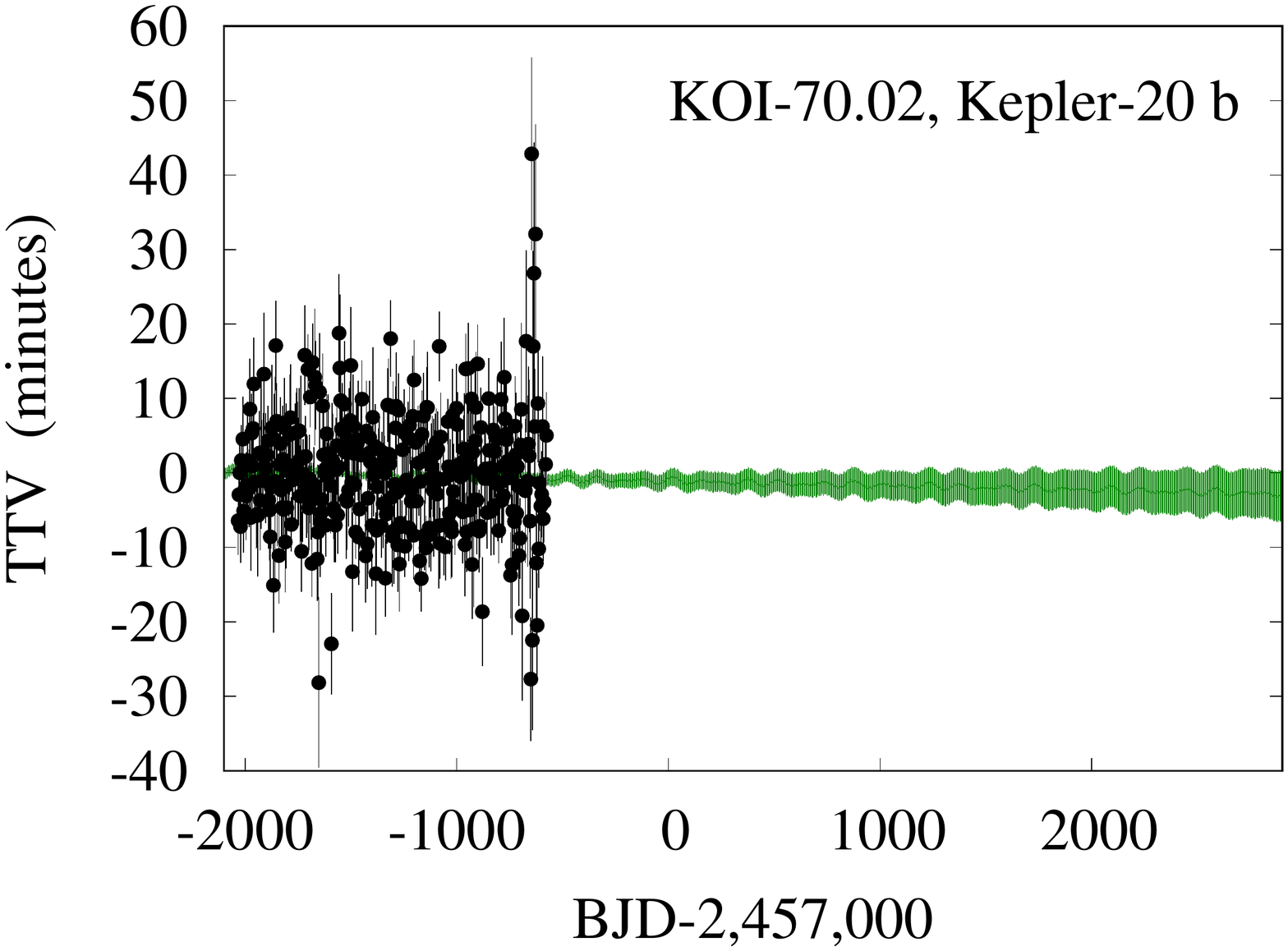}
\includegraphics [width = 2.3 in]{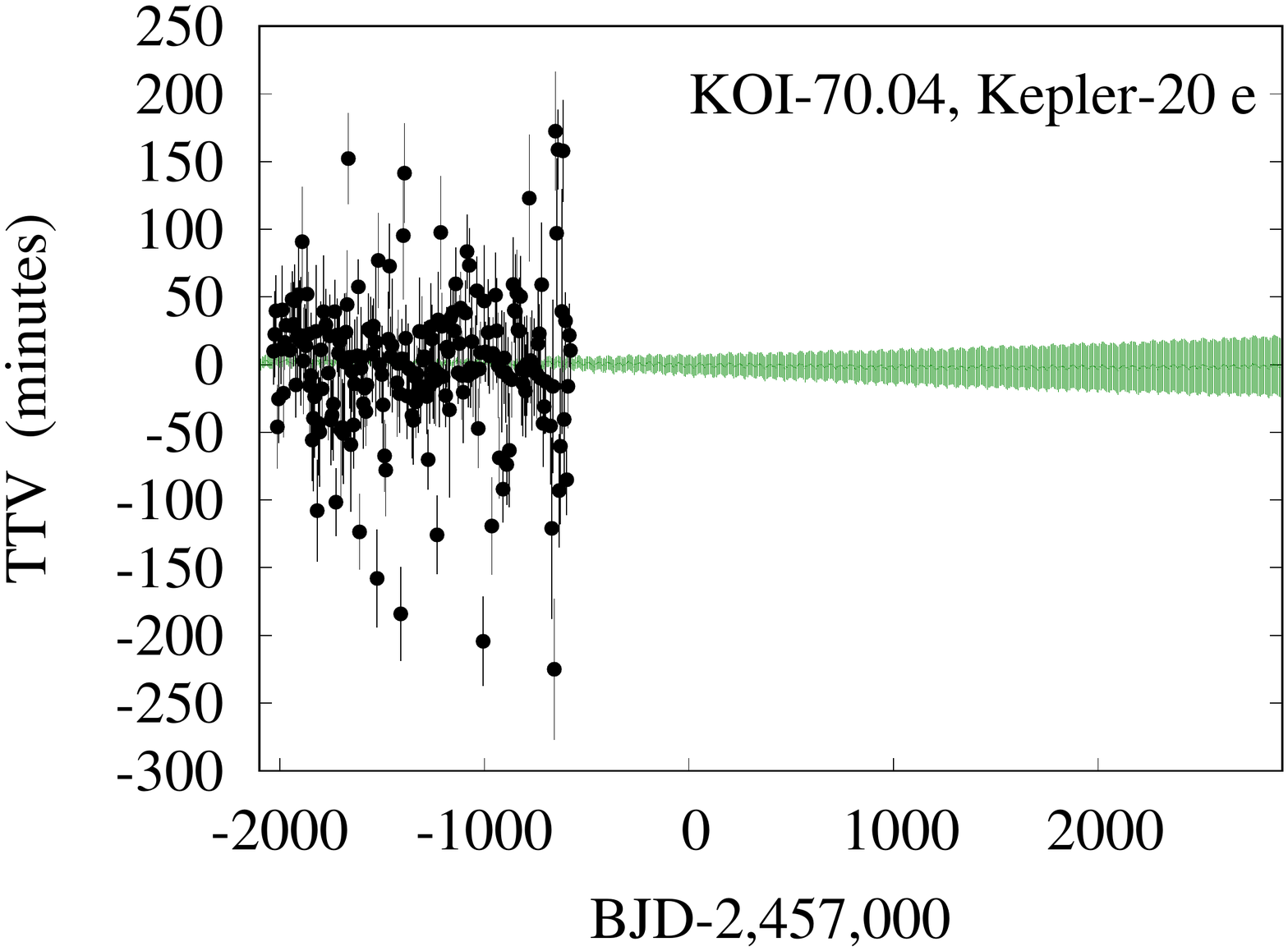}
\includegraphics [width = 2.3 in]{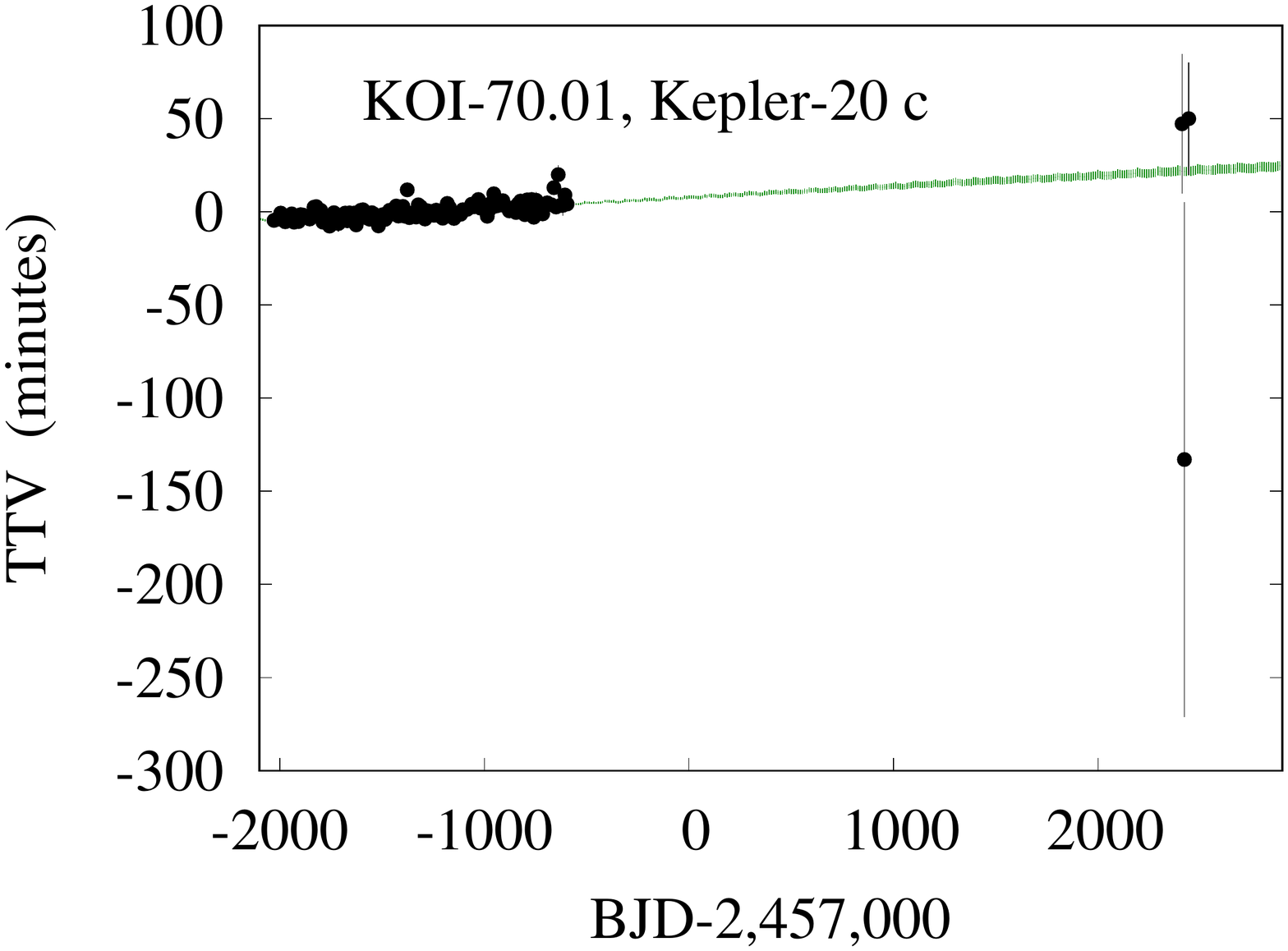}
\newline
\includegraphics [width = 2.3 in]{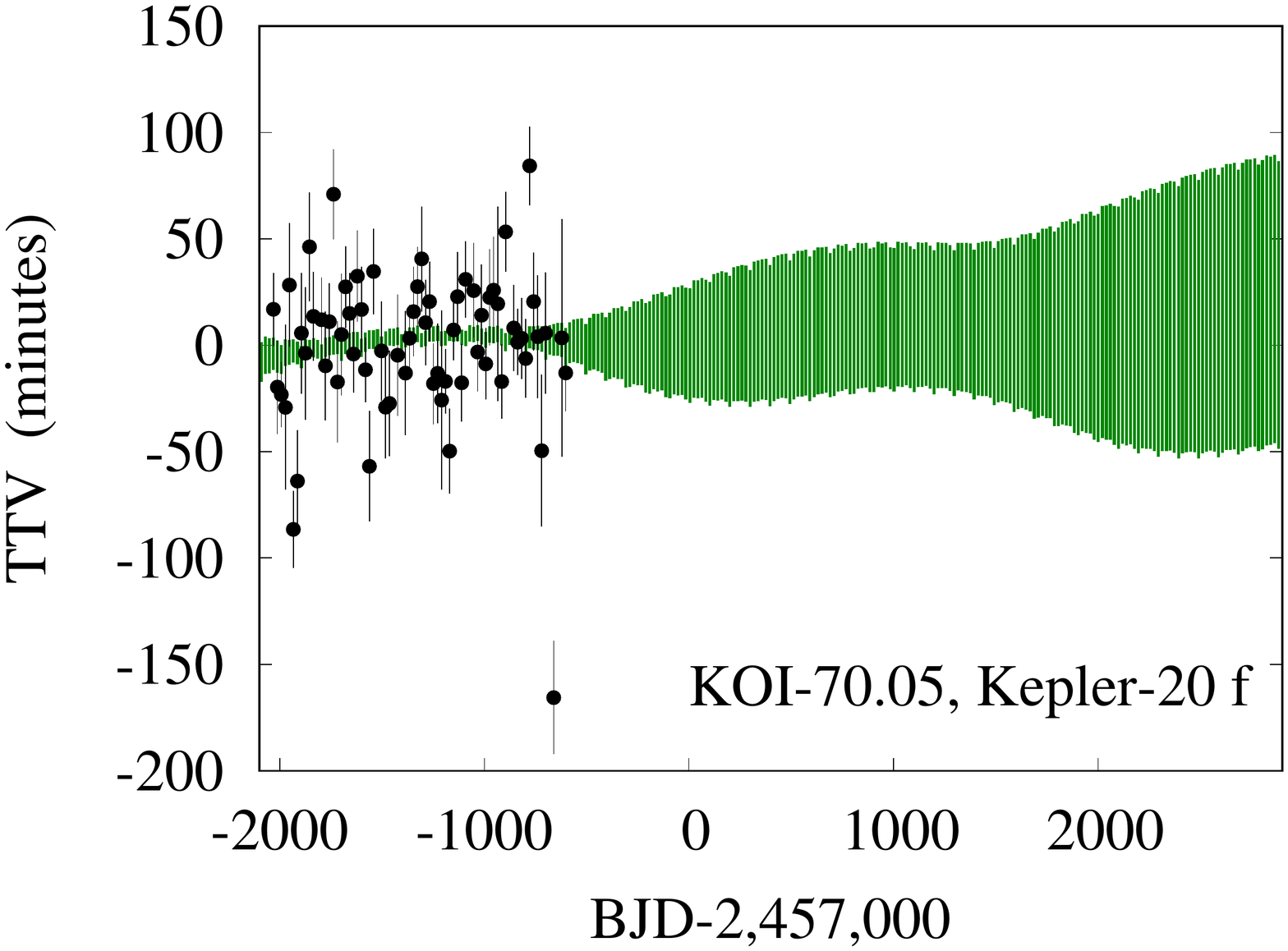}
\includegraphics [width = 2.3 in]{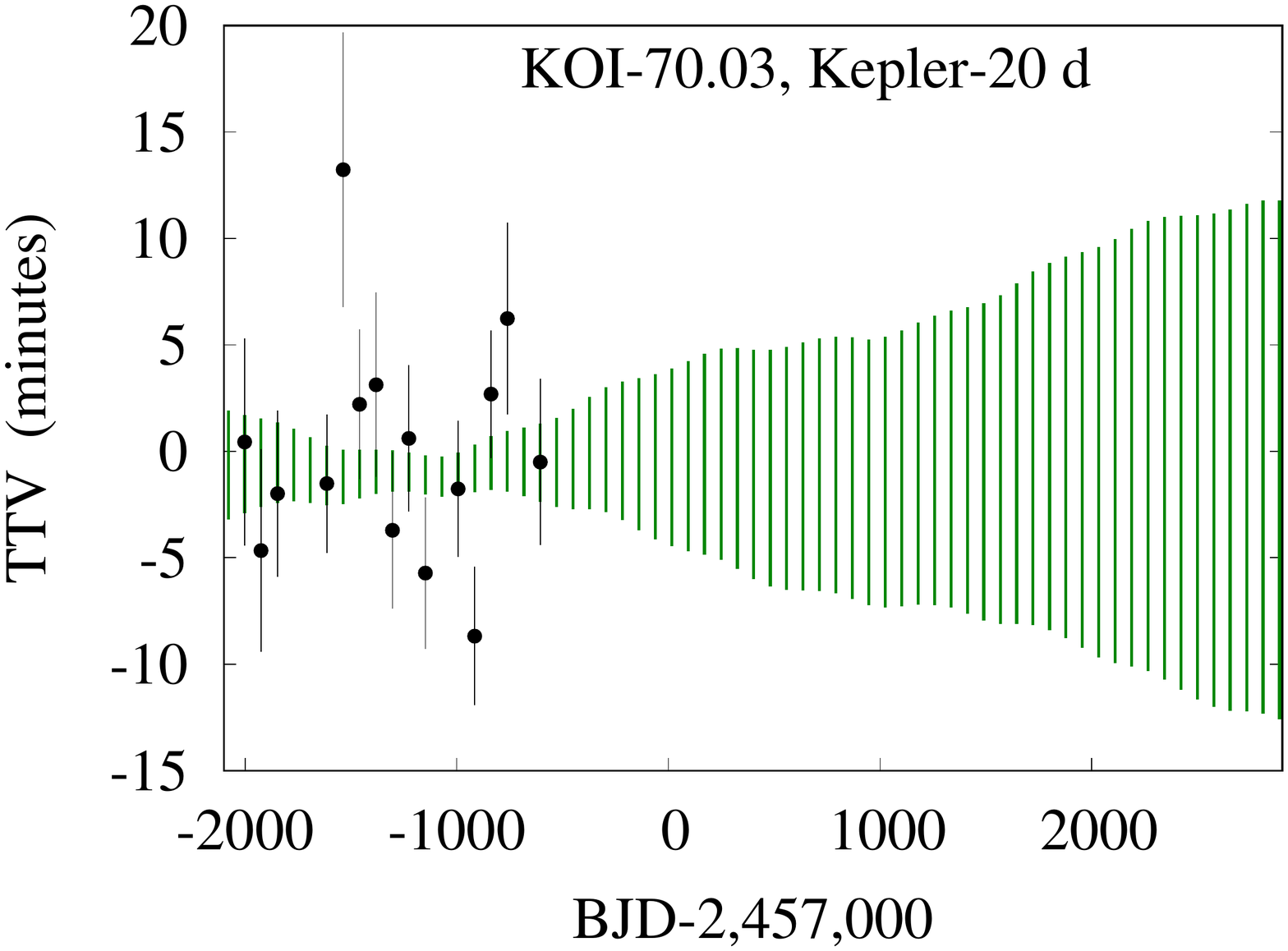}
\newline
\includegraphics [width = 2.3 in]{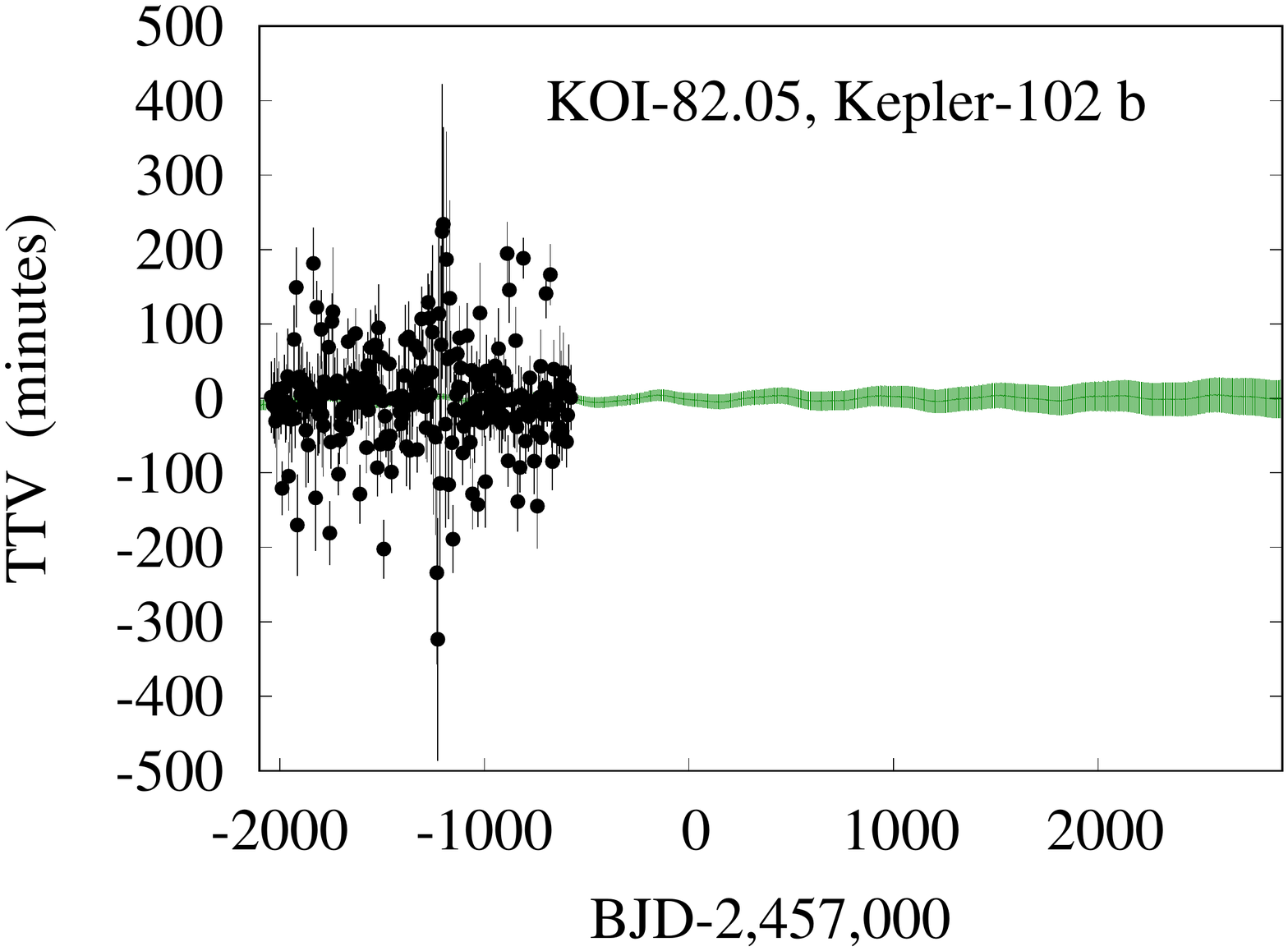}
\includegraphics [width = 2.3 in]{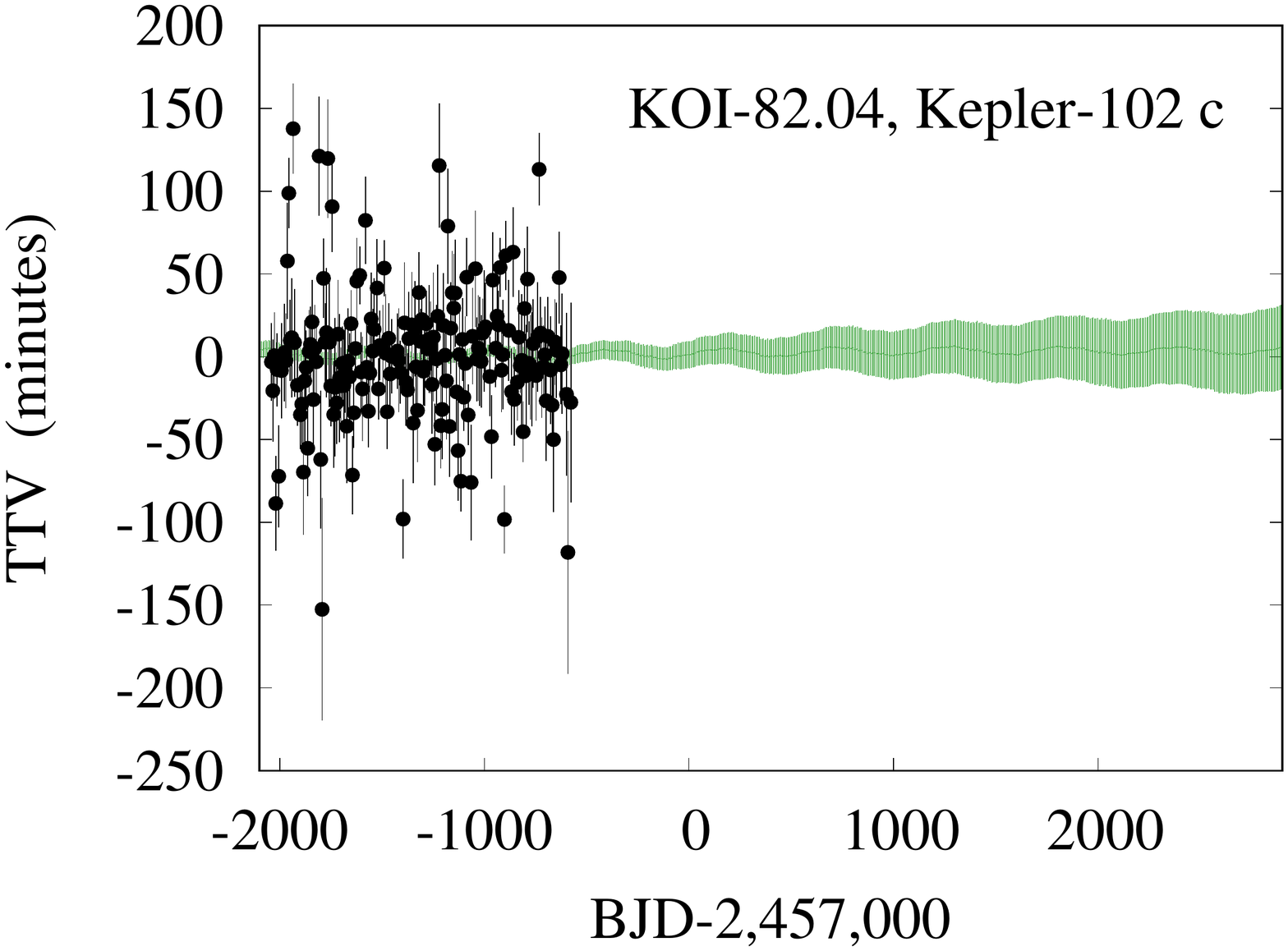}
\includegraphics [width = 2.3 in]{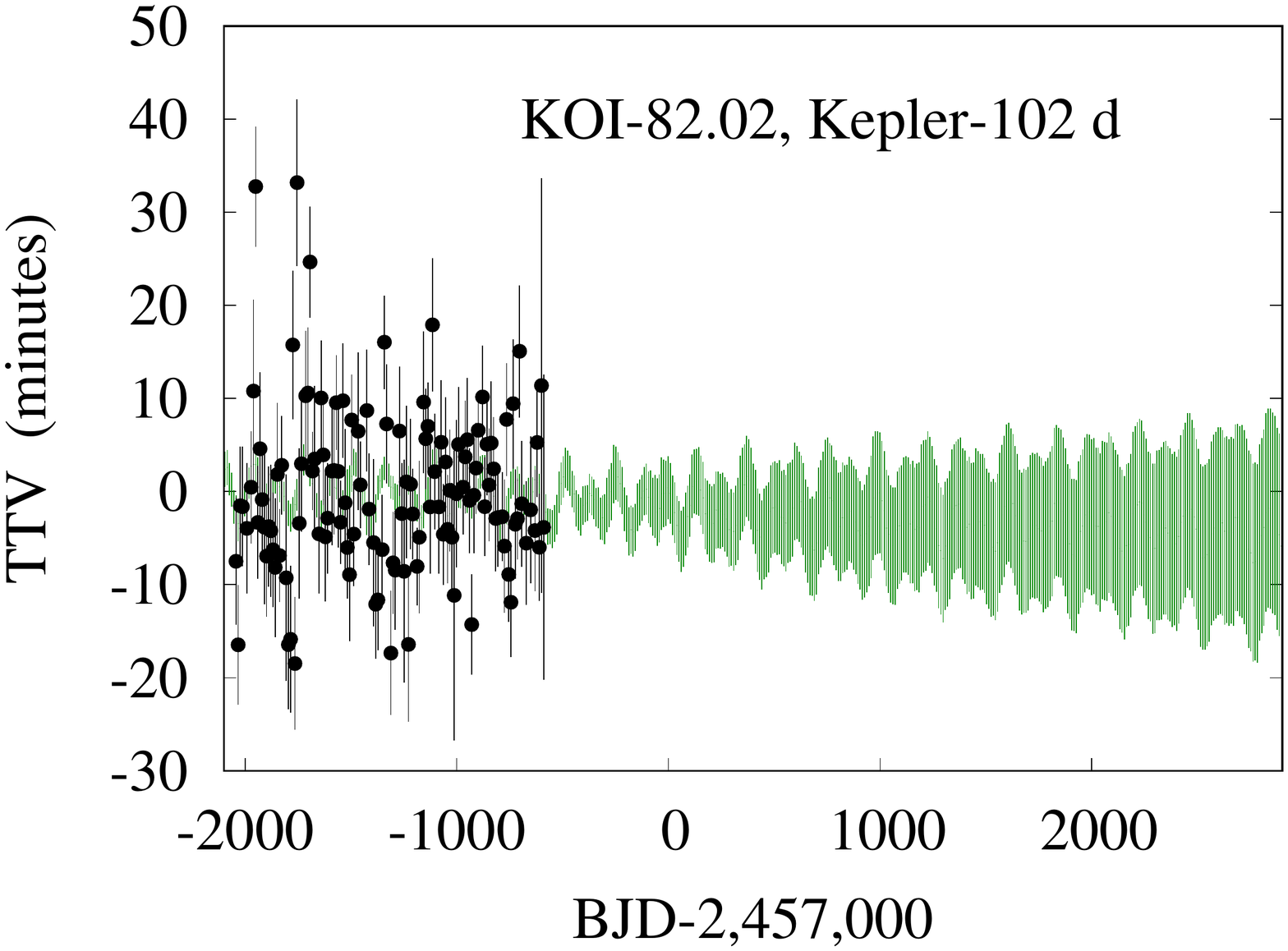}
\newline
\includegraphics [width = 2.3 in]{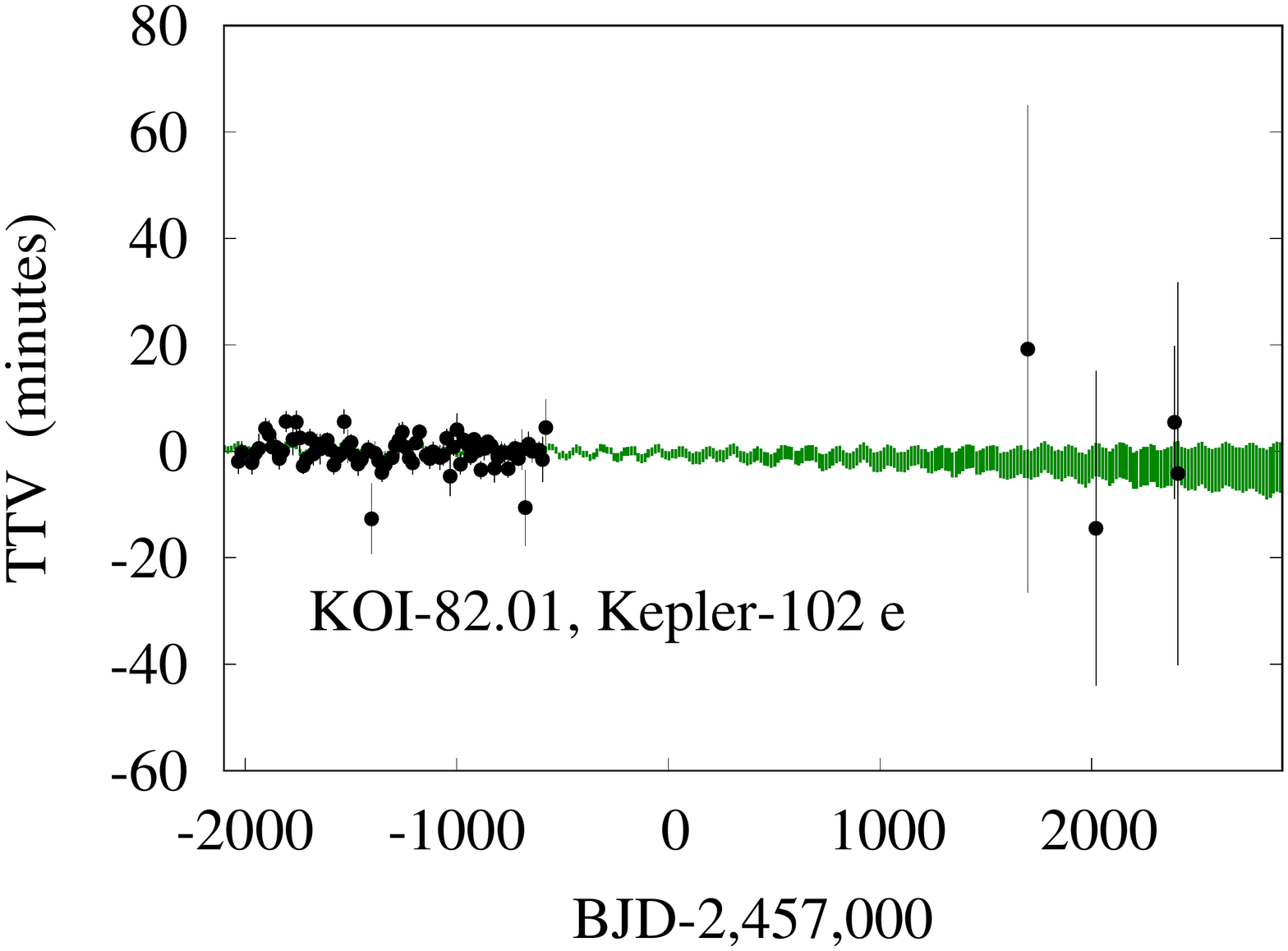}
\includegraphics [width = 2.3 in]{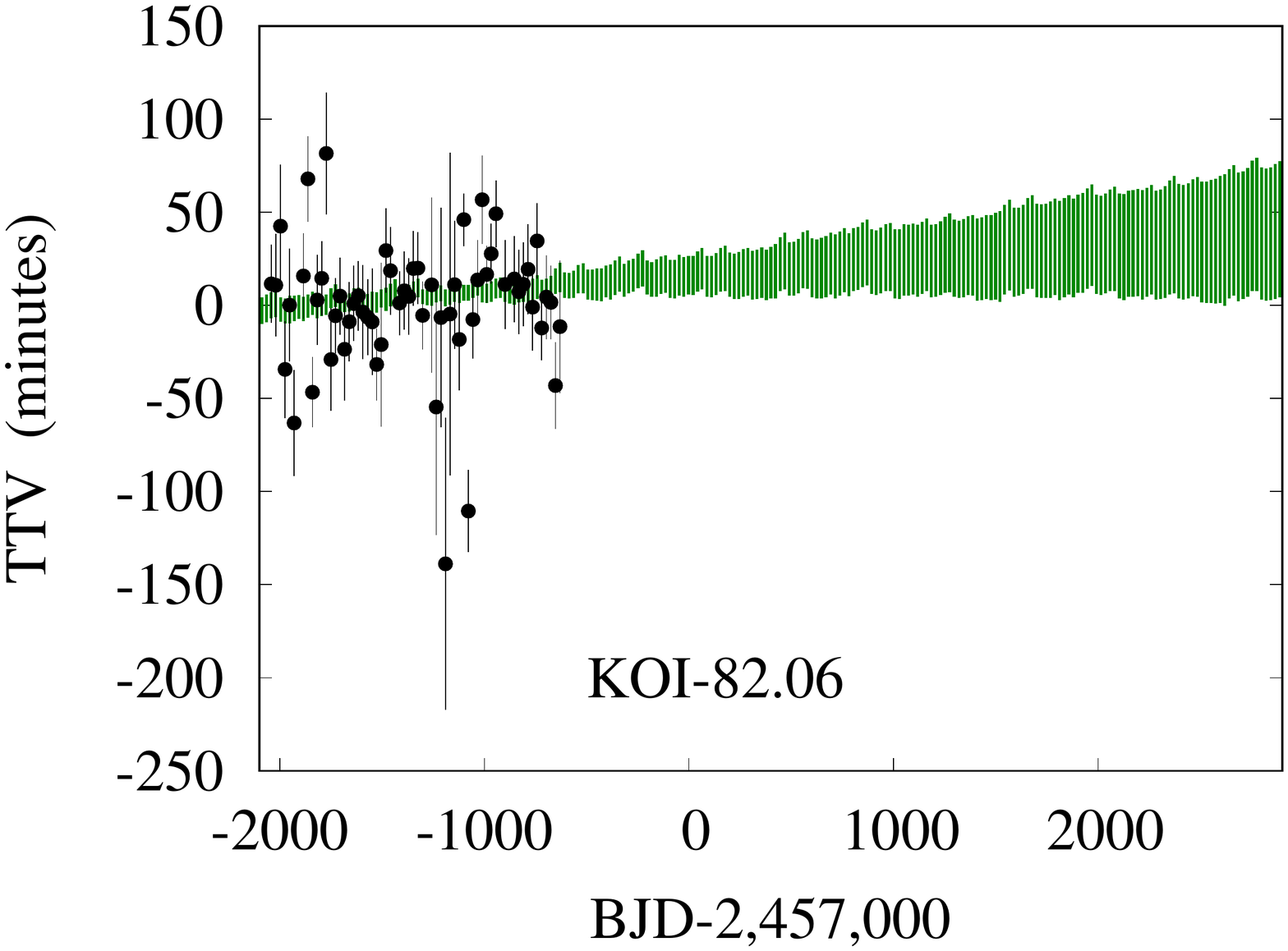}
\includegraphics [width = 2.3 in]{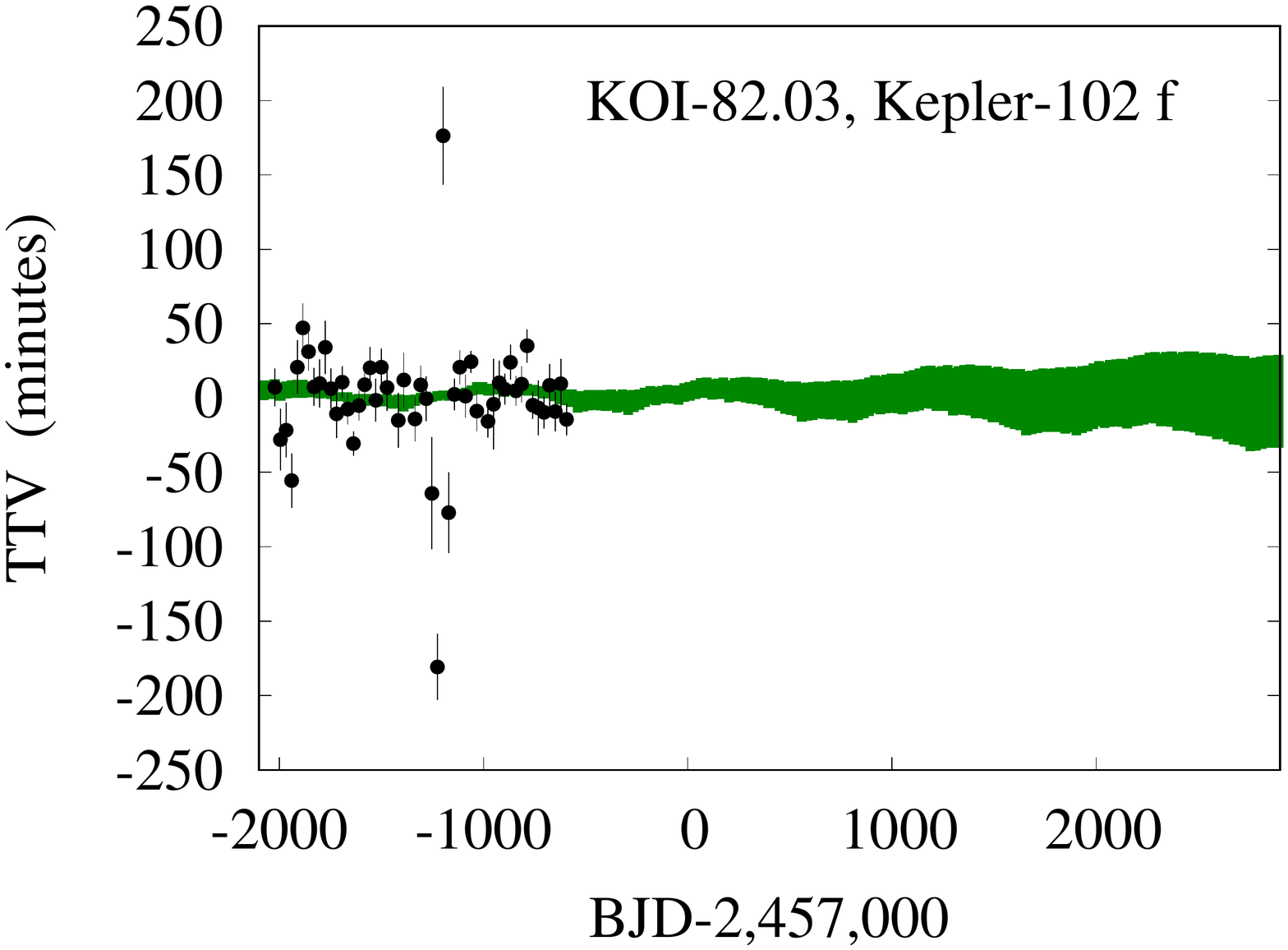}
%\newline
\caption{TTV models ($O-C$, in minutes) for planet candidates identified within each panel, with the combined \kepler\ /\tess\ transit times in black, and 68\% confidence intervals with equal weight in the tails on simulated transit times in green. (Part 1 of 3.)}
\label{fig:ttv1} 
\end{figure}

\begin{figure}[ht!]
\includegraphics [width = 2.3 in]{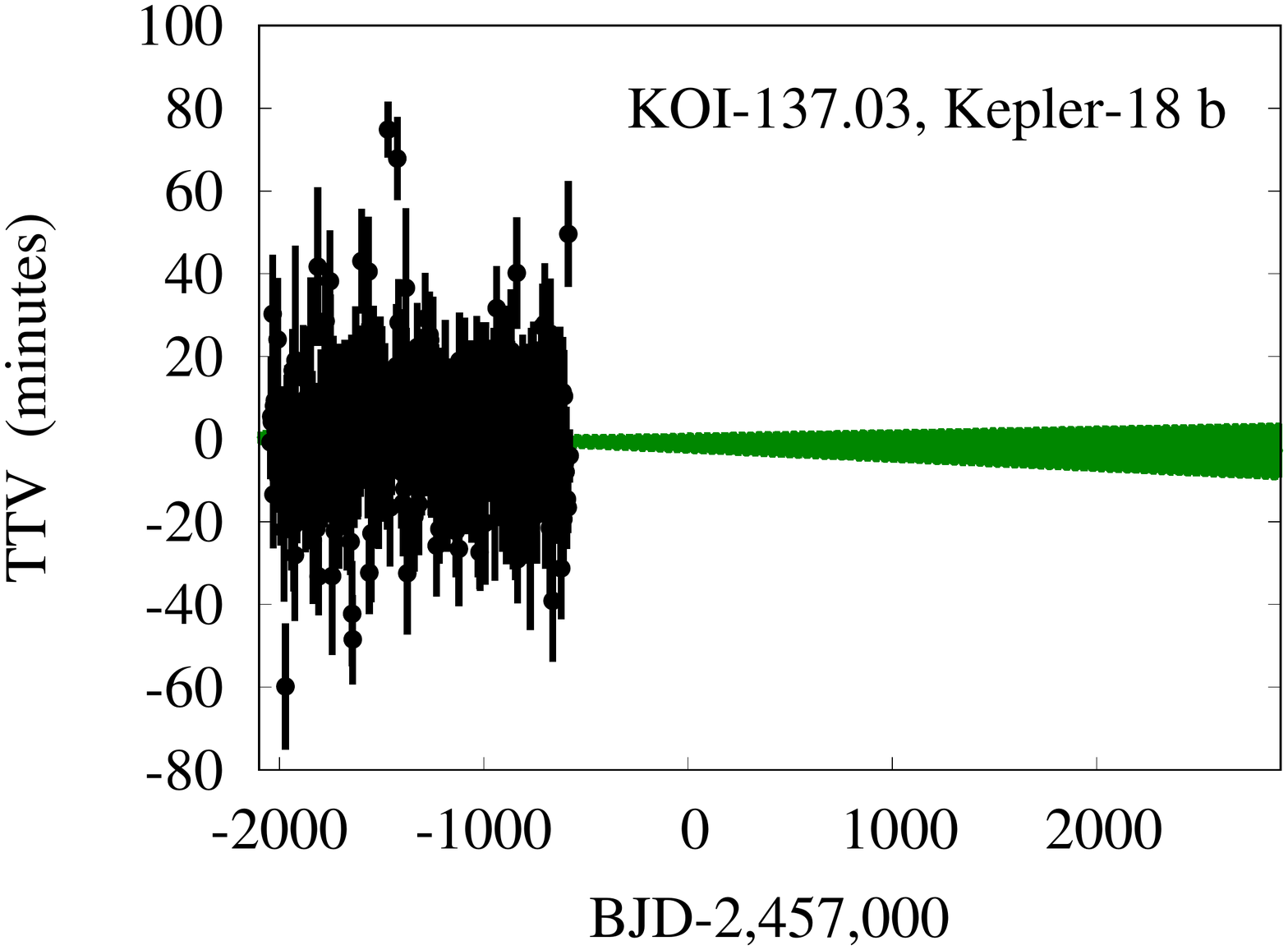}
\includegraphics [width = 2.3 in]{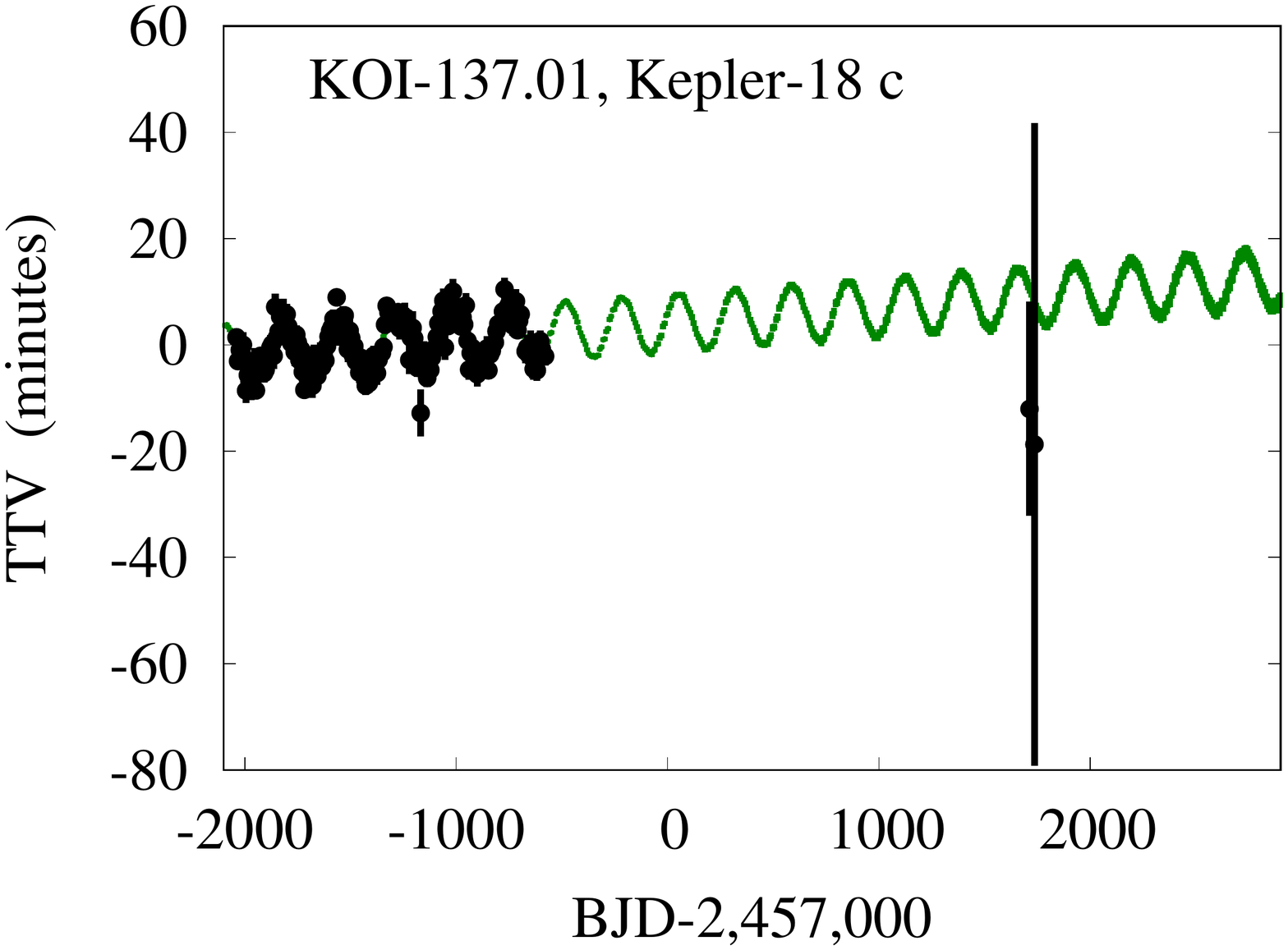}
\includegraphics [width = 2.3 in]{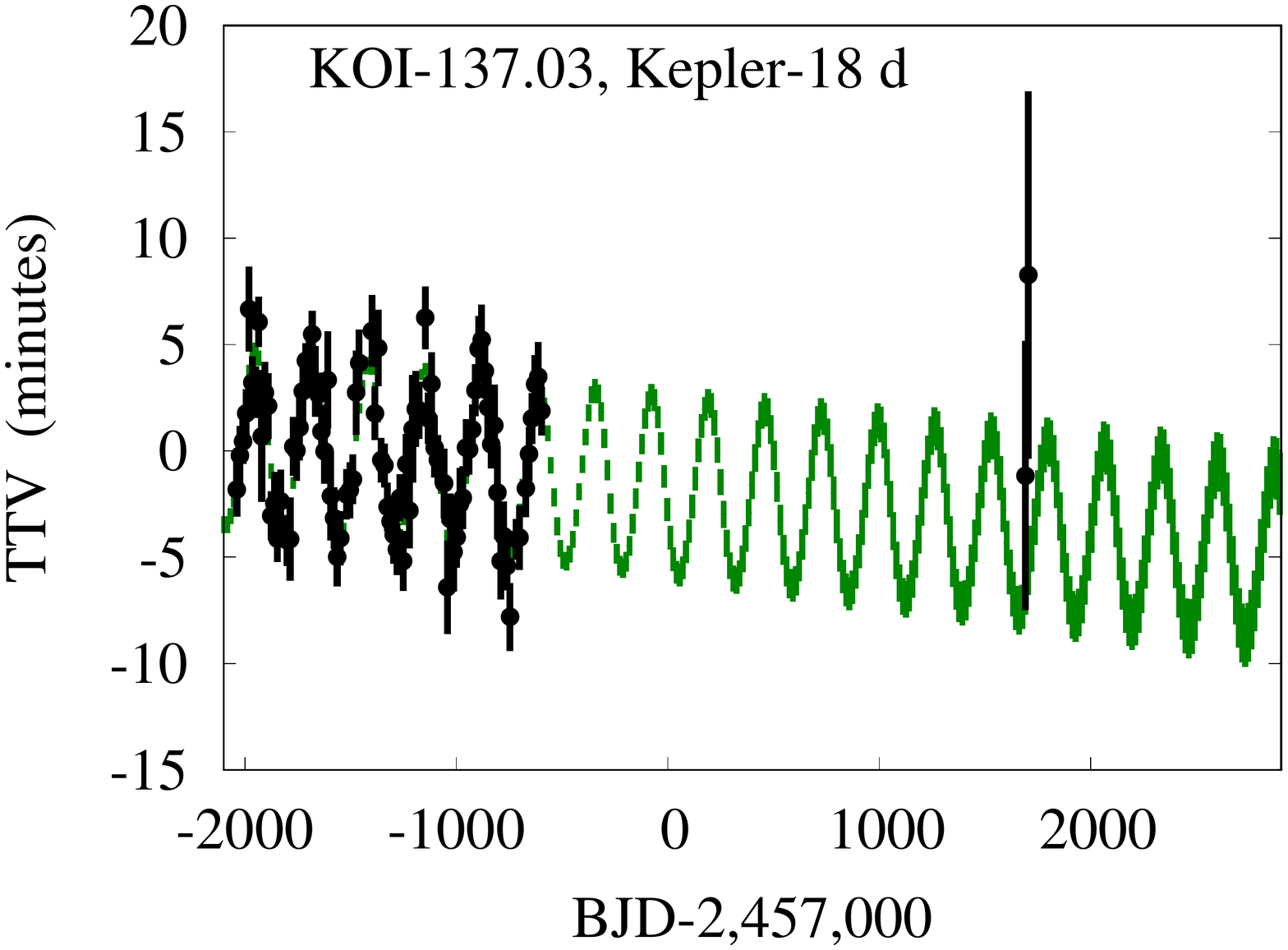}
\newline
\includegraphics [width = 2.3 in]{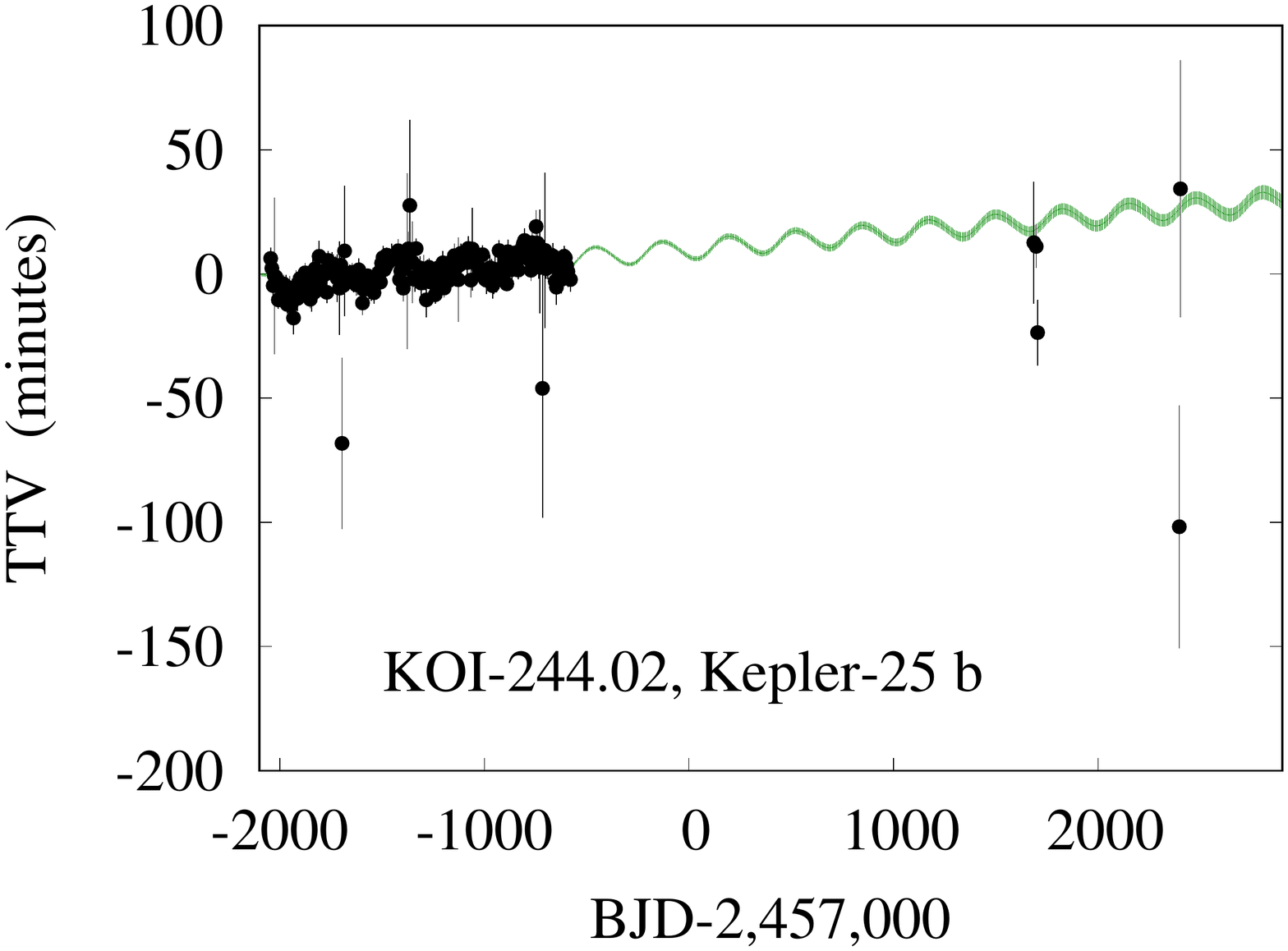}
\includegraphics [width = 2.3 in]{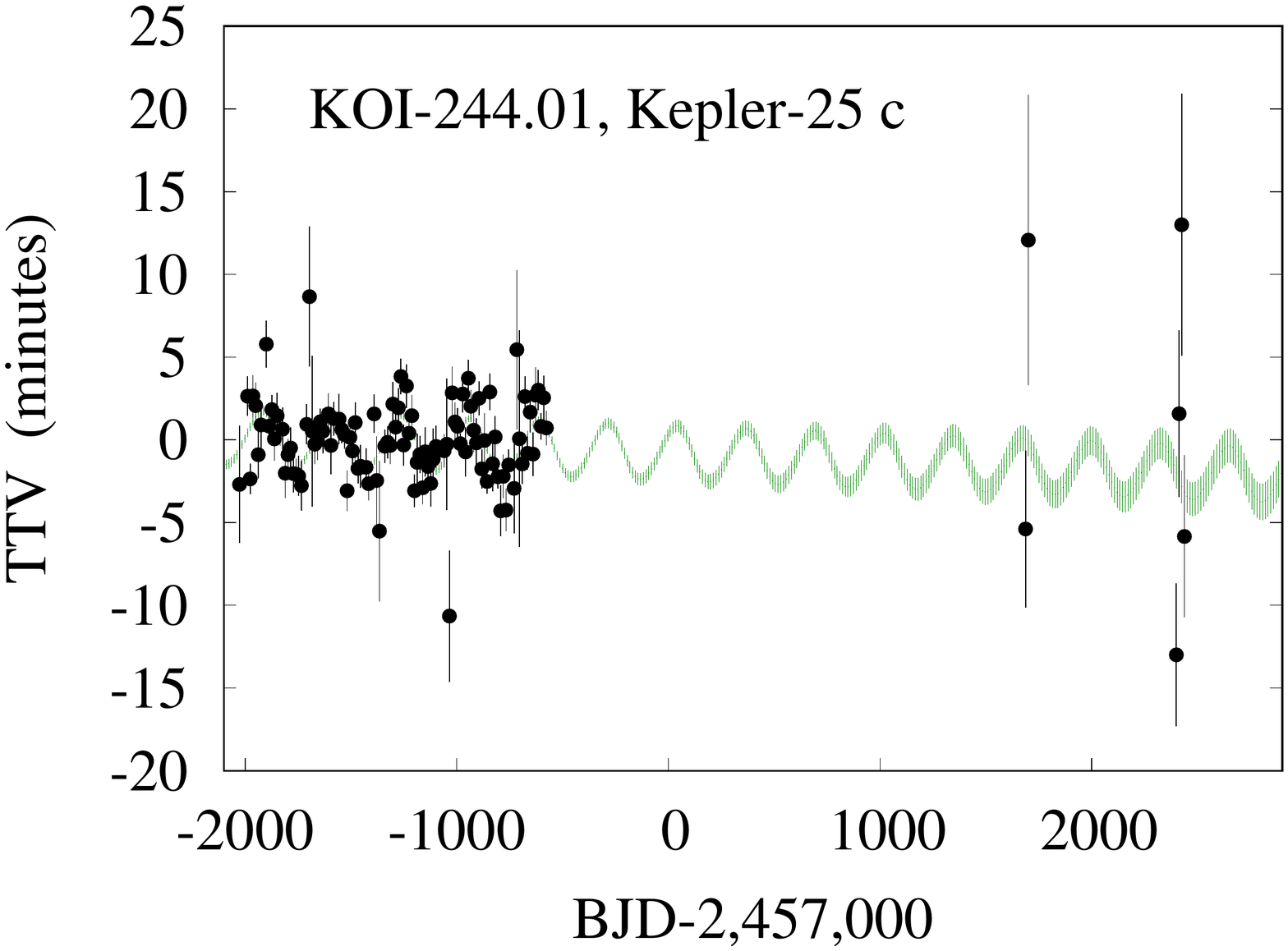}
\newline
\includegraphics [width = 2.3 in]{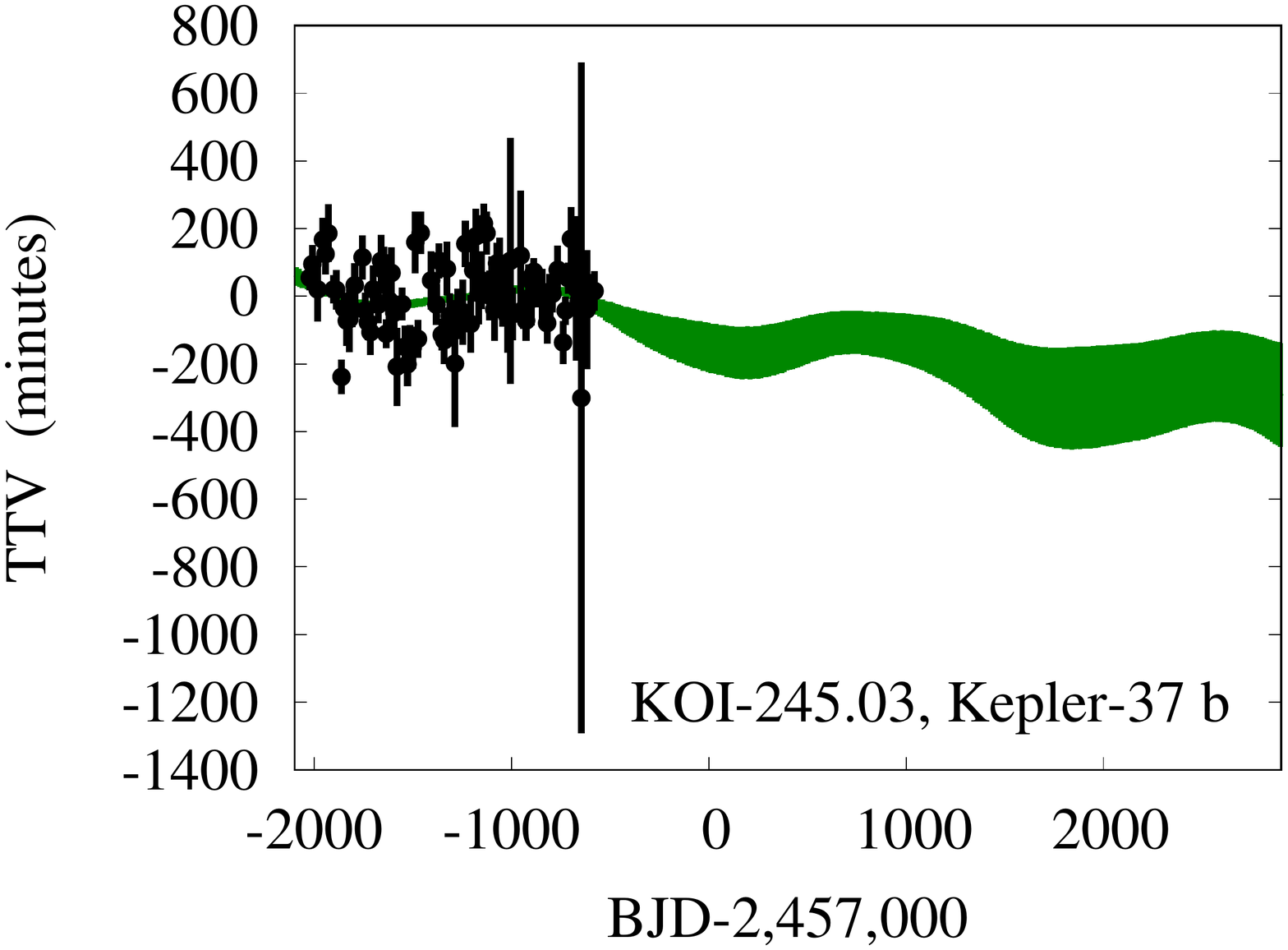}
\includegraphics [width = 2.3 in]{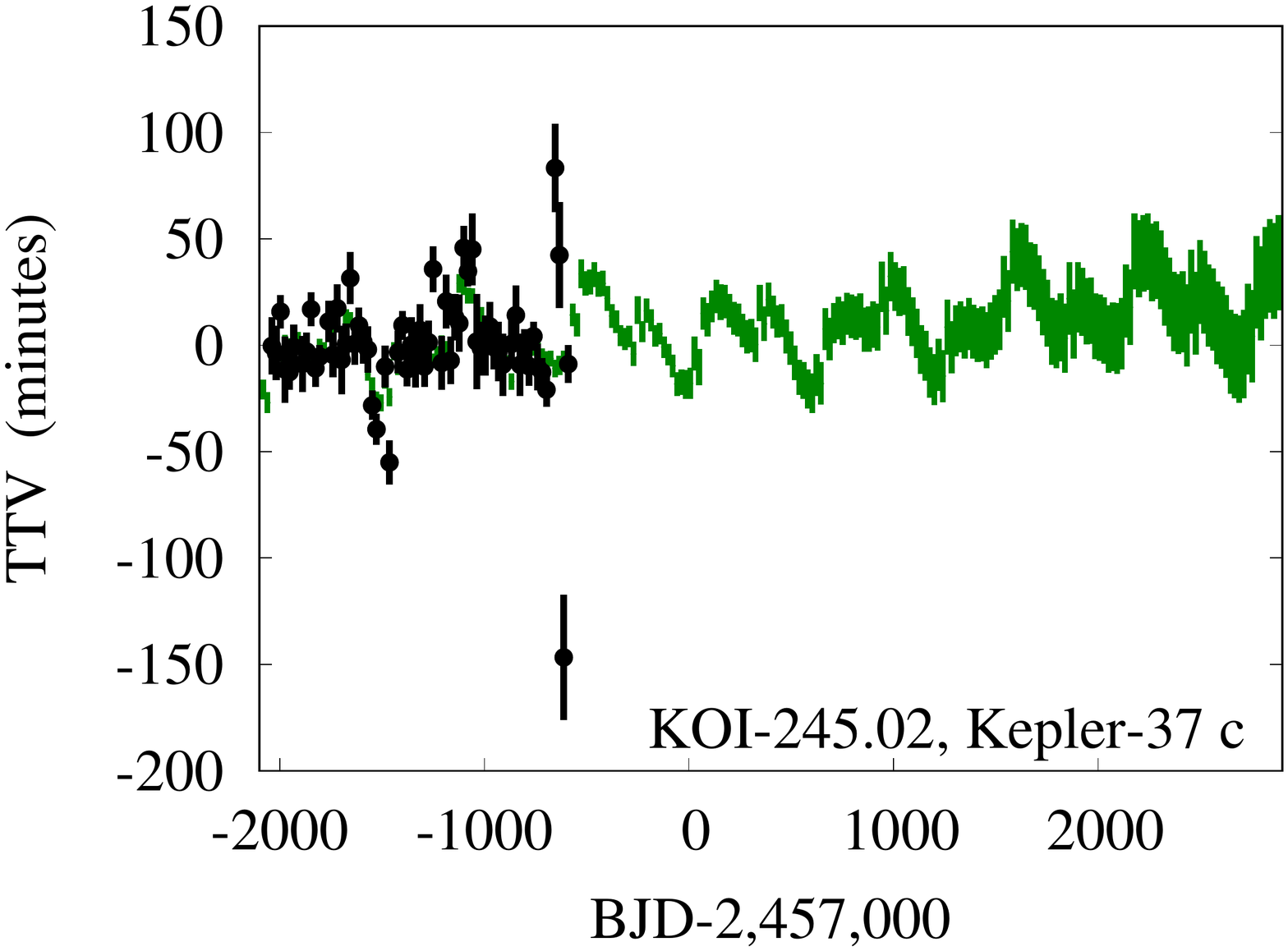}
\includegraphics [width = 2.3 in]{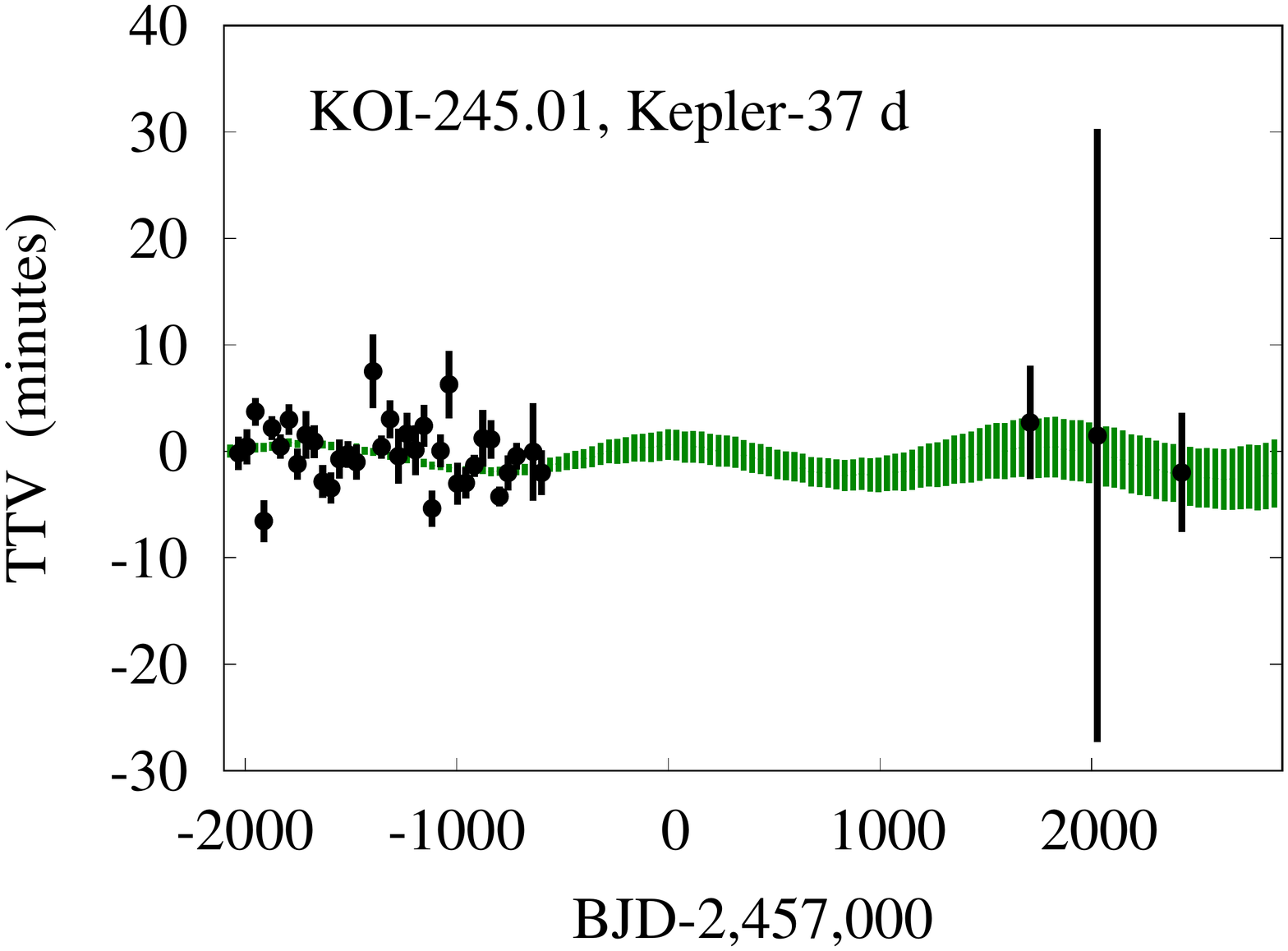}
\includegraphics [width = 2.3 in]{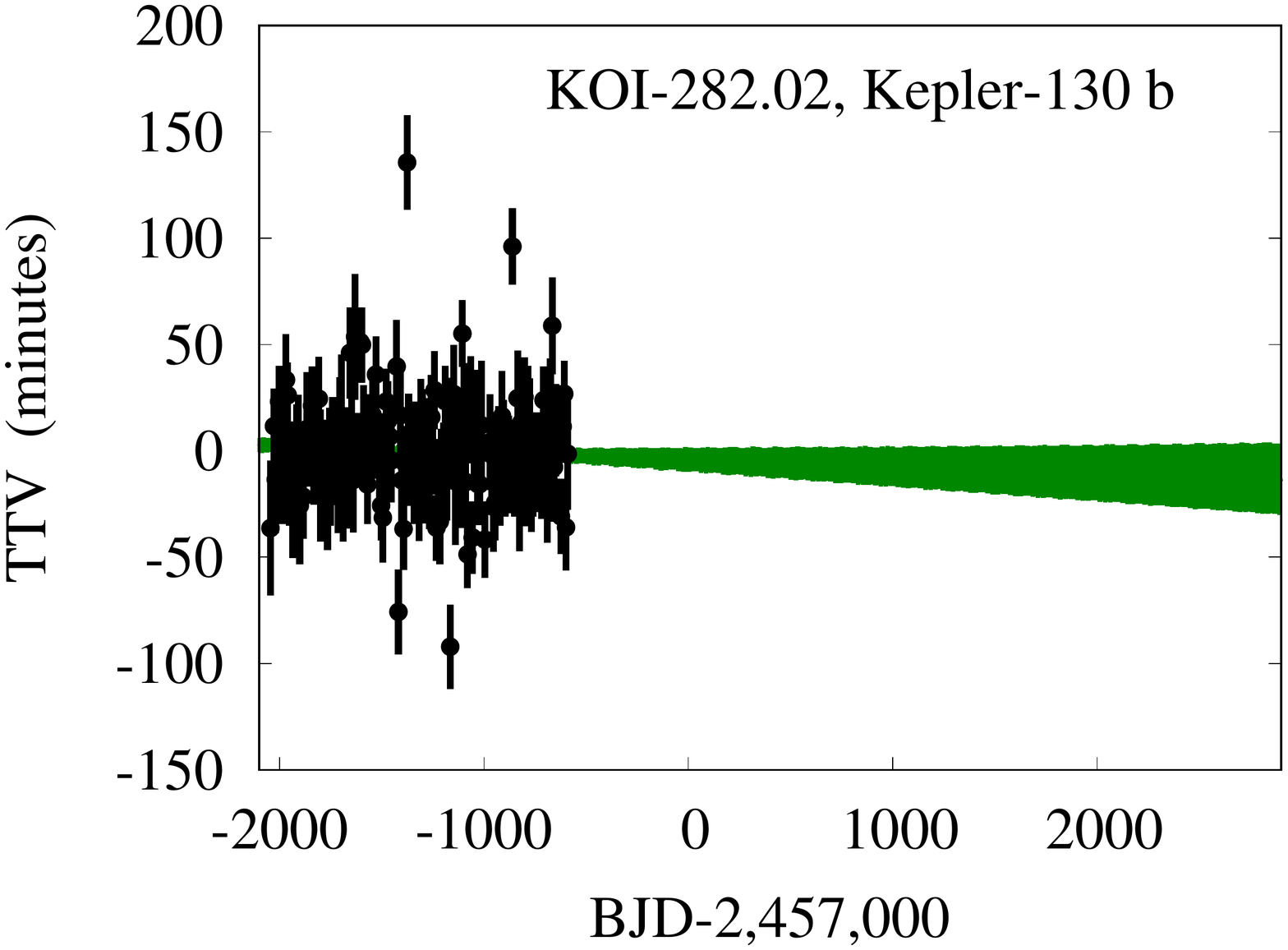}
\includegraphics [width = 2.3 in]{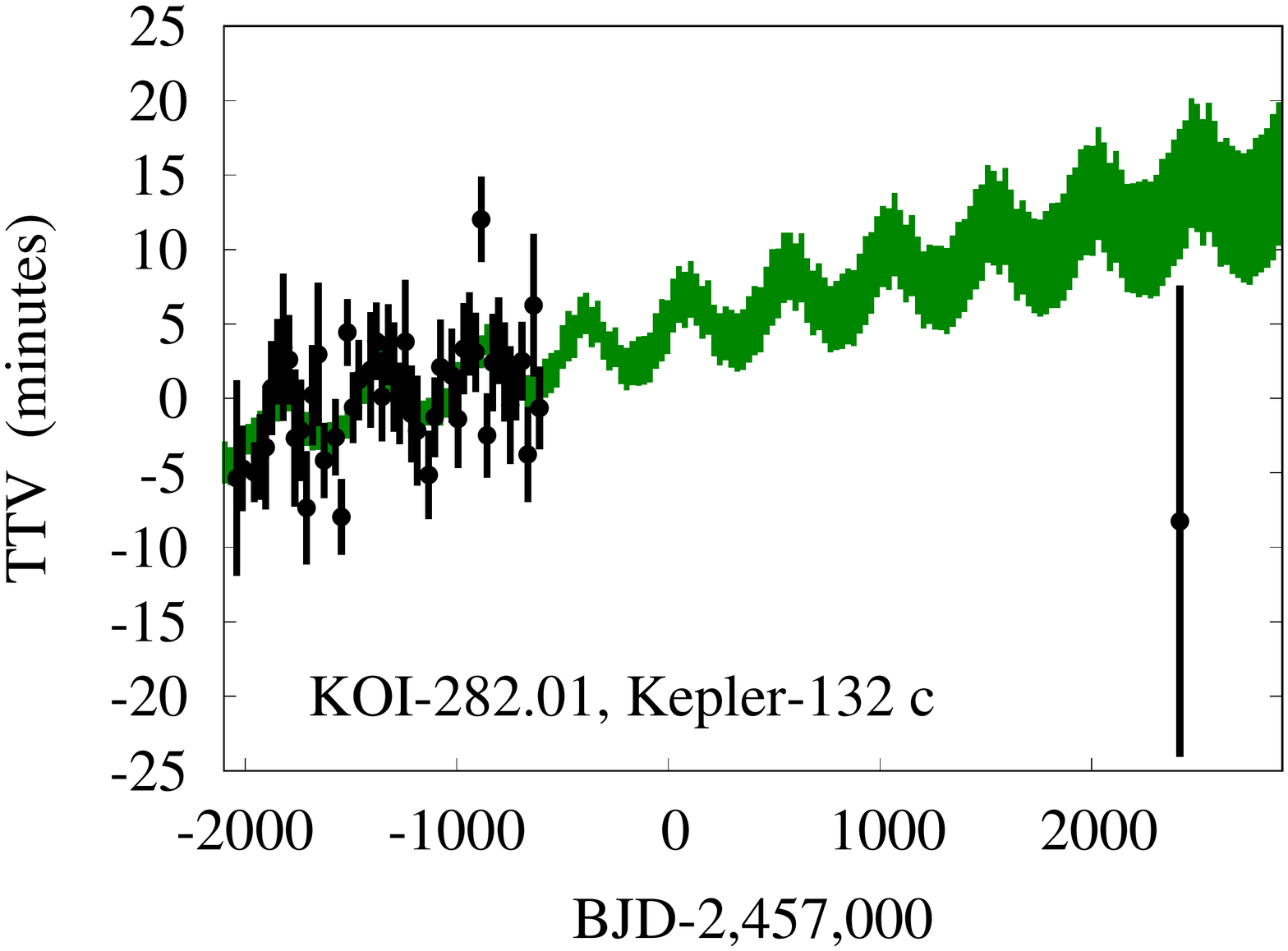}
\includegraphics [width = 2.3 in]{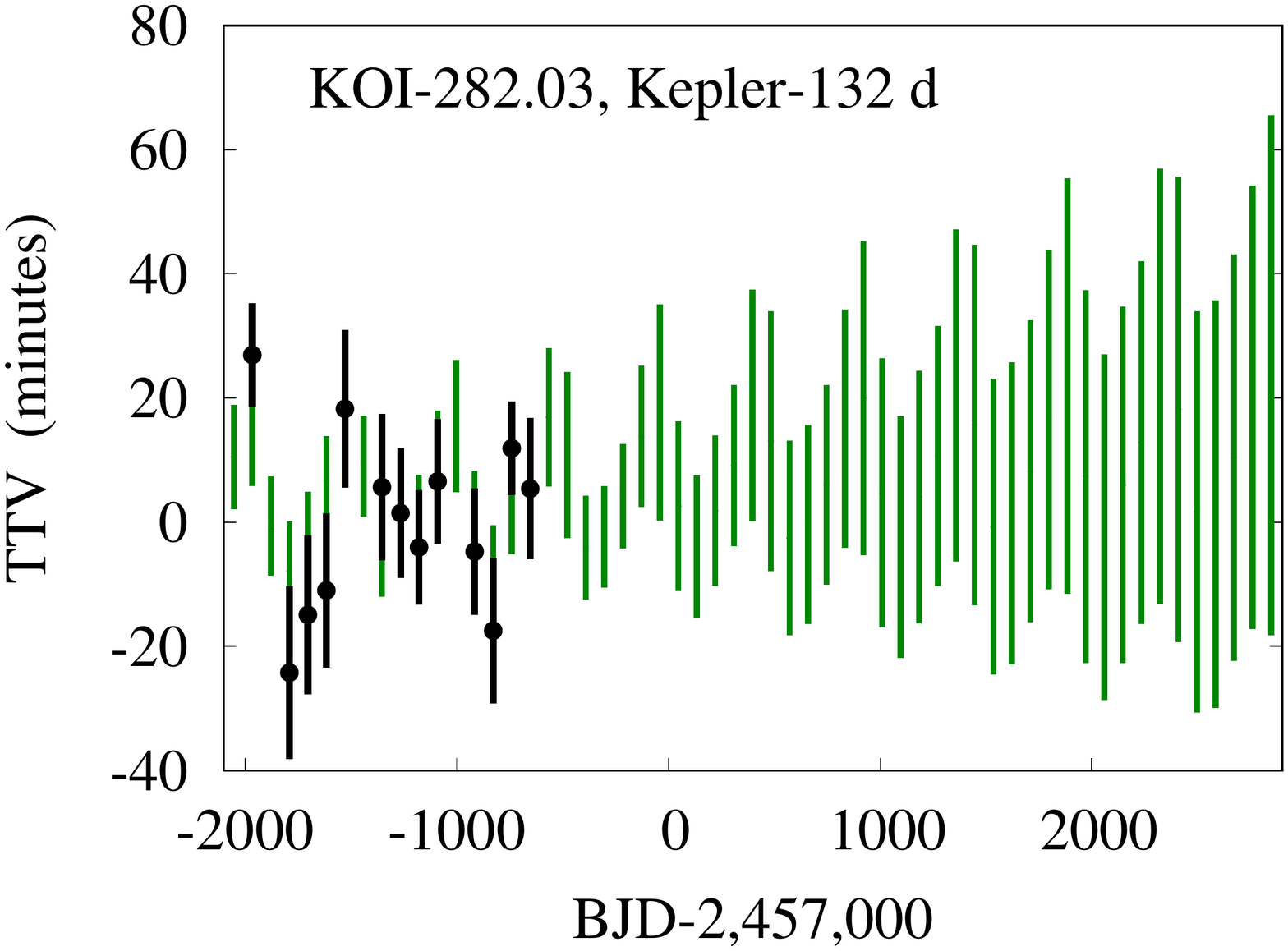}
\caption{TTV models ($O-C$, in minutes) for planet candidates identified within each panel, with the combined \kepler\ /\tess\ transit times in black, and 68\% confidence intervals with equal weight in the tails on simulated transit times in green. (Part 2 of 3.)}
\label{fig:ttv2} 
\end{figure}

\begin{figure}[ht!]
\includegraphics [width = 2.29 in]{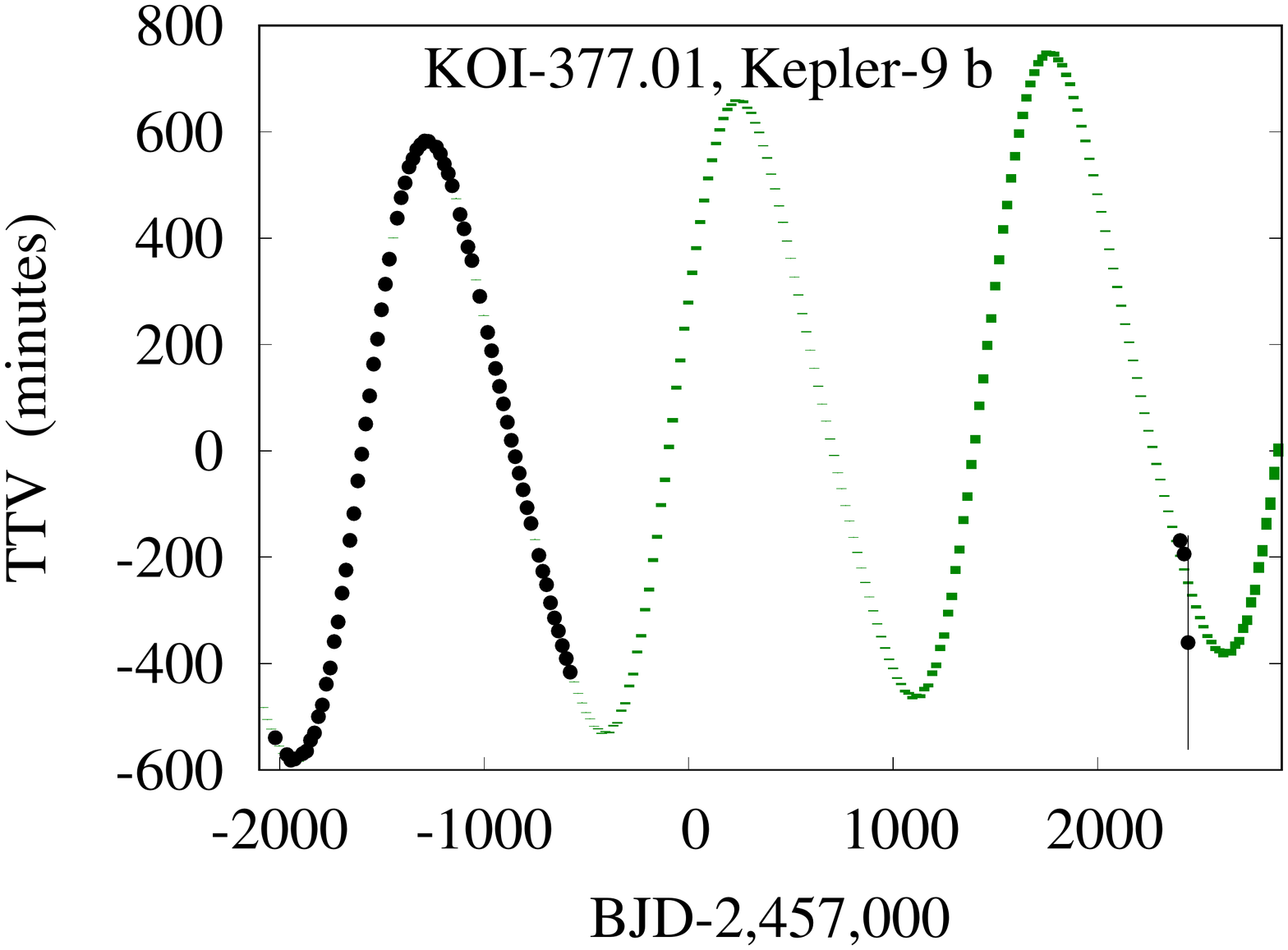}
\includegraphics [width = 2.29 in]{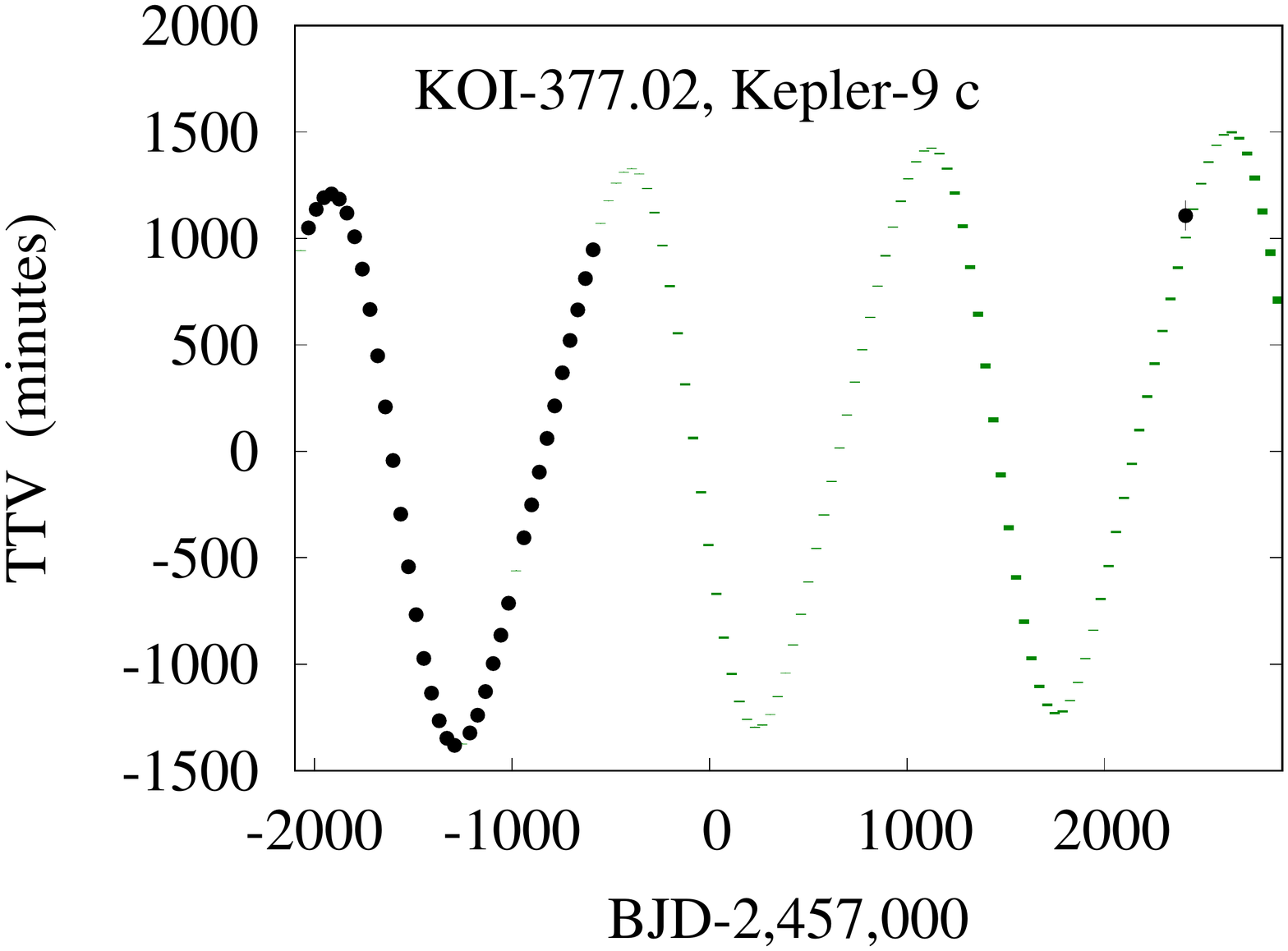}
\newline
\includegraphics [width = 2.29 in]{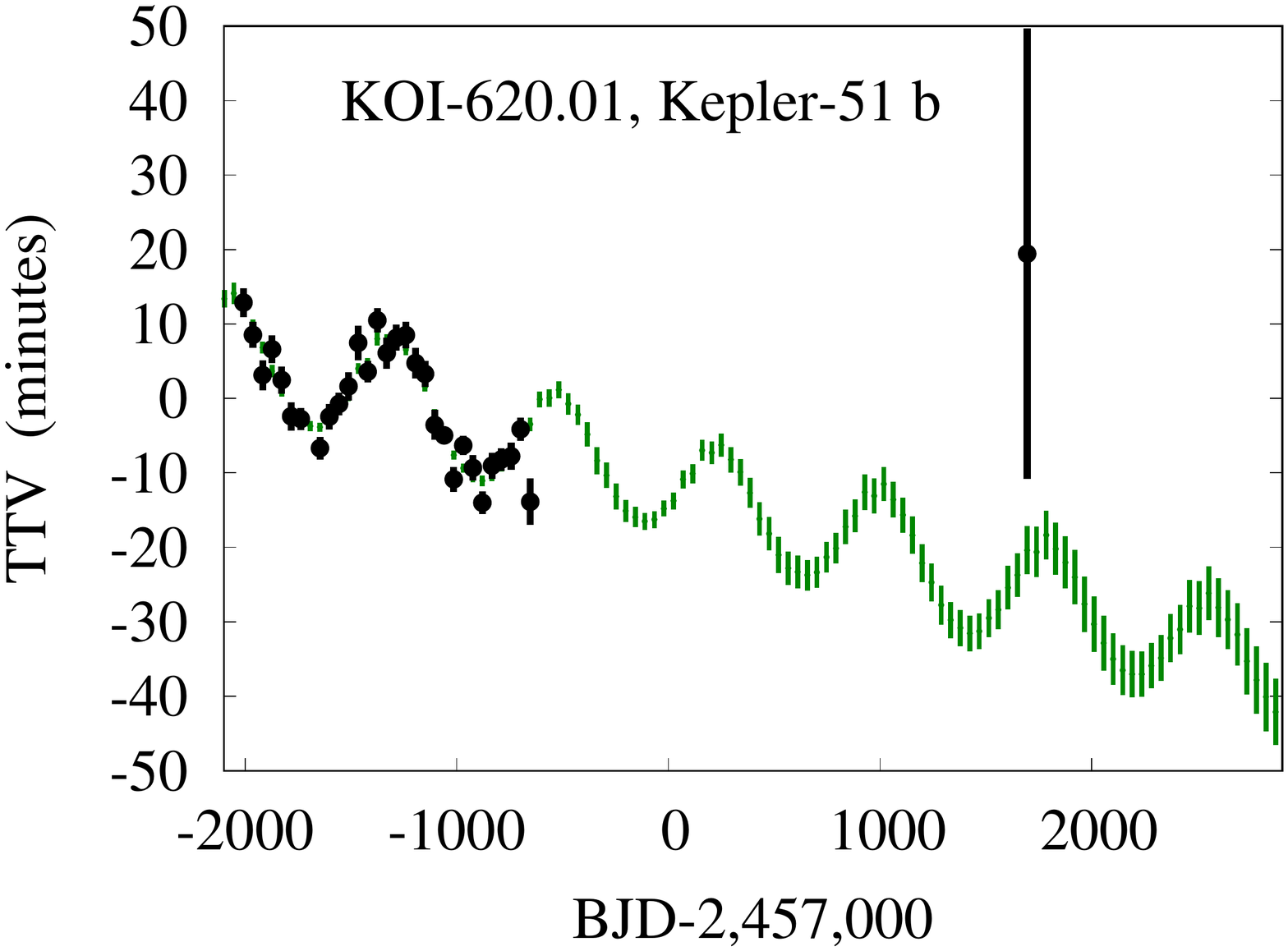}
\includegraphics [width = 2.29 in]{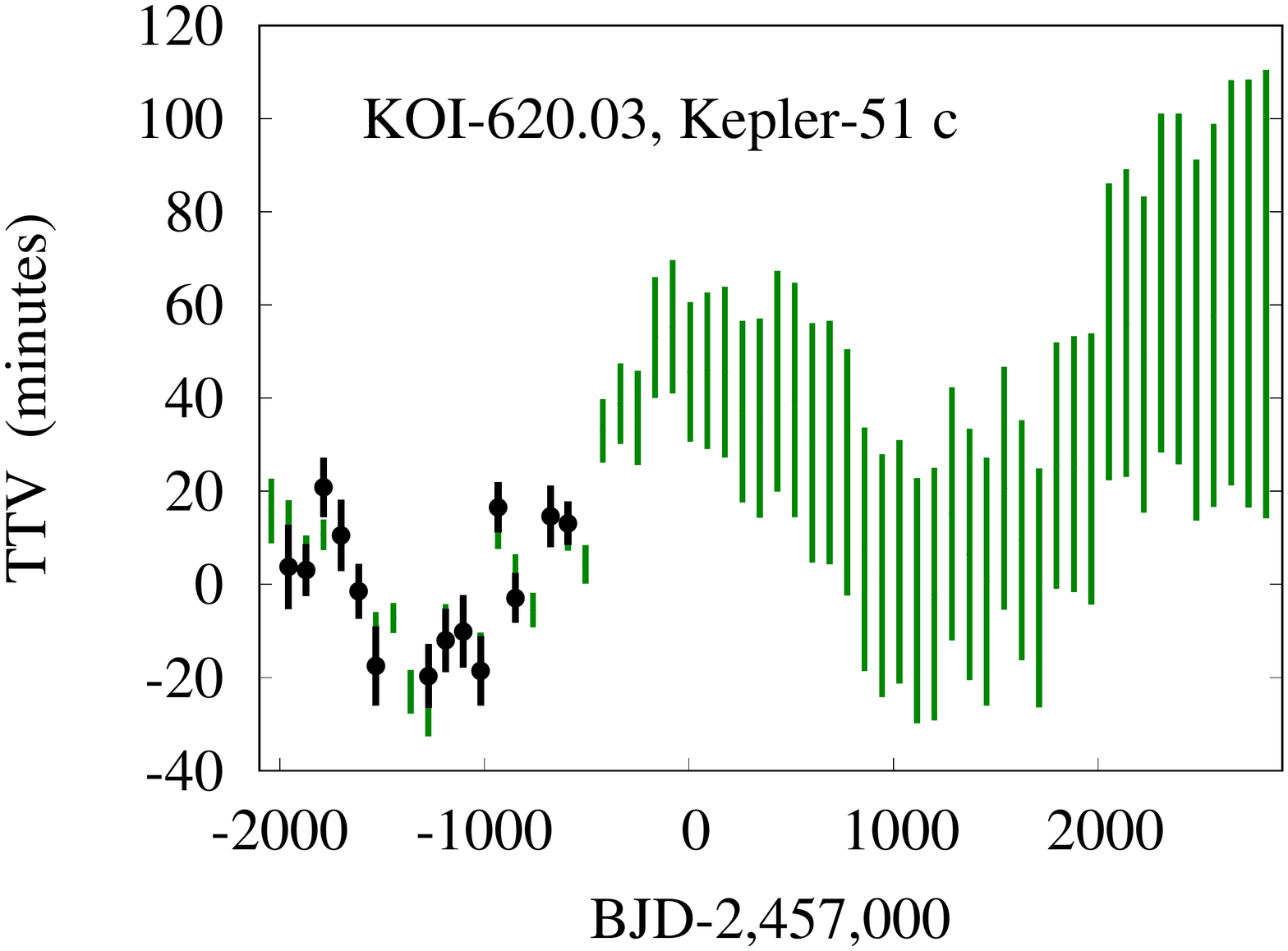}
\includegraphics [width = 2.29 in]{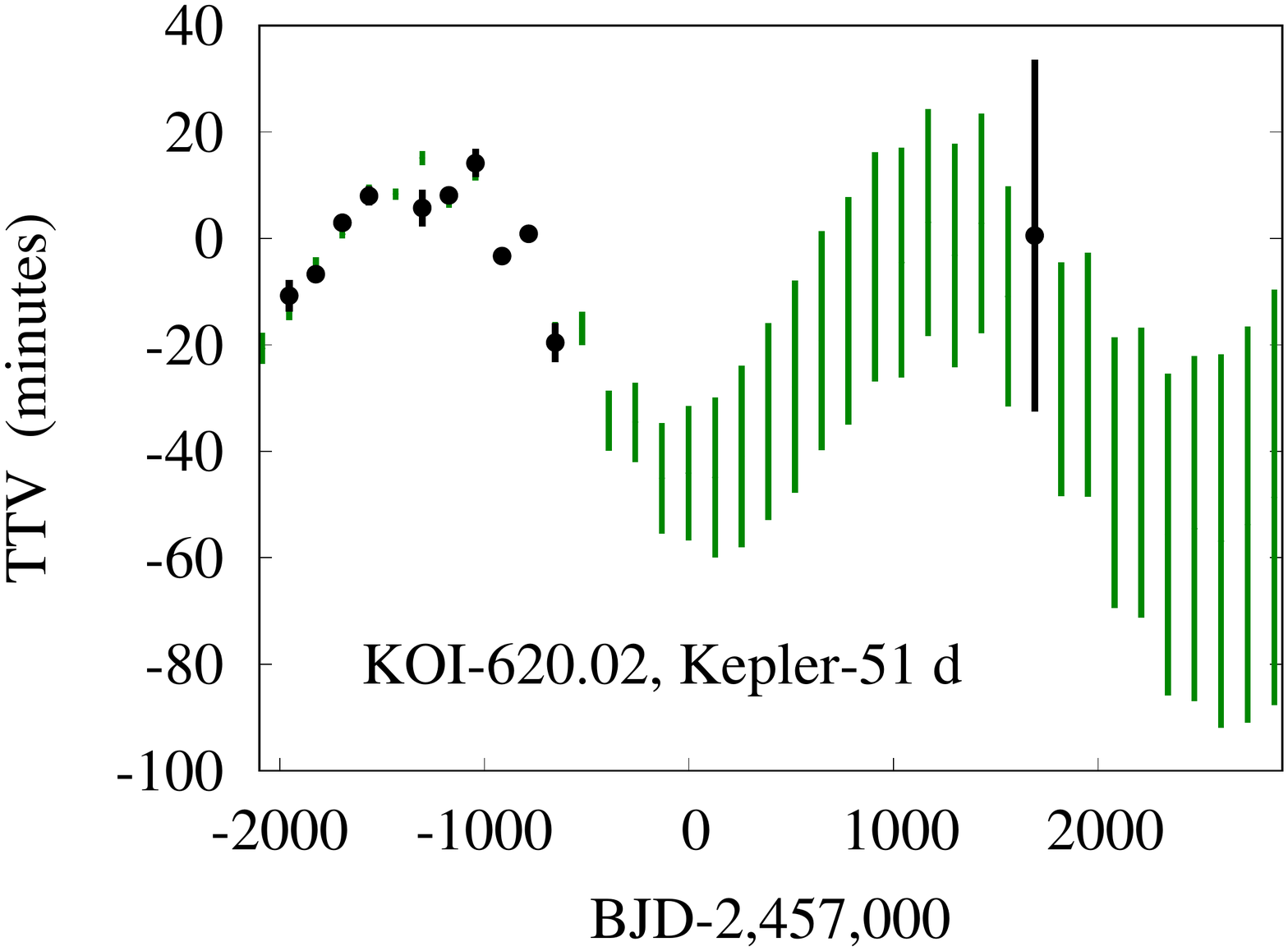}
\includegraphics [width = 2.29 in]{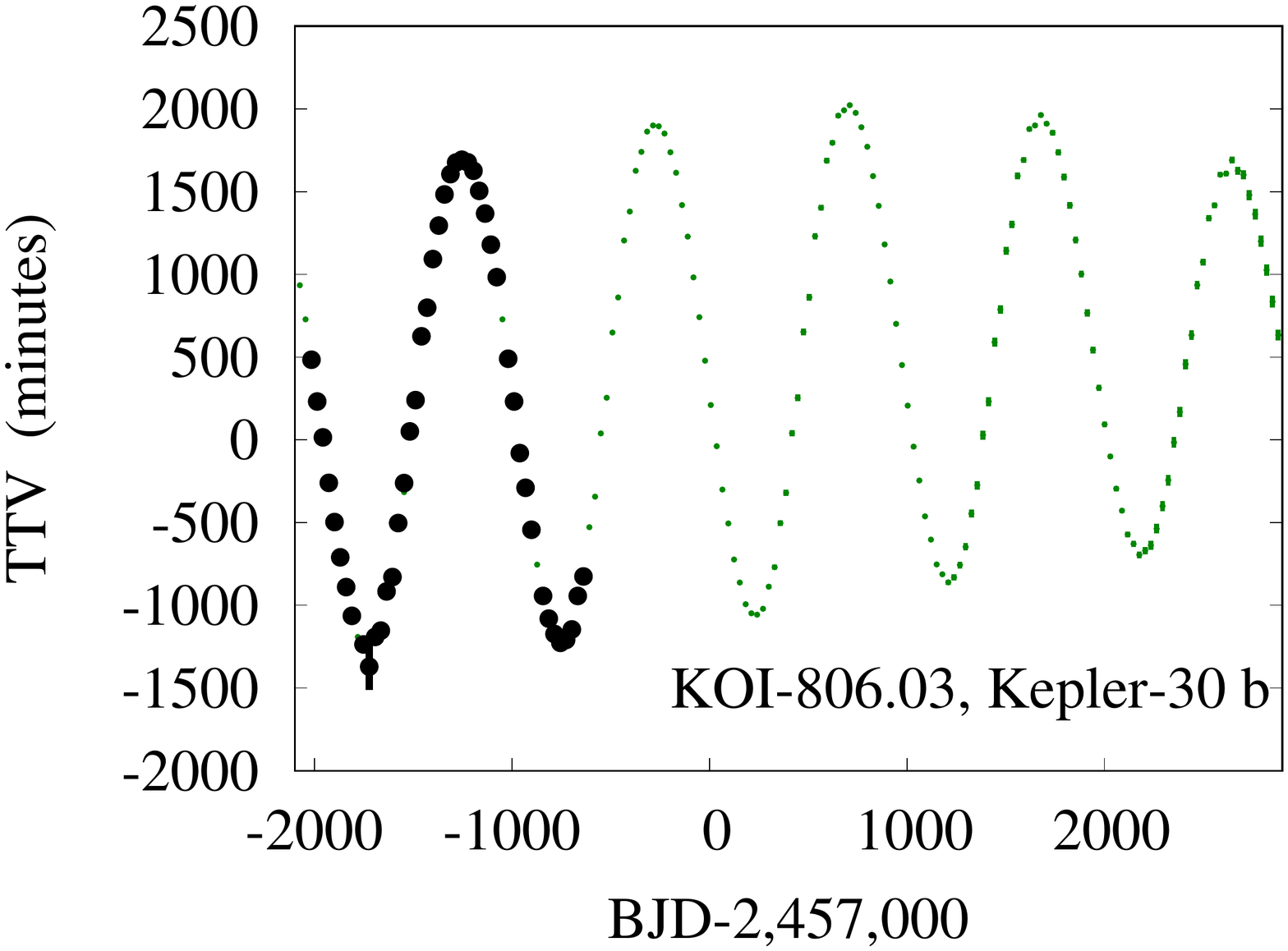}
\includegraphics [width = 2.29 in]{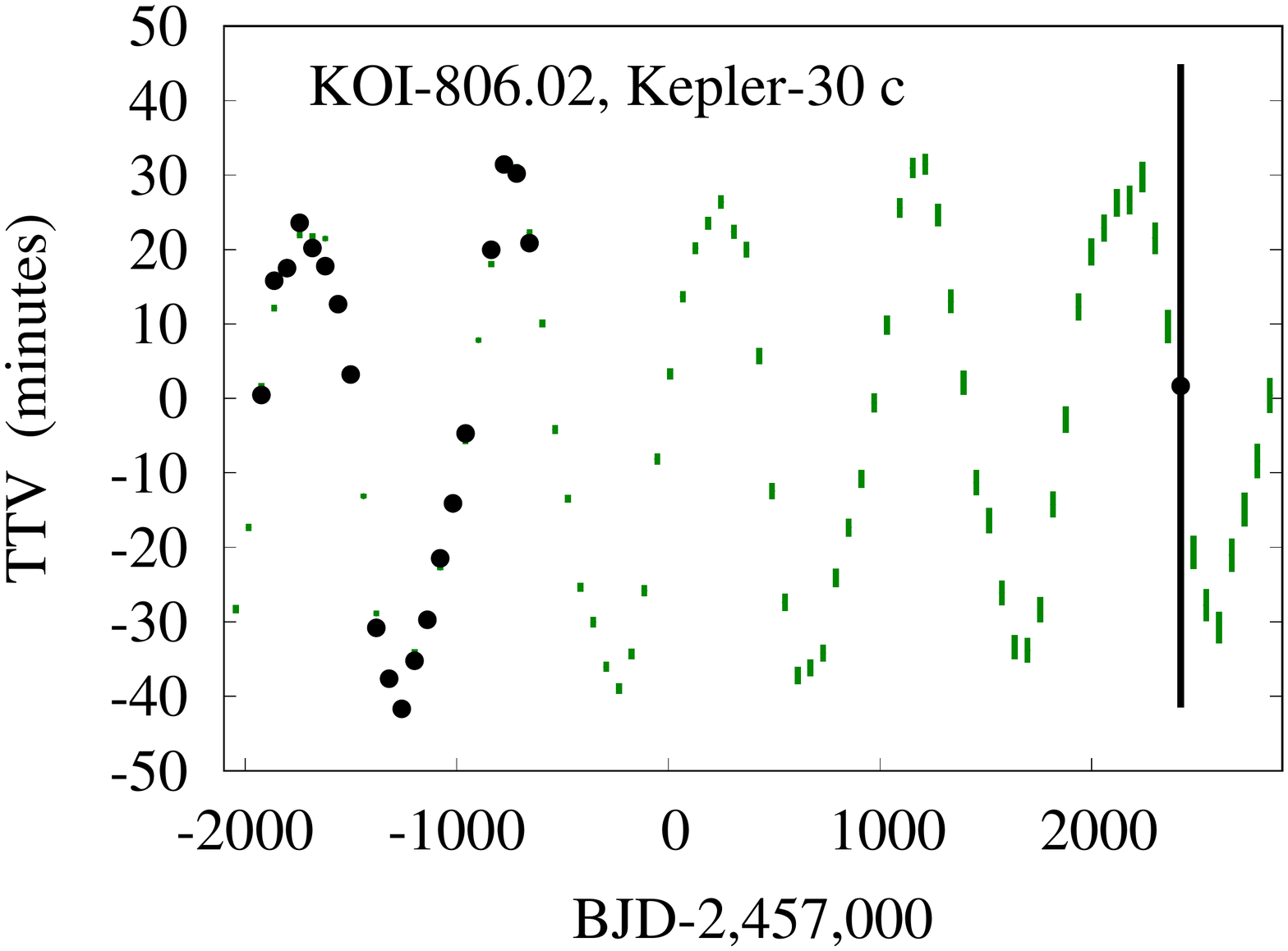}
\includegraphics [width = 2.29 in]{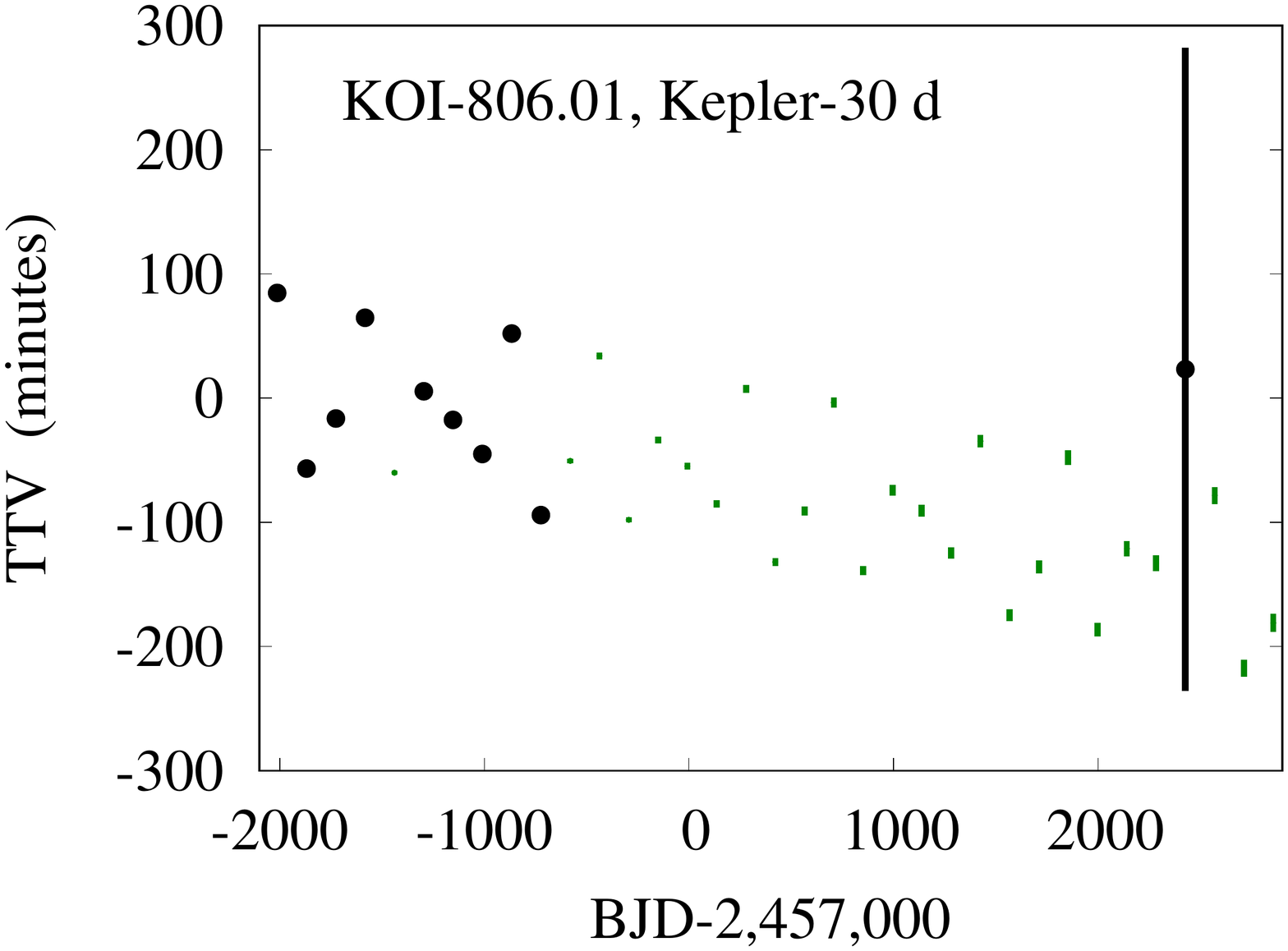}
\includegraphics [width = 2.29 in]{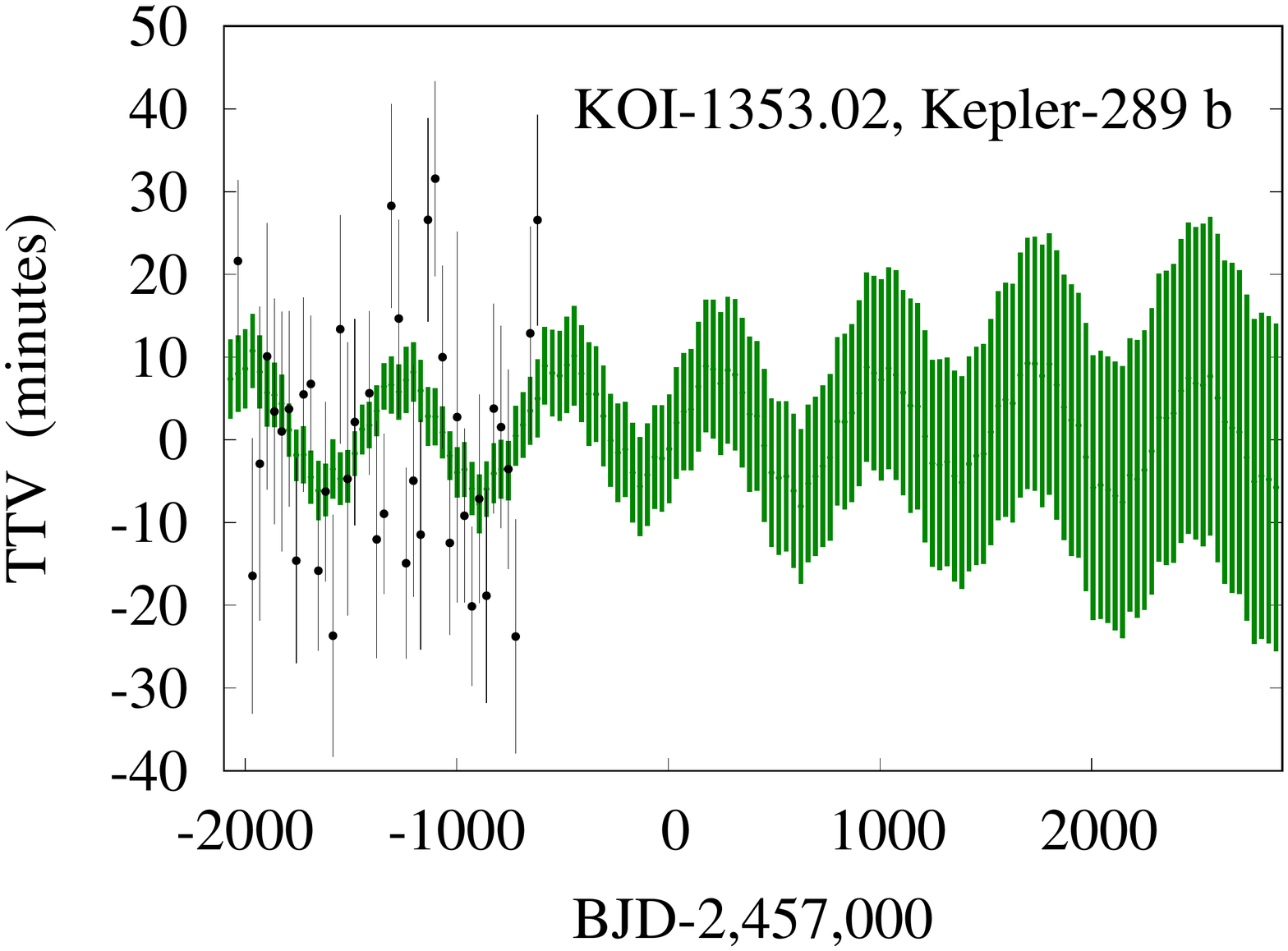}
\includegraphics [width = 2.29 in]{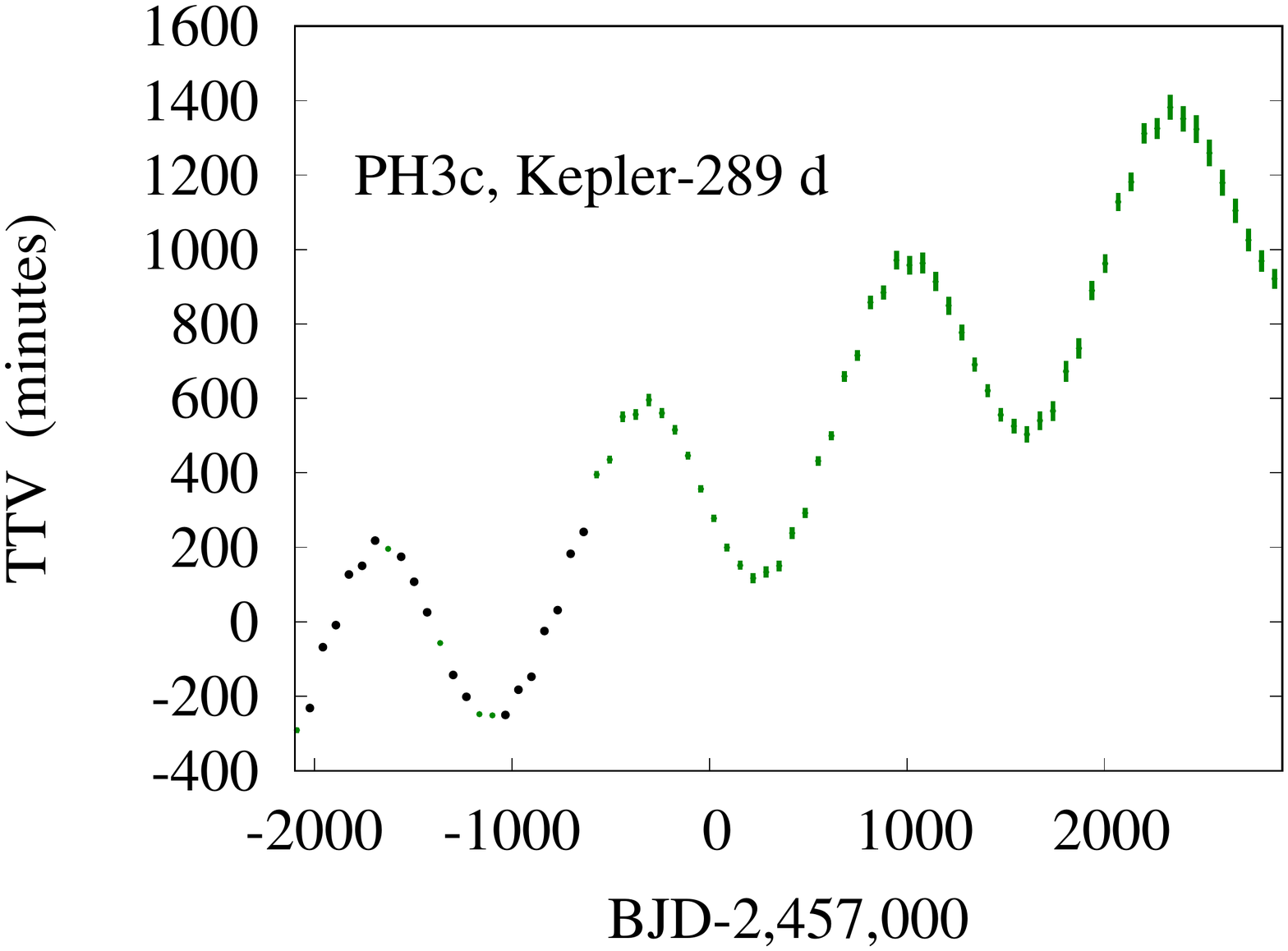}
\includegraphics [width = 2.29 in]{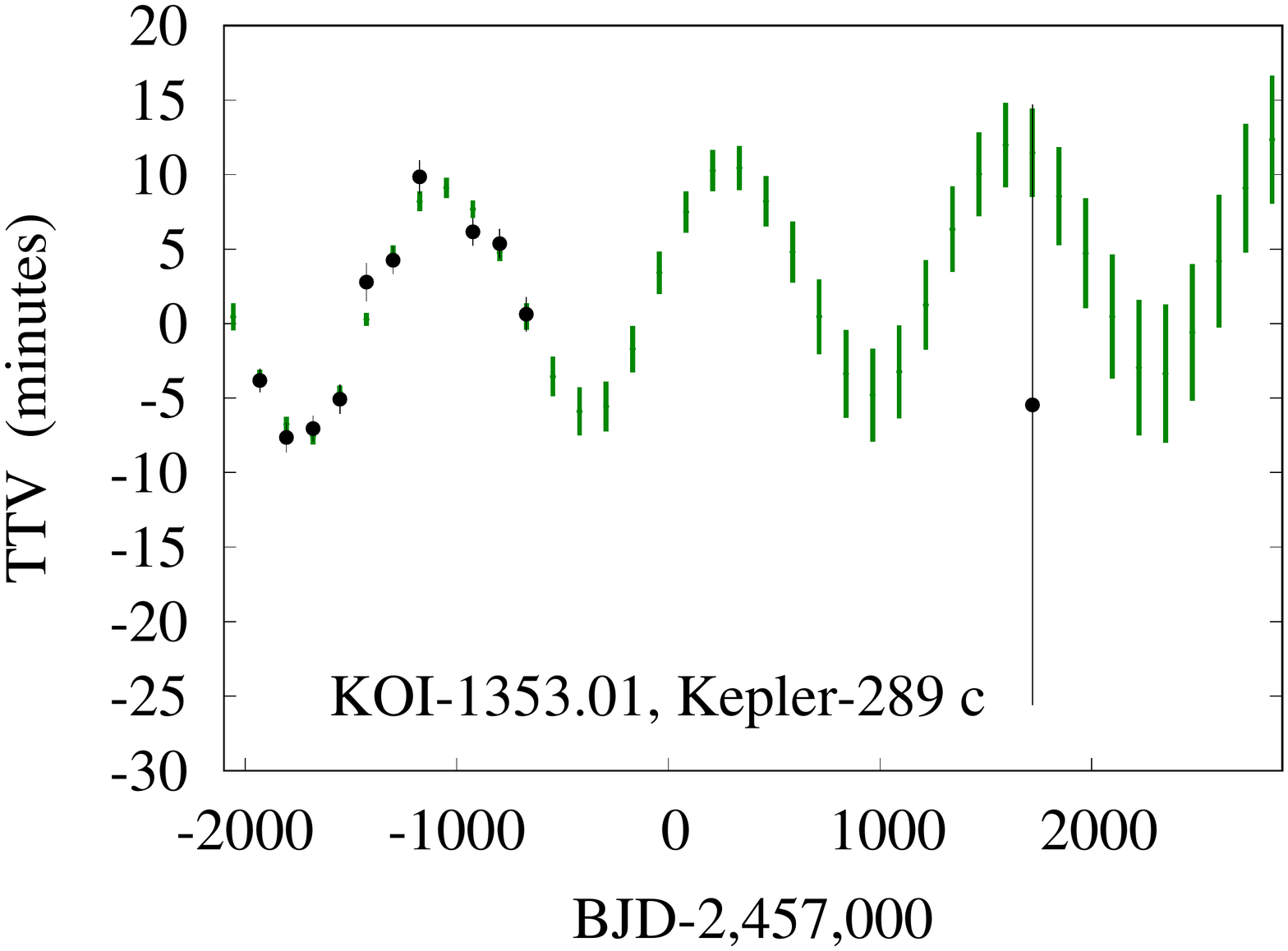}
\includegraphics [width = 2.29 in]{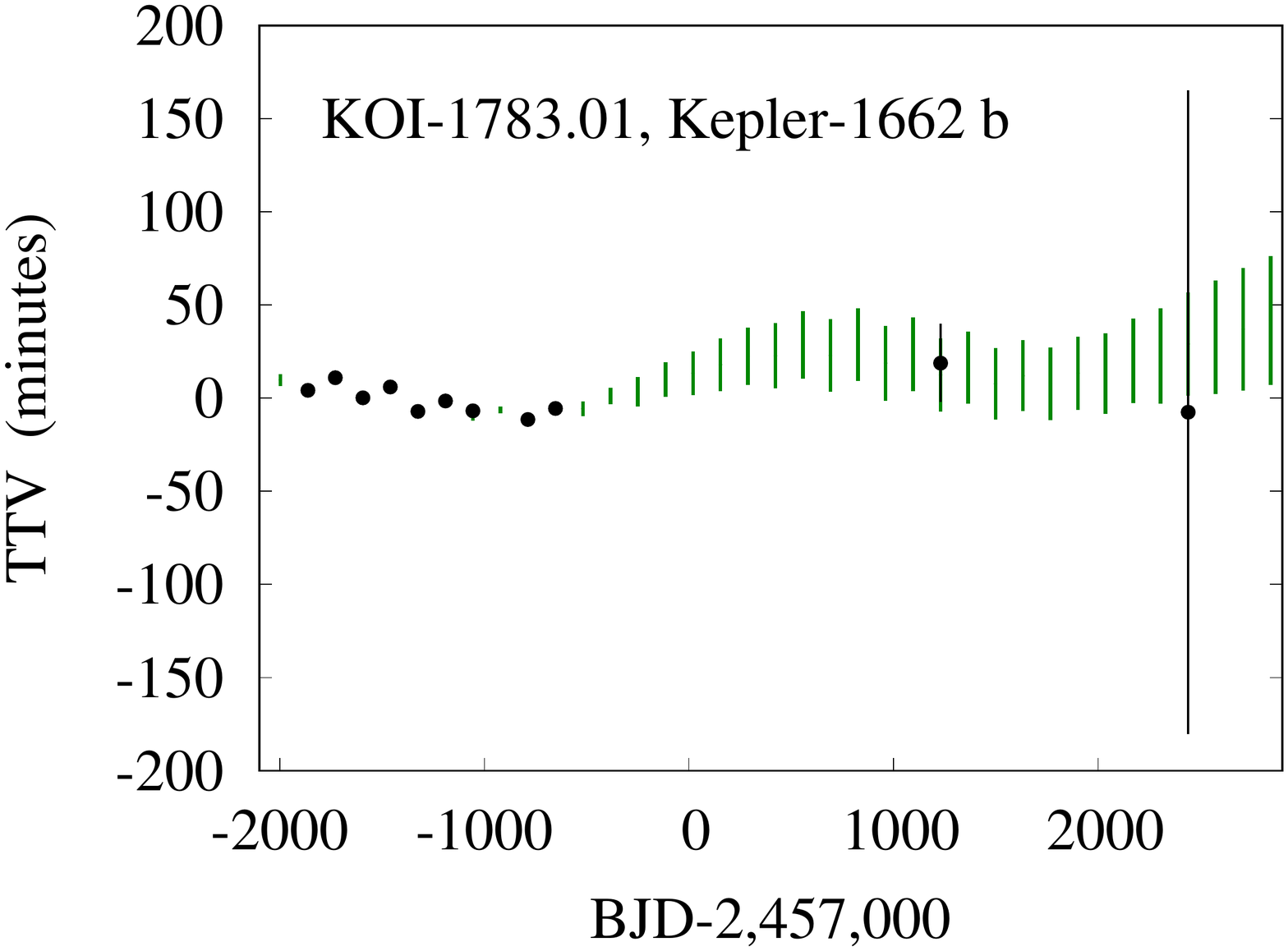}
\includegraphics [width = 2.29 in]{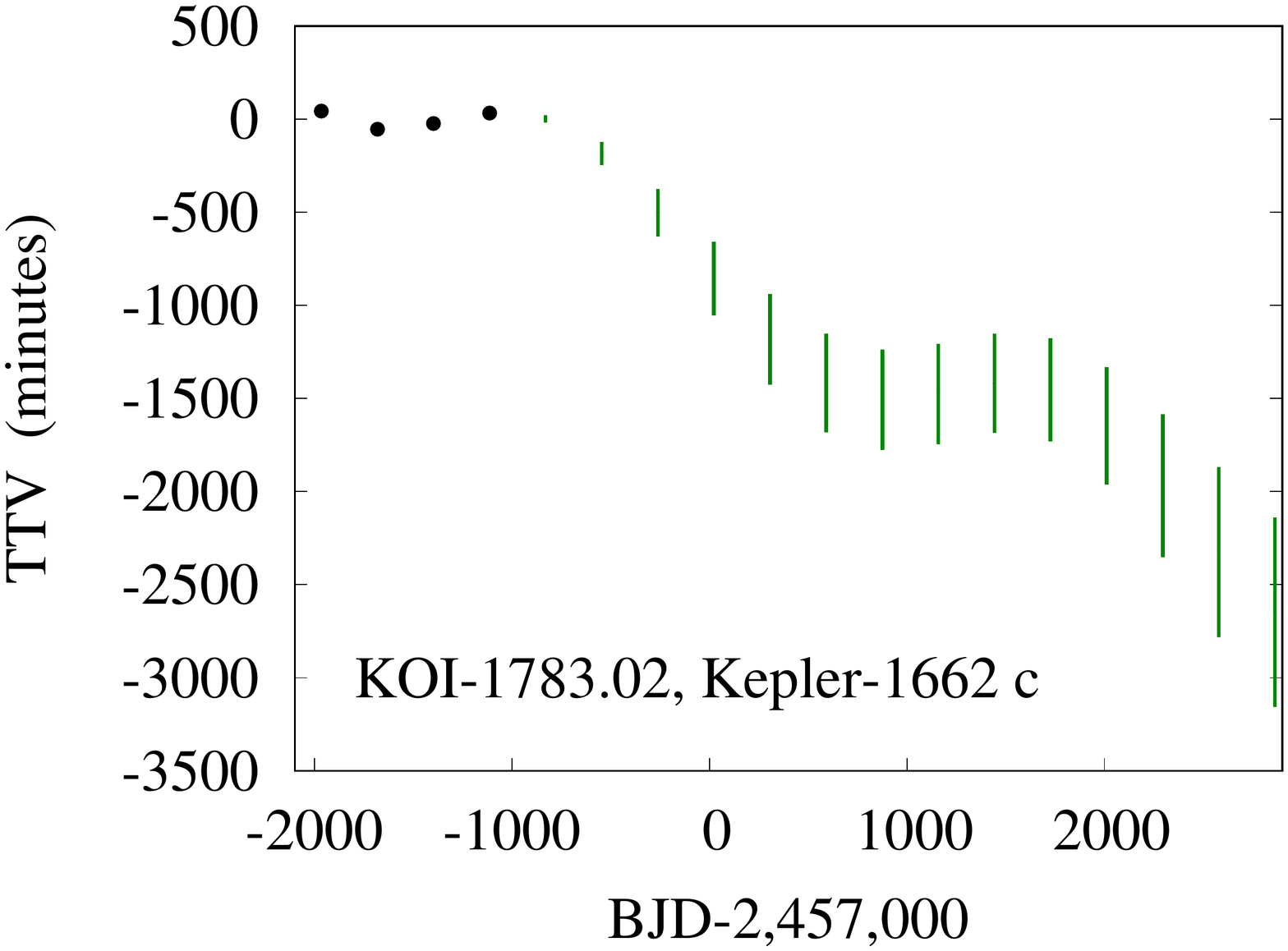}
\caption{TTV models ($O-C$, in minutes) for planet candidates identified within each panel, with the combined \kepler\ /\tess\ transit times in black, and 68\% confidence intervals with equal weight in the tails on simulated transit times in green. (Part 3 of 3.)}
\label{fig:ttv3} 
\end{figure}

We list our results for the remaining systems in Table~\ref{tbl-ttv} and display their TTVs in Figures~\ref{fig:ttv1}, ~\ref{fig:ttv2} and~\ref{fig:ttv3}. 

 \begin{table}[h!]
 \tiny
  \begin{center}
    \begin{tabular}{|c|c|c|c|c|c|c|c|c|}
      \hline
 KOI  & $P$ (days)  &      $T_0$ (days)    &      $e\cos\omega$   &  $e\sin\omega$     &     $ \frac{ M_p  }{M_{ \star} } \frac{M_{\odot} } {M_{\oplus} }  $ &  M$_{p}$ (M$_{\oplus}$) &  R$_{p}$ (R$_{\oplus}$) &  $\rho_{p}$ (g cm$^{-3}$) \\ 
\hline
\hline
70.02   & 3.69605  $^{ + 0.00005 }_{ - 0.00008 }$ & -1319.1472 $^{ + 0.0004  }_{ - 0.0004 }$   & -0.012 $^{ + 0.037  }_{ - 0.034 }$  & -0.013 $^{ + 0.025  }_{ - 0.032 }$  & $<$ 17 & $<$ 16 &  1.71  $^{ + 0.05 }_{ - 0.03 }$  & $<$ 18  \\ 
 70.04   &  6.09844  $^{ + 0.00041 }_{ - 0.00041 }$ &  -1317.5396  $^{ + 0.0017 }_{ - 0.0017 }$ & 0.005  $^{ + 0.030 }_{ - 0.031 }$ & -0.003  $^{ + 0.029 }_{ - 0.030 }$ & $<$ 6.6 & $<$ 6.0 &  0.81  $^{ + 0.04 }_{ - 0.03 }$  & $<$ 68  \\ 
 70.01   & 10.85434  $^{ + 0.00018 }_{ - 0.00015 }$  & -1312.0217  $^{ + 0.0003 }_{ - 0.0003 }$  & -0.035  $^{ + 0.036 }_{ - 0.030 }$ &  0.005  $^{ + 0.027 }_{ - 0.021 }$  & $<$ 36 & $<$ 34 &  2.87  $^{ + 0.15 }_{ - 0.07 }$   & $<$ 8.3  \\ 
 70.05   & 19.57800  $^{ + 0.00115 }_{ - 0.00086 }$  & -1307.4195  $^{ + 0.0038 }_{ - 0.0027 }$  & -0.013  $^{ + 0.029 }_{ - 0.030 }$ &  -0.008  $^{ + 0.024 }_{ - 0.026 }$  & $<$ 6.0 & $<$ 5.5 &  0.86  $^{ + 0.05 }_{ - 0.03 }$   & $<$ 49  \\
 70.03   & 77.61201  $^{ + 0.00029 }_{ - 0.00028 }$  & -1303.7691  $^{ + 0.0007 }_{ - 0.0007 }$  & -0.029  $^{ + 0.068 }_{ - 0.065 }$ &  0.036  $^{ + 0.066 }_{ - 0.086 }$  & $<$ 145 & $<$ 133 &  2.49  $^{ + 0.07 }_{ - 0.05 }$   & $<$ 48  \\ 
\hline
  82.05   & 5.2868  $^{ + 0.0001 }_{ - 0.0001 }$ & -1317.4130 $^{ + 0.0037  }_{ - 0.0027 }$   & -0.053 $^{ + 0.046  }_{ - 0.053 }$  & 0.076 $^{ + 0.046  }_{ - 0.050 }$  & $<$ 4.2 & $<$ 3.2 &  0.51  $^{ + 0.10 }_{ - 0.05 }$  & $<$ 246  \\ 
 82.04   &  7.0713  $^{ + 0.0003 }_{ - 0.0002 }$ &  -1319.8768  $^{ + 0.0023 }_{ - 0.0020 }$ & -0.044  $^{ + 0.040 }_{ - 0.045 }$ & 0.067  $^{ + 0.039 }_{ - 0.046 }$ & $<$ 4.9 & $<$ 3.7 &  0.63  $^{ + 0.18 }_{ - 0.05 }$  & $<$ 401  \\ 
 82.02 & 10.3123  $^{ + 0.0008 }_{ - 0.0005 }$ & -1311.0958  $^{ + 0.0014 }_{ - 0.0013 }$   & -0.026  $^{ + 0.057 }_{ - 0.062 }$  &  0.044  $^{ + 0.041 }_{ - 0.047 }$  & 0.76  $^{ + 0.51 }_{ - 0.36 }$  & 0.59  $^{ + 0.40 }_{ - 0.28 }$ &  1.34  $^{ + 0.03 }_{ - 0.12 }$   & 1.3  $^{ + 0.9 }_{ - 0.6 }$   \\ 
 82.01   & 16.1459  $^{ + 0.0002 }_{ - 0.0002 }$  & -1305.6906  $^{ + 0.0004 }_{ - 0.0005 }$  & 0.001  $^{ + 0.032 }_{ - 0.036 }$ &  0.001  $^{ + 0.029 }_{ - 0.038 }$  & $<$ 5.5 & $<$ 4.2 &  2.48  $^{ + 0.04 }_{ - 0.21 }$   & $<$ 1.6  \\ 
 82.06   & 22.4108  $^{ + 0.0012 }_{ - 0.0009 }$  & -1302.4960  $^{ + 0.0048 }_{ - 0.0040 }$  & -0.021  $^{ + 0.043 }_{ - 0.031 }$ &  -0.009  $^{ + 0.033 }_{ - 0.039 }$  & $<$ 0.78 & $<$ 0.60 &  0.75  $^{ + 0.05 }_{ - 0.05 }$   & $<$ 8.4 \\ 
 82.03   & 27.4546  $^{ + 0.0010 }_{ - 0.0007 }$  & -1308.1785  $^{ + 0.0024 }_{ - 0.0028 }$  & 0.003  $^{ + 0.033 }_{ - 0.034 }$ &  -0.004  $^{ + 0.031 }_{ - 0.037 }$  & $<$ 1.2 & $<$ 0.90 &  0.95  $^{ + 0.11 }_{ - 0.10 }$   & $<$ 6.3  \\ 
\hline
 137.03   & 3.50470  $^{ + 0.00008 }_{ - 0.00005 }$ & -1318.5278 $^{ + 0.0005  }_{ - 0.0005 }$   & -0.005 $^{ + 0.033  }_{ - 0.039 }$  & 0.018 $^{ + 0.045  }_{ - 0.035 }$  & $<$ 36 & $<$ 34 &  1.76  $^{ + 0.17 }_{ - 0.15 }$  & $<$ 36  \\ 
 137.01 & 7.64149  $^{ + 0.00019 }_{ - 0.00026 }$  & -1313.2815  $^{ + 0.0004 }_{ - 0.0004 }$   & 0.002  $^{ + 0.004 }_{ - 0.004 }$   & -0.031  $^{ + 0.016 }_{ - 0.014 }$  &  6.4  $^{ + 3.4 }_{ - 1.6 }$  &  6.29  $^{ + 3.29 }_{ - 1.59 }$ &  4.26  $^{ + 0.08 }_{ - 0.07 }$  & 0.45  $^{ + 0.24 }_{ - 0.11 }$ \\ 
 137.02 & 14.85905  $^{ + 0.00062 }_{ - 0.00026 }$ & -1310.7611  $^{ + 0.0006 }_{ - 0.0003 }$   & -0.004  $^{ + 0.016 }_{ - 0.016 }$  &  -0.031  $^{ + 0.023 }_{ - 0.029 }$  & 8.9  $^{ + 3.0 }_{ - 1.7 }$  & 8.73  $^{ + 2.95 }_{ - 1.66 }$ &  5.16  $^{ + 0.10 }_{ - 0.11 }$   & 0.35  $^{ + 0.12 }_{ - 0.07 }$   \\ 
 \hline
244.02   & 6.23826  $^{ + 0.00005 }_{ - 0.00005 }$ & -1315.2951 $^{ + 0.0003  }_{ - 0.0004 }$   & 0.020 $^{ + 0.033  }_{ - 0.020 }$  & 0.063 $^{ + 0.059  }_{ - 0.037 }$  & $<$ 2.5 & $<$ 2.6 &  2.78  $^{ + 0.06 }_{ - 0.06 }$  & $<$ 0.66  \\ 
 244.01 & 12.72062  $^{ + 0.00005 }_{ - 0.00005 }$  & -1314.2915  $^{ + 0.0001 }_{ - 0.0001 }$   & 0.010  $^{ + 0.042 }_{ - 0.033 }$   & 0.018  $^{ + 0.075 }_{ - 0.046 }$  &  2.4  $^{ + 2.0 }_{ - 1.2 }$  &  2.52  $^{ + 2.07 }_{ - 1.22 }$ &  5.23  $^{ + 0.10 }_{ - 0.38 }$  & 0.10  $^{ + 0.08 }_{ - 0.05 }$ \\ 
\hline
245.03 & 13.36812 $^{ + 0.00075  }_{ - 0.00074 }$  & -1314.5933 $^{ + 0.0074  }_{ - 0.0073 }$    & 0.152 $^{ + 0.052  }_{ - 0.047 }$    & -0.124 $^{ + 0.022  }_{ - 0.024 }$   & 0.68 $^{ + 0.46  }_{ - 0.23 }$  & 0.52 $^{ + 0.36  }_{ - 0.18 }$ &  0.28  $^{ + 0.03 }_{ - 0.02 }$   & 133 $^{ + 118  }_{ - 56 }$  \\ 
 245.02   &  21.30001  $^{ + 0.00072 }_{ - 0.00068 }$ &  -1314.8005  $^{ + 0.0021 }_{ - 0.0022 }$ & 0.025  $^{ + 0.010 }_{ - 0.009 }$ & 0.005  $^{ + 0.011 }_{ - 0.015 }$ & $<$ 1.6 & $<$ 1.2 &  0.74  $^{ + 0.03 }_{ - 0.02 }$  & $<$ 17  \\ 
 245.01 & 39.79212  $^{ + 0.00007 }_{ - 0.00005 }$ & -1315.2820  $^{ + 0.0003 }_{ - 0.0003 }$   & 0.040  $^{ + 0.013 }_{ - 0.012 }$  &  -0.050  $^{ + 0.009 }_{ - 0.009 }$  & 44  $^{ + 13 }_{ - 12 }$  & 34  $^{ + 11 }_{ - 9 }$ &  1.95  $^{ + 0.05 }_{ - 0.04 }$   & 25  $^{ + 8 }_{ - 7 }$   \\ 
\hline
 282.02   & 8.45738  $^{ + 0.00016 }_{ - 0.00025 }$ & -1317.5050 $^{ + 0.0012  }_{ - 0.0012 }$   & 0.027 $^{ + 0.072  }_{ - 0.066 }$  & -0.027 $^{ + 0.060  }_{ - 0.058 }$  & $<$ 190 & $<$ 193 &  1.01  $^{ + 0.02 }_{ - 0.05 }$  & $<$ 1036  \\ 
 282.01   &  27.50860  $^{ + 0.00176 }_{ - 0.00065 }$ &  -1297.1777  $^{ + 0.0019 }_{ - 0.0010 }$ & 0.033  $^{ + 0.027 }_{ - 0.033 }$ & 0.004  $^{ + 0.021 }_{ - 0.016 }$ & $<$ 302 & $<$ 309 &  2.93  $^{ + 0.18 }_{ - 0.12 }$  & $<$ 73  \\ 
 282.03 & 87.52747  $^{ + 0.01002 }_{ - 0.00568 }$ & -1265.6949  $^{ + 0.0104 }_{ - 0.0045 }$   & 0.039  $^{ + 0.103 }_{ - 0.183 }$  &  0.066  $^{ + 0.063 }_{ - 0.049 }$  & 25  $^{ + 14 }_{ - 11 }$  & 25  $^{ + 15 }_{ - 11 }$ &  1.82  $^{ + 0.15 }_{ - 0.58 }$   & 23  $^{ + 16 }_{ - 11 }$   \\ 
 \hline
  377.01 & 19.24750 $^{ + 0.00002  }_{ - 0.00002 }$  & -1309.9972 $^{ + 0.0002  }_{ - 0.0002 }$    & 0.055 $^{ + 0.001  }_{ - 0.001 }$    & 0.0123 $^{ + 0.0002  }_{ - 0.0002 }$   & 45.4 $^{ + 0.6  }_{ - 0.6 }$  & 46.5 $^{ + 2.2  }_{ - 2.3 }$ &  7.91  $^{ + 0.16 }_{ - 0.16 }$   & 0.52 $^{ + 0.04  }_{ - 0.04 }$  \\ 
 377.02 & 38.94403  $^{ + 0.00008 }_{ - 0.00008 }$  & -1292.0357  $^{ + 0.0004 }_{ - 0.0004 }$   & -0.0639  $^{ + 0.0001 }_{ - 0.0001 }$   & 0.0022  $^{ + 0.0002 }_{ - 0.0002 }$  &  31.3  $^{ + 0.4 }_{ - 0.4 }$  &  32.0  $^{ + 1.5 }_{ - 1.6 }$ &  8.14  $^{ + 0.18 }_{ - 0.18 }$  & 0.33  $^{ + 0.03 }_{ - 0.02 }$ \\ 
 \hline
  620.01 & 45.15393 $^{ + 0.00036  }_{ - 0.00038 }$  & -1285.4040 $^{ + 0.0006  }_{ - 0.0006 }$    & -0.019 $^{ + 0.009  }_{ - 0.010 }$    & -0.059 $^{ + 0.019  }_{ - 0.022 }$   & 2.8 $^{ + 1.4  }_{ - 1.2 }$  & 2.48 $^{ + 1.23  }_{ - 1.04 }$ &  6.62  $^{ + 0.19 }_{ - 0.17 }$   & 0.05 $^{ + 0.02  }_{ - 0.02 }$  \\ 
 620.03 & 85.31553  $^{ + 0.00138 }_{ - 0.00109 }$  & -1274.4886  $^{ + 0.0030 }_{ - 0.0030 }$   & 0.024  $^{ + 0.014 }_{ - 0.014 }$   & -0.048  $^{ + 0.027 }_{ - 0.029 }$  &  3.52  $^{ + 0.51 }_{ - 0.51 }$  &  3.14  $^{ + 0.50 }_{ - 0.48 }$ &  8.98  $^{ + 2.84 }_{ - 2.84 }$  & 0.02  $^{ + 0.05 }_{ - 0.01 }$ \\ 
 620.02 & 130.1827  $^{ + 0.0009 }_{ - 0.0009 }$ & -1304.0693  $^{ + 0.0009 }_{ - 0.0009 }$   & 0.014  $^{ + 0.011 }_{ - 0.011 }$  &  -0.037  $^{ + 0.022 }_{ - 0.024 }$  & 5.83  $^{ + 1.27 }_{ - 1.15 }$  & 5.22  $^{ + 1.17 }_{ - 1.07 }$ &  9.04  $^{ + 0.25 }_{ - 0.23 }$   & 0.04  $^{ + 0.01 }_{ - 0.01 }$   \\ 
 \hline
 806.03 & 29.35527 $^{ + 0.00043  }_{ - 0.00042 }$  & -1311.7788 $^{ + 0.0026  }_{ - 0.0027 }$    & -0.0079 $^{ + 0.0006  }_{ - 0.0006 }$    & 0.0326 $^{ + 0.0004  }_{ - 0.0004 }$   & 9.47 $^{ + 0.15  }_{ - 0.15 }$  & 8.63 $^{ + 0.42  }_{ - 0.41 }$ &  1.89  $^{ + 0.10 }_{ - 0.16 }$   & 7.1 $^{ + 1.3  }_{ - 1.0 }$  \\ 
 806.02 & 60.31922  $^{ + 0.00024 }_{ - 0.00023 }$  & -1319.8839  $^{ + 0.0003 }_{ - 0.0003 }$   & 0.0087  $^{ + 0.0006 }_{ - 0.0006 }$   & -0.0084  $^{ + 0.0005 }_{ - 0.0005 }$  &  546  $^{ + 5 }_{ - 5 }$  &  498  $^{ + 24 }_{ - 23 }$ &  12.03  $^{ + 0.41 }_{ - 0.40 }$  & 1.57  $^{ + 0.19 }_{ - 0.16 }$ \\ 
 806.01 & 143.4992  $^{ + 0.0036 }_{ - 0.0037 }$ & -1296.7204  $^{ + 0.0011 }_{ - 0.0011 }$   & -0.0262  $^{ + 0.0013 }_{ - 0.0013 }$  &  -0.0078  $^{ + 0.0013 }_{ - 0.0013 }$  & 20.1  $^{ + 1.4 }_{ - 1.4 }$  & 18.3  $^{ + 1.5 }_{ - 1.5 }$ &  8.79  $^{ + 0.49 }_{ - 0.31 }$   & 0.15  $^{ + 0.03 }_{ - 0.02 }$   \\ 
 \hline
  1353.02   & 34.54304  $^{ + 0.00101 }_{ - 0.00096 }$ & -1308.8883 $^{ + 0.0024  }_{ - 0.0022 }$   & -0.003 $^{ + 0.030  }_{ - 0.027 }$  & -0.061 $^{ + 0.036  }_{ - 0.040 }$  & $<$ 13.5 & $<$ 14.2 &  2.41  $^{ + 0.41 }_{ - 0.34 }$  & $<$ 8.5  \\ 
 K289d & 65.95930  $^{ + 0.00145 }_{ - 0.00168 }$  & -1298.0481  $^{ + 0.0028 }_{ - 0.0028 }$   & -0.003  $^{ + 0.001 }_{ - 0.001 }$   & -0.028  $^{ + 0.007 }_{ - 0.009 }$  &  2.89  $^{ + 0.50 }_{ - 0.44 }$  &  3.04  $^{ + 0.55 }_{ - 0.47 }$ &  2.65  $^{ + 1.08 }_{ - 0.19 }$  & 0.9  $^{ + 3.4 }_{ - 0.6 }$ \\ 
 1353.01 & 125.8697  $^{ + 0.0007 }_{ - 0.0005 }$ & -1301.0103  $^{ + 0.0004 }_{ - 0.0003 }$   & 0.009  $^{ + 0.005 }_{ - 0.005 }$  &  -0.002  $^{ + 0.008 }_{ - 0.011 }$  & 116  $^{ + 11 }_{ - 11 }$  & 122  $^{ + 12 }_{ - 13 }$ &  11.14  $^{ + 0.23 }_{ - 0.23 }$   & 0.48  $^{ + 0.06 }_{ - 0.06 }$   \\ 
 \hline
  1783.01 & 134.4629 $^{ + 0.0034  }_{ - 0.0035 }$  & -1190.8662 $^{ + 0.0039  }_{ - 0.0038 }$    & 0.007 $^{ + 0.007  }_{ - 0.005 }$    & -0.047 $^{ + 0.014  }_{ - 0.016 }$   & 78.9 $^{ + 16.2  }_{ - 13.2 }$  & 89 $^{ + 19  }_{ - 16 }$ &  9.03  $^{ + 0.23 }_{ - 0.24 }$   & 0.67 $^{ + 0.16  }_{ - 0.13 }$  \\ 
 1783.02 & 284.2162  $^{ + 0.0280 }_{ - 0.0215 }$  & -1113.8267  $^{ + 0.0292 }_{ - 0.0217 }$   & 0.018  $^{ + 0.015 }_{ - 0.014 }$   & -0.018  $^{ + 0.034 }_{ - 0.029 }$  &  16.3  $^{ + 4.8 }_{ - 3.9 }$  &  18.5  $^{ + 5.5 }_{ - 4.4 }$ &  5.46  $^{ + 0.51 }_{ - 0.30 }$  & 0.63  $^{ + 0.31 }_{ - 0.20 }$ \\ 
\hline
     \end{tabular}    
    \caption{Posteriors Summarized for Orbital Parameters, Dynamical Masses, and Bulk Planet Properties at the Epoch BJD=2,455,680 (Noted as BJD$-$2,457,000), With Measured Times from \tess\ in Addition to \kepler\ Light-curve Transit Times \citep{Rowe2015a} (with Some Exceptions Noted in \S 3.1). Note. The columns denote orbital period in days, the first transit time after epoch, eccentricity vector components, dynamical mass scaled to the Earth-Sun system, planetary masses, radii and bulk densities. Summary statistics of the posteriors include the median and 68.3\% confidence intervals with equal weight in the tails. Upper limits list the 97.7th percentile of the posterior distribution. Note Kepler-289 d (PH3 c, orbiting KOI-1353) does not have a KOI number.  
 }\label{tbl-ttv} 
  \end{center}
\end{table}

KOI-70 (Kepler-20) has five transiting planets. This system was selected because of a near 3:1 resonance between KOI-70.01 and KOI-70.02, with an expected periodicity $\approx$ 171 days, although no TTVs were reported by \citet{Holczer2016}. \citet{Buchhave2016} discovered a nontransiting planet in RV with an orbital period of 34.94 days. This additional planet, while between Kepler-20 c and Kepler-20 d, is far from any low-order resonance with other planets and is unlikely to cause detectable TTV. Indeed, no TTVs were found in this system by \citet{Buchhave2016}, and we exclude this additional planet from our analysis.

The two relatively precise measured transit times from \tess\ are consistent with a linear ephemeris for KOI-70.01, and the third transit time is too uncertain to have an effect on the models. The outer pair at KOI-70 are close to the 4:1 commensurability, with an expected TTV periodicity of $\approx$2171 days. This can be seen in the projected TTV of both planets, leading to the prospect of more meaningful upper limits on their masses with future data, although both planets are too small to be detected in \tess\ light curves.  

KOI-82 (Kepler-102) has six known planet candidates. The overall configuration is more compact than Kepler-11 \citep{Lissauer2011a}, requiring low masses for stability. However, there are few adjacent pairs near resonance, making the TTV signals weak. Our upper limits in mass for most of the planets at Kepler-102 are uninformative, given the planetary radii. However, our results for KOI-82.02 (Kepler-102 d) and KOI-82.01 (Kepler-102 e), as well as revised planet radius measurements, provide evidence for lower bulk densities than found from the weak RV signal observed by \citet{Marcy2014}. 

KOI-137 (Kepler-18) has three transiting planets with planetary masses measured by a combined RV and TTV data set \citep{Cochran2011}. Kepler-18 b has no evidence of interaction with Kepler-18 c, and its mass is poorly constrained from the TTV alone. Kepler-18 c and Kepler-18 d have anticorrelated TTVs with a periodicity $\sim$270 days. The mass constraints following \tess\ agree closely with the results of \citet{Jontof-Hutter2021}. 

KOI-244 (Kepler-25) has two planets with strongly detected anticorrelated TTVs \citep{Lithwick2012}. The TTV periodicity at $\sim$325 days is well sampled by the \kepler\ data set. The masses following \tess\ transit times are in close agreement with the masses following dynamical fits with \kepler-only transit times \citep{Jontof-Hutter2021} and lower than those found via RV \citep{Marcy2014}. \citet{Hadden2017} found that the TTV posteriors are sensitive to priors in eccentricity, and there is closer agreement between the RV and TTV masses with a low-eccentricity prior.

\citet{Jontof-Hutter2021} projected the transit times of Kepler-18 and Kepler-25, and found very little divergence in the decades following the \kepler\ mission, suggesting that \tess\ data would not significantly improve the TTV model. Our results confirm that expectation and agree closely with previous mass estimates. For both of these systems, our results are also closely consistent with \citet{Battley2021}. 

KOI-245 (Kepler-37) did not have a detection of TTVs reported in \citet{Holczer2016}. However, there is an expected TTV periodicity of $\sim$300 days, due to the near 2:1 resonance of Kepler-37 c and Kepler-37 d. There is a fourth candidate (KOI-245.04), with a period of 51.22 days, that is dispositioned in the DR25 Supplement as a false positive. It was earlier identified in the Q6 catalog \citep{Batalha2013} and appears in subsequent \kepler\ catalogs (\citealt{Burke2014,Mullally2015}). \citet{Mazeh2013} published a comprehensive catalog of measured transit times following 12 quarters of \kepler\ light curves and included KOI-245.04, although no significant TTV signals were detected in the system. The candidate was listed as ``confirmed" in the Q16 catalog of \citet{Mullally2015}, possibly based on a 1$\sigma$ detection of near-resonant TTVs by \citet{Hadden2014}, although only its inner neighbor KOI-245.01 has mass constraints derived from the TTV. Given the small size of the putative KOI-245.04 (0.37 $R_{\oplus}$, \citealt{Burke2014}), and the near 5:4 resonant period with KOI-245.01, we estimate the amplitude of TTVs in KOI-245.01 potentially caused by KOI-245.04 to be of order $\sim$10 s, which would be undetectable. Hence, we excluded KOI-245.04 from our dynamical fits. In agreement with \citet{Hadden2014}, we find the bulk density of KOI-245.01 (Kepler-37 d) to be unphysically high. One possible explanation for the high bulk density inference is significant dilution in the \kepler\ light curves causing the planet radii to be underestimated. This appears unlikely; according to the ExoFOP \kepler\ page for KOI-245, the nearest stars in the field are 2 mag fainter\footnote{\url{https://exofop.ipac.caltech.edu/kepler/welcome.php}}. \citet{Marcy2014} reported RV data on KOI-245, and there were no detections, although the upper limit they found for the mass of KOI-245.01 (Kepler-37 d), 12.2 $M_{\oplus}$, is lower than what we find via transit timing. As noted by \citet{Hadden2014} the high dynamical mass from TTV of KOI-245.01 (Kepler-37 d) given the small radius could be caused by orbital eccentricities outside of our prior. 

\citet{Holczer2016} detected TTVs at KOI-282.01 (Kepler-132 c) with a periodicity of $\approx$462 days. These can be attributed to the near 3:1 commensurability with its outer neighbor. A similar period ratio exists between KOI-282.01 and its inner neighbor KOI-282.02 (Kepler-130 b), although that pair is farther from resonance, has a shorter TTV period, and has a lower TTV amplitude. The inner two planets have uninformative upper limits only, while the outer planet appears to be weakly detected at the 2$\sigma$ level. The inferred density is significantly higher than other well-characterized exoplanets. As with KOI-245, we attribute this to either eccentricities that are excluded by our prior, in which a lower mass could cause the TTVs, or an underestimate of the planet radius due to dilution. The latter is unlikely since all the identified nearby stars for this target on ExoFOP Kepler are several magnitudes fainter. The single transit detected in the \tess\ data does little to further constrain the planetary masses. 

KOI-377 (Kepler-9) has two sub-Saturn planets that have strongly detected TTVs and a super-Earth that is dynamically isolated from the interacting pair. We included only the large planets, KOI-377.01 (Kepler-9 b) and KOI-377.02 (Kepler-9 c), in our dynamical models. We did not detect the transits of Kepler-9 in Sector 14 despite the expectation of an S/N $>$9 in the absence of correlated noise for both Kepler-9 b and Kepler-9 c. This may be due to stray-light contamination in S14/S15 (e.g. \citealt{Dalba2020}). For Sectors 40 and 41, we detected transits in the long-cadence data for Kepler-9. Our mass measurements agree closely with \citet{Freudenthal2018}. 

KOI-620 (Kepler-51) has three planets characterized from strong TTV signals \citep{Masuda2014}. All three have extremely low bulk densities, although the size of KOI-620.03 (Kepler-51 c) is poorly constrained owing to an impact parameter above unity. Nevertheless, a minimum size for Kepler-51 c implies a very low upper limit in density, like its neighbors. The detected \tess\ transit times have large uncertainties and do little to further constrain the planetary masses, as found by \citet{Battley2021}. 

KOI-806 (Kepler-30) has high-amplitude TTVs, with an amplitude of 1 day at the innermost planet KOI-806.03, (Kepler-30 b). The TTVs were initial modeled by \citet{Sanchis-Ojeda2012}, before a complete TTV period had been observed. Studies from \citet{Wu2018}, \citet{Panichi2018}, and \citet{Jontof-Hutter2021} refined the TTV model after the \kepler\ mission. Our results agree with these studies. Despite the large TTVs in this system, the \tess\ measurements do not significantly improve the mass posteriors. Both KOI-806.03 (Kepler-30 c) and KOI-806.02 (Kepler-30 d) have low S/N in the \tess\ light curves and large transit timing uncertainties.

KOI-1353 (Kepler-289) has three planets with strongly detected TTVs. We adopted the transit times and measured planet size for Kepler-289 d of \citet{Schmitt2014}, with additional transit times for the other planets from \citet{Jontof-Hutter2021}. The additional transit time during the \tess\ mission is consistent with the projected transit times following the \kepler\ mission. The measured masses are in close agreement with the results following \kepler\ data only of \citet{Jontof-Hutter2021}. 

For KOI-1783 (Kepler-1662), our results are closely consistent with the corresponding models of \citet{Vissapragada2020}. In that study, a single ground-based observation of KOI-1783.01 supplemented the \kepler\ dataset. Here we combined all three datasets. 


\section{Discussion and Conclusions}\label{sec:disc}
For several systems, we have detected transits in \tess\  data, at a significantly lower S/N than in \kepler\ data, with PLD. In all cases, the \kepler\ data are more precise and there are more measurements than in the \tess\ data set. Hence, the dynamical constraints we have found here are mostly driven by the \kepler\ data. Nevertheless, we find strong evidence of an additional perturber at KOI-2672 (Kepler-396). For two very different configurations involving an additional planet, we find closely consistent masses for the two transiting planets, results that are inconsistent with the posteriors for a two-planet model for the system. The transits detected by \tess\ at this system, alongside the \kepler\ data, are a significant improvement over the \kepler\ data alone for the purposes of TTV modeling. If the apparent robustness of the mass measurements of the transiting planets to different configurations with a nontransiting perturber persists with future data and additional modeling, we note that KOI-2672.01 (Kepler-396 c) has a very low density. Its orbital period and low density place it in a similar regime to the lowest density planets characterized via TTVs (e.g. Kepler-51 (\citealt{Masuda2014,Libby-Roberts2020}), Kepler-79 d \citealt{Jontof-Hutter2014,Chachan2020}), but with a brighter host.

With several transits of KOI-94 (Kepler-89) observed by \tess\, we found the one transit of KOI-94.03 (Kepler-89 e) inconsistent at 2$\sigma$ with projections from a range of TTV models that we explored here, with the four known planets, as well as with an additional planet. While the additional planet improved the model fits substantially, we did not find solutions that persisted between different five-planet models, nor did we find solutions that were consistent with the strong detection of KOI-94.01 (Kepler-89 d) in RV.  

The choice to apply PLD to \tess\ photometry borrows from extensive experience with the noise properties of photometry taken by the {\it Spitzer Space Telescope} \citep[e.g.,][]{Ingalls2012,Deming2015}. The large pixels in the \tess\ spacecraft's detectors result in undersampled PSFs for relatively faint stars. This allows any variation in pixel sensitivity, coupled with random pointing jitter and systematic drift of the spacecraft, to produce time-correlated noise that obscures astrophysical signals. The \kepler\ systems analyzed here are among the faintest stars for which \tess\ can achieve a reasonable photometric precision, making them optimal test cases for the application of PLD. 
Although in many cases our analysis yielded nondetections, in many others PLD enabled low S/N detections of transits that would have otherwise been hidden in the noise. This powerful and simple tool thereby expands the discovery space beyond the demographic of bright stars for which the \tess\ mission is optimized. Here, we have shown the benefits of this technique for the sample of \kepler\ planets exhibiting TTVs. Beyond these planets, though, the application of PLD to extract transits from noisy \tess\ data has several useful applications. Staying with \kepler, the longest-period transiting systems only exhibit a few transits \citep[e.g.,][]{Kipping2016b,Dalba2021c}, precluding robust TTV analyses. If the \tess\ extended mission continues, transits of these planets will occasionally occur during \tess\ observations. These will identify (or rule out) the presence of TTVs and  eventually enable new dynamical investigations \citep[e.g.,][]{Dalba2019c}. The benefits of PLD extend to planets discovered by \tess\ as well. Transits of those with orbital periods greater than the observational baseline will not show the periodicity that is usually necessary to confidently claim a planet candidate detection (\citealt{Diaz2020,Dalba2022}). For faint stars and/or shallow transits (i.e., small planets), these planets may go undetected or at least not validated. In these cases, PLD will enable careful searches for additional transit signals that will fill in the parameter space of planet discoveries made by the \tess\ mission.

Despite different host star selection functions, the \kepler\ and \tess\ missions can be highly complementary. With PLD, we have unlocked \tess's ability to detect transits in noisy data and refined the properties of some of \kepler's most interesting planetary systems.

%\acknowledgements
We thank an anonymous referee for comments that improved this paper. D.J.-H. acknowledges grant No. 80NSSC20K1008 from the \tess\ GI program. P.D. is supported by a National Science Foundation (NSF) Astronomy and Astrophysics Postdoctoral Fellowship under award AST-1903811. This paper includes data collected with the \tess\ mission, obtained from the MAST data archive at the Space Telescope Science Institute (STScI). Funding for the \tess\ mission is provided by the NASA Explorer Program. STScI is operated by the Association of Universities for Research in Astronomy, Inc., under NASA contract NAS 5–26555.

This work made use of \texttt{tpfplotter} by J. Lillo-Box (publicly available in www.github.com/jlillo/tpfplotter), which also made use of the Python packages \texttt{astropy}, \texttt{lightkurve}, \texttt{matplotlib}, and \texttt{numpy}.

\vspace{5mm}
\facilities{\tess, \kepler }
%facilities list https://journals.aas.org/facility-keywords/

\software{   \texttt{astropy} \citep{astropy2013,astropy2018},\\
                %\texttt{BATMAN} \citep{Kreidberg2015}, \\
                %\texttt{corner} \citep{ForemanMackey2016a},\\
                \texttt{emcee} \citep{ForemanMackey2013}, \\
                %\texttt{EXOFASTv2} \citep{Eastman2013,Eastman2017,Eastman2019}, \\
                %\texttt{Exo-Transmit} \citep{Kempton2017}, \\
                %\texttt{forecaster} \citep{Chen2017}
                \texttt{Lightkurve} \citep{Lightkurve2018}, \\
                %\texttt{PyKE} \citep{Still2012}, \\
                %\texttt{RadVel} \citep{Fulton2018}
                %\texttt{RETrO} \citep{Dalba2017b}, \\
                %\texttt{SpecMatch} \citep{Petigura2015,Petigura2017}, \\
                %\texttt{SpecMatch-Emp} \citep{Yee2017}, \\
                %\texttt{The Joker} \citep{PriceWhelan2017},
                \texttt{tpfplotter} \citep{Aller2020}
                }

\clearpage
\section*{Appendix: Projected Transit Times for (KOI-2672) Kepler-396.}
We list projected transit times and uncertainties for Kepler-396 b and c in Table~\ref{tbl-predictions}. 
 \begin{table}[h!]
 \tiny
  \begin{center}
    \begin{tabular}{|c|c|c|c|c|c|}
 \hline
 2638.8191 $\pm$       0.0066  &     3498.6522 $\pm$       0.0067 &     4358.5536 $\pm$       0.0093   &     5218.4140 $\pm$       0.0096   &     6078.2823 $\pm$       0.0124   &     6938.1826 $\pm$       0.0141  \\ 
        2681.8124 $\pm$       0.0066  &     3541.6482 $\pm$       0.0066 &     4401.5431 $\pm$       0.0094   &     5261.4116 $\pm$       0.0097   &     6121.2726 $\pm$       0.0124   &     6981.1784 $\pm$       0.0144  \\ 
        2724.8033 $\pm$       0.0067  &     3584.6441 $\pm$       0.0065 &     4444.5326 $\pm$       0.0094   &     5304.4087 $\pm$       0.0099   &     6164.2635 $\pm$       0.0125   &     7024.1760 $\pm$       0.0148  \\ 
        2767.7946 $\pm$       0.0066  &     3627.6407 $\pm$       0.0065 &     4487.5224 $\pm$       0.0095   &     5347.4073 $\pm$       0.0103   &     6207.2549 $\pm$       0.0126   &     7067.1702 $\pm$       0.0150  \\ 
        2810.7848 $\pm$       0.0067  &     3670.6376 $\pm$       0.0065 &     4530.5119 $\pm$       0.0095   &     5390.4039 $\pm$       0.0105   &     6250.2468 $\pm$       0.0126   &     7110.1659 $\pm$       0.0152  \\ 
        2853.7747 $\pm$       0.0066  &     3713.6345 $\pm$       0.0066 &     4573.5020 $\pm$       0.0095   &     5433.4024 $\pm$       0.0110   &     6293.2394 $\pm$       0.0126   &     7153.1584 $\pm$       0.0153  \\ 
        2896.7643 $\pm$       0.0067  &     3756.6321 $\pm$       0.0067 &     4616.4922 $\pm$       0.0096   &     5476.3983 $\pm$       0.0113   &     6336.2328 $\pm$       0.0127   &     7196.1518 $\pm$       0.0154  \\ 
        2939.7539 $\pm$       0.0067  &     3799.6294 $\pm$       0.0069 &     4659.4832 $\pm$       0.0096   &     5519.3957 $\pm$       0.0116   &     6379.2267 $\pm$       0.0127   &     7239.1428 $\pm$       0.0154  \\ 
        2982.7434 $\pm$       0.0067  &     3842.6277 $\pm$       0.0072 &     4702.4745 $\pm$       0.0096   &     5562.3899 $\pm$       0.0118   &     6422.2210 $\pm$       0.0126   &     7282.1341 $\pm$       0.0154  \\ 
        3025.7330 $\pm$       0.0067  &     3885.6244 $\pm$       0.0075 &     4745.4665 $\pm$       0.0096   &     5605.3855 $\pm$       0.0120   &     6465.2163 $\pm$       0.0127   &     7325.1244 $\pm$       0.0155  \\ 
        3068.7231 $\pm$       0.0067  &     3928.6229 $\pm$       0.0080 &     4788.4594 $\pm$       0.0097   &     5648.3780 $\pm$       0.0121   &     6508.2114 $\pm$       0.0126   &     7368.1144 $\pm$       0.0155  \\ 
        3111.7133 $\pm$       0.0067  &     3971.6185 $\pm$       0.0082 &     4831.4525 $\pm$       0.0097   &     5691.3713 $\pm$       0.0121   &     6551.2075 $\pm$       0.0126   &     7411.1040 $\pm$       0.0156  \\ 
        3154.7042 $\pm$       0.0067  &     4014.6160 $\pm$       0.0086 &     4874.4466 $\pm$       0.0097   &     5734.3622 $\pm$       0.0121   &     6594.2035 $\pm$       0.0125   &     7454.0937 $\pm$       0.0156  \\ 
        3197.6954 $\pm$       0.0067  &     4057.6101 $\pm$       0.0088 &     4917.4410 $\pm$       0.0097   &     5777.3537 $\pm$       0.0121   &     6637.2002 $\pm$       0.0125   &     7497.0833 $\pm$       0.0157  \\ 
        3240.6876 $\pm$       0.0068  &     4100.6056 $\pm$       0.0090 &     4960.4361 $\pm$       0.0097   &     5820.3437 $\pm$       0.0122   &     6680.1970 $\pm$       0.0125   &     7540.0730 $\pm$       0.0157  \\ 
        3283.6802 $\pm$       0.0068  &     4143.5980 $\pm$       0.0091 &     5003.4313 $\pm$       0.0096   &     5863.3337 $\pm$       0.0121   &     6723.1940 $\pm$       0.0126   &     7583.0632 $\pm$       0.0158  \\ 
        3326.6734 $\pm$       0.0068  &     4186.5911 $\pm$       0.0092 &     5046.4274 $\pm$       0.0096   &     5906.3235 $\pm$       0.0123   &     6766.1919 $\pm$       0.0128   &     7626.0534 $\pm$       0.0159  \\ 
        3369.6676 $\pm$       0.0068  &     4229.5822 $\pm$       0.0093 &     5089.4234 $\pm$       0.0095   &     5949.3129 $\pm$       0.0122   &     6809.1889 $\pm$       0.0130   &     7669.0444 $\pm$       0.0159  \\ 
        3412.6618 $\pm$       0.0067  &     4272.5735 $\pm$       0.0093 &     5132.4201 $\pm$       0.0095   &     5992.3026 $\pm$       0.0123   &     6852.1873 $\pm$       0.0133   &     7712.0356 $\pm$       0.0159  \\ 
        3455.6570 $\pm$       0.0067  &     4315.5634 $\pm$       0.0093 &     5175.4168 $\pm$       0.0095   &     6035.2923 $\pm$       0.0124   &     6895.1842 $\pm$       0.0137   &     7755.0278 $\pm$       0.0160  \\ 
\hline
  2618.3110 $\pm$       0.0154   &     3503.5568 $\pm$       0.0930   &     4388.6804 $\pm$       0.0922   &     5273.7746 $\pm$       0.0737   &     6158.9297 $\pm$       0.1586   &     7043.9941 $\pm$       0.2790  \\ 
        2706.8330 $\pm$       0.0221   &     3592.0599 $\pm$       0.0945   &     4477.2046 $\pm$       0.0939   &     5362.2726 $\pm$       0.0723   &     6247.4458 $\pm$       0.1701   &     7132.5141 $\pm$       0.2927  \\ 
        2795.3549 $\pm$       0.0327   &     3680.5607 $\pm$       0.0947   &     4565.7288 $\pm$       0.0948   &     5450.7678 $\pm$       0.0750   &     6335.9734 $\pm$       0.1819   &     7221.0397 $\pm$       0.3078  \\ 
        2883.8846 $\pm$       0.0447   &     3769.0571 $\pm$       0.0966   &     4654.2630 $\pm$       0.0856   &     5539.2667 $\pm$       0.0799   &     6424.4880 $\pm$       0.1951   &     7309.5599 $\pm$       0.3199  \\ 
        2972.4171 $\pm$       0.0555   &     3857.5549 $\pm$       0.0986   &     4742.7859 $\pm$       0.0799   &     5627.7837 $\pm$       0.0911   &     6512.9948 $\pm$       0.2091   &     7398.0872 $\pm$       0.3310  \\ 
        3060.9444 $\pm$       0.0679   &     3946.0744 $\pm$       0.0913   &     4831.3012 $\pm$       0.0758   &     5716.3039 $\pm$       0.1017   &     6601.4898 $\pm$       0.2220   &     7486.6160 $\pm$       0.3380  \\ 
        3149.4678 $\pm$       0.0789   &     4034.5913 $\pm$       0.0901   &     4919.8064 $\pm$       0.0755   &     5804.8281 $\pm$       0.1129   &     6689.9860 $\pm$       0.2344   &     7575.1392 $\pm$       0.3447  \\ 
        3238.0094 $\pm$       0.0797   &     4123.1132 $\pm$       0.0923   &     5008.3058 $\pm$       0.0734   &     5893.3507 $\pm$       0.1247   &     6778.4835 $\pm$       0.2448   &     7663.6595 $\pm$       0.3512  \\ 
        3326.5359 $\pm$       0.0841   &     4211.6287 $\pm$       0.0923   &     5096.7993 $\pm$       0.0722   &     5981.8774 $\pm$       0.1369   &     6866.9775 $\pm$       0.2560   &     7752.1916 $\pm$       0.3548  \\ 
        3415.0538 $\pm$       0.0893   &     4300.1534 $\pm$       0.0920   &     5185.2862 $\pm$       0.0726   &     6070.4075 $\pm$       0.1472   &     6955.4738 $\pm$       0.2677   &     7840.7109 $\pm$       0.3603  \\ 
  \hline
  \hline
   2638.7853 $\pm$       0.0071  &     3498.6611 $\pm$       0.0253 &     4358.7198 $\pm$       0.0630   &     5218.7414 $\pm$       0.1088   &     6078.6147 $\pm$       0.2510   &     6938.4049 $\pm$       0.4725  \\ 
        2681.7773 $\pm$       0.0073  &     3541.6617 $\pm$       0.0269 &     4401.7200 $\pm$       0.0651   &     5261.7443 $\pm$       0.1114   &     6121.6008 $\pm$       0.2651   &     6981.4006 $\pm$       0.4903  \\ 
        2724.7685 $\pm$       0.0076  &     3584.6642 $\pm$       0.0286 &     4444.7203 $\pm$       0.0671   &     5304.7456 $\pm$       0.1141   &     6164.5858 $\pm$       0.2797   &     7024.3974 $\pm$       0.5097  \\ 
        2767.7603 $\pm$       0.0080  &     3627.6665 $\pm$       0.0302 &     4487.7197 $\pm$       0.0692   &     5347.7476 $\pm$       0.1166   &     6207.5718 $\pm$       0.2945   &     7067.3940 $\pm$       0.5307  \\ 
        2810.7512 $\pm$       0.0084  &     3670.6704 $\pm$       0.0319 &     4530.7196 $\pm$       0.0713   &     5390.7466 $\pm$       0.1185   &     6250.5573 $\pm$       0.3097   &     7110.3912 $\pm$       0.5533  \\ 
        2853.7421 $\pm$       0.0087  &     3713.6728 $\pm$       0.0337 &     4573.7188 $\pm$       0.0733   &     5433.7468 $\pm$       0.1205   &     6293.5439 $\pm$       0.3193   &     7153.3872 $\pm$       0.5773  \\ 
        2896.7326 $\pm$       0.0092  &     3756.6771 $\pm$       0.0355 &     4616.7187 $\pm$       0.0754   &     5476.7459 $\pm$       0.1215   &     6336.5302 $\pm$       0.3266   &     7196.3842 $\pm$       0.6024  \\ 
        2939.7240 $\pm$       0.0098  &     3799.6804 $\pm$       0.0372 &     4659.7179 $\pm$       0.0775   &     5519.7449 $\pm$       0.1232   &     6379.5176 $\pm$       0.3352   &     7239.3809 $\pm$       0.6284  \\ 
        2982.7147 $\pm$       0.0105  &     3842.6853 $\pm$       0.0391 &     4702.7181 $\pm$       0.0796   &     5562.7398 $\pm$       0.1256   &     6422.5054 $\pm$       0.3438   &     7282.3778 $\pm$       0.6555  \\ 
        3025.7060 $\pm$       0.0111  &     3885.6886 $\pm$       0.0410 &     4745.7180 $\pm$       0.0818   &     5605.7352 $\pm$       0.1300   &     6465.4938 $\pm$       0.3527   &     7325.3736 $\pm$       0.6833  \\ 
        3068.6972 $\pm$       0.0120  &     3928.6931 $\pm$       0.0429 &     4788.7189 $\pm$       0.0840   &     5648.7293 $\pm$       0.1362   &     6508.4829 $\pm$       0.3617   &     7368.3703 $\pm$       0.7120  \\ 
        3111.6901 $\pm$       0.0130  &     3971.6967 $\pm$       0.0448 &     4831.7197 $\pm$       0.0862   &     5691.7229 $\pm$       0.1439   &     6551.4719 $\pm$       0.3708   &     7411.3673 $\pm$       0.7415  \\ 
        3154.6827 $\pm$       0.0140  &     4014.7011 $\pm$       0.0468 &     4874.7213 $\pm$       0.0886   &     5734.7134 $\pm$       0.1526   &     6594.4624 $\pm$       0.3801   &     7454.3654 $\pm$       0.7720  \\ 
        3197.6765 $\pm$       0.0151  &     4057.7044 $\pm$       0.0487 &     4917.7233 $\pm$       0.0909   &     5777.7040 $\pm$       0.1625   &     6637.4518 $\pm$       0.3897   &     7497.3624 $\pm$       0.8037  \\ 
        3240.6707 $\pm$       0.0163  &     4100.7080 $\pm$       0.0508 &     4960.7257 $\pm$       0.0934   &     5820.6935 $\pm$       0.1734   &     6680.4430 $\pm$       0.3993   &     7540.3609 $\pm$       0.8361  \\ 
        3283.6671 $\pm$       0.0177  &     4143.7107 $\pm$       0.0528 &     5003.7286 $\pm$       0.0959   &     5863.6826 $\pm$       0.1850   &     6723.4339 $\pm$       0.4092   &     7583.3596 $\pm$       0.8688  \\ 
        3326.6637 $\pm$       0.0191  &     4186.7137 $\pm$       0.0548 &     5046.7308 $\pm$       0.0984   &     5906.6696 $\pm$       0.1973   &     6766.4264 $\pm$       0.4196   &     7626.3596 $\pm$       0.9025  \\ 
        3369.6613 $\pm$       0.0206  &     4229.7157 $\pm$       0.0569 &     5089.7343 $\pm$       0.1010   &     5949.6569 $\pm$       0.2101   &     6809.4174 $\pm$       0.4304   &     7669.3590 $\pm$       0.9383  \\ 
        3412.6600 $\pm$       0.0221  &     4272.7176 $\pm$       0.0589 &     5132.7367 $\pm$       0.1036   &     5992.6431 $\pm$       0.2234   &     6852.4121 $\pm$       0.4423   &     7712.3604 $\pm$       0.9751  \\ 
        3455.6600 $\pm$       0.0237  &     4315.7186 $\pm$       0.0610 &     5175.7402 $\pm$       0.1063   &     6035.6296 $\pm$       0.2371   &     6895.4080 $\pm$       0.4564   &     7755.3622 $\pm$       1.0133  \\ 
\hline
 2618.2716 $\pm$       0.0126   &      3503.3184 $\pm$       0.0438   &      4388.2456 $\pm$       0.1261   &      5273.2409 $\pm$       0.1844   &      6158.3415 $\pm$       0.2330   &      7043.3611 $\pm$       0.2725 \\ 
        2706.7752 $\pm$       0.0154   &      3591.8029 $\pm$       0.0515   &      4476.7559 $\pm$       0.1325   &      5361.7370 $\pm$       0.1884   &      6246.8475 $\pm$       0.2371   &      7131.8698 $\pm$       0.2774 \\ 
        2795.2828 $\pm$       0.0183   &      3680.2889 $\pm$       0.0594   &      4565.2671 $\pm$       0.1387   &      5450.2369 $\pm$       0.1930   &      6335.3491 $\pm$       0.2413   &      7220.3814 $\pm$       0.2821 \\ 
        2883.7937 $\pm$       0.0211   &      3768.7756 $\pm$       0.0676   &      4653.7776 $\pm$       0.1448   &      5538.7398 $\pm$       0.1979   &      6423.8542 $\pm$       0.2445   &      7308.8966 $\pm$       0.2868 \\ 
        2972.3067 $\pm$       0.0239   &      3857.2642 $\pm$       0.0761   &      4742.2849 $\pm$       0.1507   &      5627.2461 $\pm$       0.2031   &      6512.3543 $\pm$       0.2481   &      7397.4119 $\pm$       0.2910 \\ 
        3060.8198 $\pm$       0.0267   &      3945.7551 $\pm$       0.0847   &      4830.7874 $\pm$       0.1566   &      5715.7563 $\pm$       0.2084   &      6600.8539 $\pm$       0.2514   &      7485.9296 $\pm$       0.2953 \\ 
        3149.3306 $\pm$       0.0293   &      4034.2487 $\pm$       0.0934   &      4919.2843 $\pm$       0.1625   &      5804.2701 $\pm$       0.2137   &      6689.3524 $\pm$       0.2552   &      7574.4523 $\pm$       0.3013 \\ 
        3237.8376 $\pm$       0.0320   &      4122.7450 $\pm$       0.1021   &      5007.7759 $\pm$       0.1684   &      5892.7870 $\pm$       0.2189   &      6777.8517 $\pm$       0.2592   &      7662.9742 $\pm$       0.3071 \\ 
        3326.3372 $\pm$       0.0343   &      4211.2450 $\pm$       0.1109   &      5096.2618 $\pm$       0.1744   &      5981.3061 $\pm$       0.2240   &      6866.3526 $\pm$       0.2634   &      7751.4948 $\pm$       0.3127 \\ 
        3414.8314 $\pm$       0.0366   &      4299.7436 $\pm$       0.1185   &      5184.7463 $\pm$       0.1807   &      6069.8244 $\pm$       0.2286   &      6954.8555 $\pm$       0.2679   &      7840.0111 $\pm$       0.3183 \\ 
\hline
    \end{tabular}    
    \caption{Predicted transit timing posteriors (BJD-2,457,000) for KOI-2672.02 (Kepler-396 b) and KOI-2672.01 (Kepler-396 c) for two possible three-planet models. The upper (lower) two panels correspond to a model with a non-transiting planet orbiting near 141 (57) days. (See Section 3.1 for more detail.)}\label{tbl-predictions}
  \end{center}
\end{table}

\clearpage
\bibliography{references}

\end{document}